\pdfoutput=1
\documentclass[acmsmall,screen,nonacm]{acmart}
\newcommand{\sym}{\mathit{sym}}

\newcommand{\cid}[1]{\texttt{#1}}

\newcommand{\lframe}{\eta}

\newcommand{\vptr}[2]{\mathsf{ptr}(#1, #2)}
\newcommand{\vnull}{\mathsf{null}}
\newcommand{\vundef}{\mathsf{undef}}
\newcommand{\vuptr}{\mathsf{uptr}}

\newcommand{\sem}[1]{[\![#1]\!]}

\newcommand{\rulename}[1]{\textsf{#1}}

\newcommand{\nostepj}[1]{#1\not\leadsto}
\newcommand{\stepj}[2]{#1\leadsto #2}

\newcommand{\lstepj}[4]{#2\xrightarrow{#3}_{#1}#4}
\newcommand{\lstepsj}[4]{#2\xrightarrow{#3\,*}_{#1}#4}
\newcommand{\lplusj}[4]{#2\xrightarrow{#3\,+}_{#1}#4}
\newcommand{\behavesj}[3]{#2\Downarrow_{#1}#3}

\newcommand{\asg}{\rho}
\newcommand{\Cn}{\mathrm{Cn}}

\newcommand{\cert}{\pi}
\newcommand{\pool}{\Pi}
\newcommand{\concl}{\Delta}
\newcommand{\cfg}[2]{\langle #1 \,;\, #2 \rangle}
\newcommand{\stepto}{\longrightarrow}
\newcommand{\replayto}{\longrightarrow^{*}}

\newcommand{\rep}[1]{\mathrel{\sim_{#1}}}
\newcommand{\crep}{\mathrel{\approx}}
\newcommand{\poolat}[2][]{\cid{pool}_{#1}\langle #2\rangle}
\newcommand{\checkC}{\mathit{check}_{\mathrm{C}}}
\newcommand{\checkL}{\mathit{check}_{\mathrm{Lean}}}
\newcommand{\accept}{\mathsf{accept}}
\newcommand{\reject}{\mathsf{reject}}

\newcommand{\tool}{\texttt{Keel}\xspace}
\newcommand{\code}[1]{\texttt{#1}}
\newcommand{\Rule}[1]{\textsc{#1}}
\newcommand{\BV}{\mathrm{BV}}
\newcommand{\Bool}{\mathrm{Bool}}
\newcommand{\Any}{\mathrm{Any}}
\newcommand{\List}{\mathrm{List}}

\newcommand{\eg}{\emph{e.g.}\xspace}
\newcommand{\ie}{\emph{i.e.}\xspace}

\newcommand{\embc}{\texttt{EmbC}\xspace}
\newcommand{\semr}[2]{\llbracket #1 \rrbracket_{#2}}

\usepackage{longtable}
\usepackage{wrapfig}
\usepackage{flafter}
\makeatletter
\newenvironment{embcfigure}{%
  \par\addvspace{\intextsep}%
  \begingroup
  \let\@vspace\@vspace@orig
  \let\@vspacer\@vspacer@orig
  \captionsetup{type=figure,hypcap=false}%
  \small
  \allowdisplaybreaks[2]%
}{%
  \par\endgroup
  \addvspace{\intextsep}%
}
\makeatother

\usepackage{xspace}
\usepackage{tikz}
\usetikzlibrary{positioning,fit,arrows.meta,backgrounds}
\usepackage{pgfplots}
\pgfplotsset{compat=1.18}
\usepgfplotslibrary{groupplots}
\definecolor{keelblue}{RGB}{42,120,214}
\definecolor{bcforange}{RGB}{235,104,52}
\definecolor{offgray}{RGB}{137,135,129}
\definecolor{thirdgreen}{RGB}{27,175,122}
\usepackage{array}
\usepackage{colortbl}
\usepackage{listings}
\usepackage{bussproofs}
\usepackage{needspace}
\usepackage{algpseudocode}

\algnewcommand{\IfThen}[2]{\State \algorithmicif\ #1\ \algorithmicthen\ #2}
\algnewcommand{\Fail}{\textbf{return}\ $\mathsf{fail}$}
\algnewcommand{\ForDo}[2]{\State \algorithmicfor\ #1\ \algorithmicdo\ #2}
\algrenewcommand\algorithmicrequire{\textbf{Input:}}
\algrenewcommand\algorithmicensure{\textbf{Output:}}
\lstdefinestyle{paperc}{
  language=C,
  basicstyle=\ttfamily\footnotesize,
  columns=fullflexible, keepspaces=true, upquote=true,
  keywordstyle=\color{blue!65!black}\bfseries,
  commentstyle=\color{green!40!black},
  emph={bpf_get_stack}, emphstyle=\color{violet!80!black},
  emph={[2]USER,KERNEL}, emphstyle={[2]\color{orange!70!black}},
  morekeywords={u32,u64},
  aboveskip=2pt, belowskip=0pt,
  xleftmargin=0pt, showstringspaces=false}
\lstdefinestyle{paperlean}{
  basicstyle=\ttfamily\footnotesize,
  columns=fullflexible, keepspaces=true, upquote=true,
  morekeywords={def,match,with,some,none,if,then,else,true,false,fun},
  keywordstyle=\color{blue!65!black}\bfseries,
  comment=[l]{--}, commentstyle=\color{green!40!black},
  extendedchars=true, inputencoding=utf8,
  literate={⟨}{{$\langle$}}1 {⟩}{{$\rangle$}}1
           {→}{{$\rightarrow$}}1 {×}{{$\times$}}1
           {≤}{{$\le$}}1 {∧}{{$\wedge$}}1,
  aboveskip=2pt, belowskip=0pt,
  xleftmargin=0pt, showstringspaces=false}
\newcommand{\Keel}{Keel\xspace}

\newcommand{\LeanCheckerLines}{20{,}290}
\newcommand{\NumRules}{50}
\newcommand{\NumRewritesBCF}{151}
\newcommand{\NumRewritesAdded}{13}
\newcommand{\NumRewritesKeel}{164}
\newcommand{\EmbCProgramLines}{4{,}757}
\newcommand{\EmbCFrameworkLines}{4{,}425}

\newcommand{\PrintedCLines}{6{,}616}
\newcommand{\PrinterLines}{268}
\newcommand{\PrintedCLinesK}{6.6}
\newcommand{\PrintedCFunctions}{174}

\newcommand{\SoundnessProofLines}{52{,}398}
\newcommand{\CostProofLines}{12{,}863}

\newcommand{\ProofToProgramRatio}{fourteen}

\newcommand{\BCFCheckerTotalLines}{5{,}535}
\newcommand{\KeelText}{35.9}
\newcommand{\BCFText}{40.3}

\newcommand{\CostR}{5{,}563{,}068}
\newcommand{\CostRuleAllow}{5{,}411{,}000}
\newcommand{\CostPremAllow}{500}
\newcommand{\CostArgAllow}{80}
\newcommand{\CostStepAllow}{4{,}000}
\newcommand{\CostLoopAllow}{40}
\newcommand{\CostDriverAllow}{128}
\newcommand{\WorkPerByte}{32}
\newcommand{\WorkBase}{2^{20}}
\newcommand{\WorkBaseNeedK}{362}
\newcommand{\MemPerByte}{32}
\newcommand{\MemBase}{2^{20}}
\newcommand{\MemPerByteProved}{33}
\newcommand{\HashTableBytes}{4{,}096}
\newcommand{\PremArrayBytes}{1{,}024}
\newcommand{\StructBytes}{152}
\newcommand{\MemFixed}{5{,}272}
\newcommand{\WorkUsedMax}{10.2}
\newcommand{\MemUsedMax}{9.4}

\newcommand{\CorpusAccepted}{1{,}538}

\newcommand{\CorpusShared}{1{,}522}

\newcommand{\BigMemKeel}{9.54}
\newcommand{\BigMemBCF}{8.48}
\newcommand{\CorpusSpeedup}{8.1}
\newcommand{\CorpusSpeedupGeo}{3.6}
\newcommand{\BigSpeedup}{8.1}
\newcommand{\MemMedKeel}{13}
\newcommand{\MemMedBCF}{10}
\newcommand{\WhyHtabKB}{4}
\newcommand{\BigMemExtraPct}{13}

\newcommand{\ProgSetTrustedPct}{7.2}

\newcommand{\XCPU}{AMD EPYC 7302P}

\newcommand{\XVMs}{14}
\newcommand{\XCertMedBytes}{744}
\newcommand{\XCertMaxKB}{24}

\newcommand{\CorpusShipped}{1{,}891}
\newcommand{\CorpusOnlyKeel}{16}

\newcommand{\BigCertMB}{5.6}

\newcommand{\CrashSites}{7}
\newcommand{\InterpAllocRuns}{16{,}770}

\newcommand{\BudgetUsedFrac}{a third}
\newcommand{\BoundSlack}{two million}

\newcommand{\BuildDefaultSec}{63}
\newcommand{\BuildProofsMin}{4.7}

\newcommand{\BuildTotalMin}{6}

\newcommand{\AddedRuleLeanLines}{633}
\newcommand{\MedianLinesPerRule}{35}
\newcommand{\RebuildMinutes}{4.7}

\newcommand{\FuzzInputsM}{64}
\newcommand{\FuzzCorpusSeeds}{1{,}697}

\newcommand{\BCFDefects}{13}
\newcommand{\CrashMinBytes}{180}
\newcommand{\WarnSites}{7}

\newcommand{\StackKiB}{7.5}
\newcommand{\MaxDepth}{32}
\newcommand{\MaxSortTerms}{1{,}024}
\newcommand{\MaxPoolGiB}{4}

\newcommand{\IdxObjects}{1{,}588}

\newcommand{\IdxBcfEval}{361}
\newcommand{\IdxCapMin}{30}
\newcommand{\IdxSingleRun}{48}
\newcommand{\IdxLoaded}{1{,}226}
\newcommand{\IdxRejected}{362}
\newcommand{\IdxCertsPerLoad}{276{,}604}

\newcommand{\IdxLoadPct}{99.86}

\newcommand{\IdxShareMaxBCF}{1.2}
\newcommand{\IdxShareMaxKeel}{0.44}
\newcommand{\IdxCheckSpeedup}{1.3}
\newcommand{\IdxFastPct}{98.6}
\newcommand{\IdxFastMedUs}{12--13}
\newcommand{\IdxSlowPct}{1.4}
\newcommand{\IdxSlowMedKB}{5.4}
\newcommand{\IdxSlowMedBCF}{110}
\newcommand{\IdxSlowMedKeel}{55}
\newcommand{\IdxSlowSpeedup}{2.0}

\newcommand{\ExSteps}{33}
\newcommand{\ExBytes}{852}

\newcommand{\CnfXorSteps}{31{,}756}

\newcommand{\ForgeTableSteps}{13}

\newcommand{\AbsorbDepth}{65}

\newcommand{\RulesSimple}{42}
\newcommand{\RulesNonSimple}{8}
\newcommand{\RulesComputation}{7}
\newcommand{\LeanCheckerCodeLines}{4{,}000}
\newcommand{\LeanCheckerProofLines}{16{,}300}

\newcommand{\BitblastProofLines}{2{,}000}
\newcommand{\RwZeroSteps}{1{,}829}
\newcommand{\RwZeroShapes}{1{,}139}
\newcommand{\RwMaxWidth}{255}

\newcommand{\WhyKernelLargeSpeedup}{3.4}
\newcommand{\WhyTopTenLo}{6}
\newcommand{\WhyTopTenHi}{18}

\newcommand{\WhyEquivCallsM}{11}

\newcommand{\WhyVarMapKB}{1}
\newcommand{\PfBcfEquiv}{58}
\newcommand{\PfBcfEquivK}{12}
\newcommand{\PfBcfXarray}{20}
\newcommand{\PfBcfXarrayK}{46}
\newcommand{\PfBcfAlloc}{6}
\newcommand{\PfBcfAllocK}{20}

\newcommand{\PfBcfRulesAll}{12}
\newcommand{\PfBcfRulesAllK}{15}
\newcommand{\WhyEquivVsKeelLo}{two}
\newcommand{\WhyEquivVsKeelHi}{eight}
\newcommand{\PfBcfInput}{2}
\newcommand{\PfBcfInputK}{4}

\newcommand{\PfKernelSizedLo}{16}
\newcommand{\PfKernelSizedHi}{32}

\newcommand{\CompExtraMinMB}{0.7}
\newcommand{\CompExtraMaxMB}{1.1}
\newcommand{\CompMemOff}{28.9}

\newcommand{\CountersCostHi}{2}

\newcommand{\CompRuns}{27}
\newcommand{\CompRunsMinKB}{285}
\newcommand{\CompBigN}{seven}
\newcommand{\CompBigKeelBCF}{1.2}
\newcommand{\CompBigOffBCF}{3.4}
\newcommand{\CompNetSavePct}{4}
\newcommand{\CompSharedBarN}{18}
\newcommand{\CompMidN}{20}
\newcommand{\CompMidExtraMs}{0.5}
\newcommand{\CompMidKeelMs}{10}
\newcommand{\CompMidSpeedup}{5.8}
\newcommand{\CountersSmallPct}{1}
\newcommand{\RecompilePct}{2}

\begin{document}
\raggedbottom

\title[\tool: A Verified In-Kernel Proof Checker for Proof-Carrying BPF]{\tool: A Verified In-Kernel Proof Checker for Proof-Carrying BPF
  \\ {\small \textbf{Rutgers Department of Computer Science Technical
      Report DCS-TR-765}}}

\author{Yalu Cai}
\affiliation{%
  \institution{Rutgers University}
  \country{USA}}
\email{yc1795@cs.rutgers.edu}
\author{Tung-Che Chang}
\affiliation{%
  \institution{Rutgers University}
  \country{USA}}
\email{tc.chang@rutgers.edu}
\author{Santosh Nagarakatte}
\affiliation{%
  \institution{Rutgers University}
  \country{USA}}
\email{santosh.nagarakatte@cs.rutgers.edu}

\begin{abstract}
Recent systems such as the eBPF Certificate Framework~(BCF) and
Proof-Carrying BPF address the imprecision in the Linux eBPF verifier
with proof-carrying code. When the verifier's abstract domains cannot
prove a safety condition, an in-kernel proof checker checks an
untrusted user-space refutation proof. A single bug in this checker
can compromise the security of the entire kernel.  While formalizing
BCF's in-kernel checker, we discovered \BCFDefects{} bugs such as
unsound rewrite rules, silently unchecked proof steps, and memory
safety errors. We demonstrate that forged certificates can trick the
kernel to accept arbitrary unsafe eBPF programs.
To address this problem, we propose \tool, a formally verified
in-kernel proof checker for the Linux eBPF verifier. First, we specify
the checker as a pure functional program in Lean and prove its
soundness against bit-vector semantics.  We decouple the steps of the
proof certificate checker from the mathematical rewrite rules, which
enables us to extend the rewrite rules without the need to change the
checker.  Second, we implement the checker in \embc, a deeply embedded
C fragment, and prove that our C implementation refines the functional
specification without undefined behavior or memory leaks. Third, using
an amortized potential argument, we prove linear time and memory
bounds for the C checker.  \tool is correct-by-construction while also
being faster than the BCF checker inside the kernel,
\IdxCheckSpeedup$\times$ in total and \IdxSlowSpeedup$\times$ on
the slowest \IdxSlowPct\% of certificates.
\end{abstract}

\maketitle

\section{Introduction}
\label{sec:intro}

\textbf{The eBPF verifier and its imprecision.} The eBPF ecosystem
enables the users to extend the functionality of the Linux kernel. It
has applications to numerous domains including cloud-native
environments and networking~\cite{rice:learning-ebpf:book:2023}. The
kernel checks the safety of the eBPF bytecode when such programs are
loaded through a system call. The Linux kernel has a static analyzer,
which is known as the eBPF verifier, that checks that the eBPF
bytecode is memory safe, has bounded control flow, and terminates on
all inputs.  Subsequently, the safe programs are just-in-time compiled
to the target and attached to the appropriate kernel hooks. The
verifier explores all paths symbolically with simple non-relational
domains such as intervals and tristate numbers~\cite{agni-cav23}. The
inherent imprecision and overapproximation in the non-relational
domains used by the verifier can result in the rejection of safe eBPF
programs. Further, small perturbations to the code with equivalent
transformations either by the compiler backend or by the user can
cause the program to be rejected~\cite{cai:pcb:ebpf:2026,
  sun:bcf:sosp:2025}.

\textbf{Addressing the imprecision}. Some mechanisms that are being
explored to improve the precision include the use of expressive
relational domains in the eBPF
verifier~\cite{gershuni:prevail:pldi:2019} and writing the extensions
in a safe language such as
Rust~\cite{jiaRexClosingLanguageverifier2025}. However, the use of
relational domains in the kernel increases the complexity of the code
and the performance cost. Use of safe languages such as Rust requires
trusting the Rust runtime. These approaches significantly increase the
trusted computing base.

\textbf{BCF and Proof-Carrying BPF}. Projects such as the eBPF
certificate framework (BCF)~\cite{sun:bcf:sosp:2025} and the
Proof-Carrying BPF (PCB) framework~\cite{cai:pcb:ebpf:2026} improve
the precision of the eBPF verifier with a limited form of
proof-carrying code~\cite{necula:pcc:popl:1997,
  necula:pcc:osdi:1996}. The essence of the idea in these projects is
to use the existing eBPF verifier with its default analyses and use
the user-provided proofs only when the verification fails with the
default domains. It increases the expressivity of the verifier without
increasing the trusted components of the kernel. When the kernel's
default analyses fail for a particular safety condition $C$ on a path
with path condition $P$, the verifier constructs a quantifier-free
refinement condition ($P \wedge \neg C$), which asserts that the path
is taken and the safety condition fails. If this refinement condition
is unsatisfiable, then the safety condition holds for all inputs.
These projects use the refutation proofs from an untrusted user-space
cvc5 SMT solver~\cite{barbosa:cvc5:tacas:2022, reynolds:cpc:cav:2026}
to demonstrate the unsatisfiability of the refinement condition. In
the case of BCF~\cite{sun:bcf:sosp:2025}, the proofs are generated on
demand by pausing the kernel verifier and transitioning to the user
space to solve using an SMT solver. In Proof-Carrying
BPF~\cite{cai:pcb:ebpf:2026}, the compiler backend with a mirrored
version of the verifier generates the proofs a priori. The in-kernel
proof checker validates the provided proofs and updates the abstract
states when the proofs are valid. The BCF and PCB projects have
demonstrated that the use of refutation proofs can allow the
acceptance of thousands of safe eBPF programs, which were previously
rejected by the kernel~\cite{sun:bcf:sosp:2025, cai:pcb:ebpf:2026}.

\textbf{Discovered multiple bugs in the in-kernel checker}. The proofs
from the user space are untrusted. Hence, the security of the BCF and
PCB frameworks is dependent on the in-kernel checker. During our
effort to formalize the in-kernel proof checkers, we
\emph{surprisingly discovered \BCFDefects{} previously unknown
bugs}~(see \S\ref{sec:eval}) in the in-kernel checker, some of
which can result in the acceptance of unsafe eBPF programs. First, we
discovered some of the rewrite rules perform unsound bit-vector and
algebraic manipulation.  Second, we discovered some of the proof steps
are unchecked, which just parse the proof's premises and immediately
accept the user-supplied conclusion without verification. Third, we
discovered implementation errors and missed checks that cause the
kernel to perform out-of-bounds memory accesses. We also demonstrate
that a potential malicious actor can use these bugs to forge a
certificate that refutes a satisfiable refinement condition, which
would let an unsafe eBPF program pass a safety check.

\textbf{\tool and its design goals}. To replace BCF's unverified
checker, we built \tool, a formally verified in-kernel proof checker
in Lean~4 that accepts the same binary certificates and uses the same
Linux eBPF verifier hooks while meeting four goals. First, the
in-kernel checker should be \emph{end-to-end logically sound} under
standard bit-vector semantics. Second, it should be \emph{extensible},
allowing kernel developers to add new rewrite rules without
re-verifying the checker. Third, the checker should be \emph{memory
and undefined-behavior safe}, freeing all dynamically allocated memory
along every path under arbitrary kernel API outcomes. Fourth, the
checker should have \emph{linear time and memory complexity} to
prevent denial-of-service attacks.

\textbf{Functional specification of logical soundness}.
We specify \tool as a pure functional checker in Lean over a growing
pool of well-typed terms and derived conclusions, proving that
deriving \texttt{false} ($\bot$) guarantees unsatisfiability of the
refinement condition (Theorem~\ref{thm:sound:formal}, \S\ref{sec:checker}).
Rather than hard-coding mathematical rewrites as stateful functions,
\code{Keel} stores all rewrite rules as data in a pattern language
with explicit width and overflow guards.  A single generic
meta-theorem (Lemma~\ref{lem:rwstep}) proves that instantiating any
sound entry preserves the checker's invariant, so new rules require
only stand-alone identity proofs.

\textbf{Deep embedding of a subset of C and the refinement}.
Because the Linux kernel cannot link against the Lean runtime, we
implement the checker in \embc (\S\ref{sec:embc}), a deeply embedded C fragment in
Lean with typed memory blocks and a non-deterministic kernel API
oracle, validated via forward and backward simulations against
CompCert Clight (Theorem~\ref{thm:embc-clight}).  Our \embc
implementation realizes the expression pool with flat \code{u32}
arrays, open-addressing based hash-table insertion, and in-place
mark-and-slide compaction, and refines the functional checker without
undefined behavior or leaks (Theorem~\ref{thm:refine:formal}).

\textbf{Time and memory bounds via potentials.}  To bound kernel
resource usage, the C checker initializes two runtime counters for
time and memory linearly in the input byte size and decrements them
before every input-dependent loop, hash probe, and allocation (\S\ref{sec:cost}).
Lifting these counters into amortized potential functions on \embc's
operational semantics guarantees termination in linear time and memory
(Theorem~\ref{thm:cost}).

Our tool is open source and publicly available~\cite{keel-repo}. We evaluate \tool
using BCF's suite of real-world eBPF object files (Calico, Cilium,
Inspektor Gadget, BCC) and \CorpusShipped{} stand-alone certificates, plus \FuzzInputsM{}
million differentially fuzzed certificates. \tool loads exactly the
object files that BCF's unverified checker loads. On the stand-alone
certificates, it accepts all that BCF's checker accepts and also
\CorpusOnlyKeel{} additional valid certificates that BCF's checker
rejects. \tool correctly rejects the forged certificates that BCF's
checker accepts. Inside the Linux kernel, \tool checks certificates
\IdxCheckSpeedup$\times$ faster than BCF's checker in total and
\IdxSlowSpeedup$\times$ faster on the slowest \IdxSlowPct\%.

\section{Background and Motivation}
\label{sec:background}
We provide background on proof-carrying BPF, specifically focusing
on the precision gap that it addresses in the eBPF verifier, the
in-kernel proof checker, proof certificates generated, and motivate
the need for a formally verified in-kernel checker.

\begin{figure}[t]
\begin{minipage}[t]{0.4\textwidth}
\footnotesize
\textbf{(a)} Program fragment (simplified)
\begin{lstlisting}[style=paperc, basicstyle=\ttfamily\scriptsize]
max_len = 800;       
usize = bpf_get_stack(ctx, raw_data,
                      max_len, ...);  
if (usize < 0) return 0;   
ksize = bpf_get_stack(ctx, raw_data + usize,
                      max_len - usize, ...); 
\end{lstlisting}
\smallskip
\textbf{(b)} Refinement condition $\varphi$
\par\vspace{2pt}
\begin{tabular}{@{}l@{\;\;}l@{}}
  $\varphi := P \wedge \neg C$ & \\
  $P := 0 \le \sym \le 800$ & \\
  $C := t \le 800$ & \\
  $t := o + (800 - \sym)$ & \\
\multicolumn{2}{@{}l@{}}{$o := \mathrm{extract}_{31:0}(0^{64} + \mathrm{sext}_{64}(\sym))$}
\end{tabular}
\end{minipage}%
\hfill
\begin{minipage}[t]{0.6\textwidth}
\footnotesize
\textbf{(c)} The certificate's proof (\ExSteps{} steps, \ExBytes{} bytes)
\par\vspace{2pt}
\scriptsize
\setlength{\tabcolsep}{2.5pt}
\begin{tabular}{@{}r l l l@{}}
\toprule
Step & Rule, from steps & Concludes & In program terms \\
\midrule
0 & \textsc{evaluate} & $\neg(800 < 800)$ & a constant, computed \\
1 & \textsc{poly\_norm} & $\sym + (800 + {-\sym}) = 800$ & start + length is 800 \\
2 & \textsc{poly\_norm} & $800 - \sym = 800 + {-\sym}$ & the length, normalized \\
3--22 & 20 steps, 1 unchecked & $o = \sym$ & starts at byte \texttt{usize} \\
23 & \textsc{cong} 22, 2 & $t = \sym + (800 + {-\sym})$ & its end, offset replaced \\
24 & \textsc{trans} 23, 1 & $t = 800$ & ends at byte 800 \\
25 & \textsc{refl} & $800 = 800$ & \\
26 & \textsc{cong} 25, 24 & $(800 < t) \leftrightarrow (800 < 800)$ & the check, $t$ replaced \\
27 & \textsc{trans} 26, 0 & $\neg(800 < t)$ & the check cannot fail \\
28 & \textsc{rewrite} & $(t > 800) \leftrightarrow (800 < t)$ & failed check, normalized \\
29 & \textsc{trans} 28, 27 & $\neg(t > 800)$ & failed check is false \\
30 & \textsc{assume} & $\varphi$ & the verifier's rejection \\
31 & \textsc{and\_elim} 30 & $t > 800$ & failed check, from $\varphi$ \\
32 & \textsc{eq\_resolve} 31, 29 & $\bot$ & contradiction \\
\bottomrule
\end{tabular}
\end{minipage}
\caption{\small (a) A simplified eBPF program that writes two stack
  traces into the 800-byte buffer \texttt{raw\_data}.  The function
  \texttt{bpf\_get\_stack(ctx, p, n, flags)} writes a stack trace of
  at most \texttt{n} bytes to \texttt{p} and returns the number of
  bytes written, or a negative error code. The verifier rejects the
  program when it analyzes the second call. It bounds the offset
  \texttt{usize} and \texttt{max\_len - usize} separately in $[0,
    800]$. The verifier concludes that the second write may reach byte
  1600 and rejects. (b) The refinement condition is $\varphi := P
  \wedge \neg C$ ($\mathit{sym}$ is \code{usize}, $P$ the path
  condition, $C$ the safety condition check with $\neg C$ emitted as
  $t > 800$ and $o$ the offset for \code{raw\_data + usize}). All
  comparisons are signed. (c) The \ExSteps{}-step refutation proof ($\psi =
  \bot$ encodes $\neg\psi$). Steps 3--22 strip $o$'s width
  conversions. }
\label{fig:example}
\end{figure}

\textbf{The precision gap in the eBPF verifier}.
The eBPF verifier performs path-sensitive
abstract interpretation over the eBPF bytecode, mapping each register
to a non-relational abstract state, which is the product of 64/32-bit
signed intervals and a tristate number that tracks known-zero/one
bits~\cite{agni-sas24,agni-cav23}. The pointers are tracked as a base
object plus a scalar offset in the same domain.  As the verifier does
not track equalities or linear inequalities between registers, it
loses all correlations between values computed from a shared symbolic
source.

Figure~\ref{fig:example}(a) illustrates it on a safe eBPF program
that the verifier rejects.  The program packs two stack traces back to
back into an 800-byte map buffer, \texttt{raw\_data}.  The helper
\texttt{bpf\_get\_stack(ctx, p, n, flags)} writes at most \texttt{n}
bytes to \texttt{p}, returning the bytes written in $[0, \code{n}]$ or
a negative error code.  The first call writes the user trace at
\code{raw\_data} ($\texttt{max\_len} = 800$).  After analyzing
\texttt{if (usize < 0) return 0}, the verifier narrows \texttt{usize}
to $[0, 800]$.  The second call writes the kernel trace of at most
\texttt{max\_len - usize} bytes at \texttt{raw\_data + usize}, which
is the remaining space.  Specifically, for any \texttt{usize} $\in [0,
  800]$, the second write ends at byte \texttt{usize} + (800 -
\texttt{usize}) = 800.  With the non-relational domains used by the
verifier, the offset $r_{\text{off}}$ ($\code{usize}$) and size
$r_{\text{sz}}$ ($\code{800 - usize}$) are abstracted independently as
$r_{\text{off}}^\# = [0, 800]$ and $r_{\text{sz}}^\# = [800, 800] -^\#
[0, 800] = [0, 800]$.  Checking $r_{\text{off}} + r_{\text{sz}} \le
800$ via interval addition gives $r_{\text{off}}^\# +^\#
r_{\text{sz}}^\# = [0, 800] +^\# [0, 800] = [0, 1600] \not\subseteq
[0, 800]$. The verifier loses the relational invariant $r_{\text{off}}
+ r_{\text{sz}} = 800$. Since the access can exceed the bounds of the
buffer with the deduced intervals, the verifier rejects the program.

\textbf{Proof-Carrying BPF}. Using relational domains~(\eg,
octagons~\cite{gange:zones-octagons-fresh-look:toplas:2021},
zones~\cite{gershuni:prevail:pldi:2019,
  gange:zones-octagons-fresh-look:toplas:2021}, polyhedra) or an SMT
solver inside the kernel is impractical because of the large running
times and significantly increased code base. The eBPF Certificate
Framework (BCF)~\cite{sun:bcf:sosp:2025} and Proof-Carrying BPF
(PCB)~\cite{cai:pcb:ebpf:2026} projects make a case for using
proof-carrying code while using the default analyses in the kernel
verifier. When the verifier's safety check fails on an abstract path,
the verifier identifies a backward slice from the failing instruction
and includes all instructions influencing it. It provides symbolic
variables to the leaves of the collection of directed acyclic graphs
obtained from this analysis.  Subsequently, it performs forward
symbolic execution to build the formula for the safety condition $C$
and the path conditions $P$. The refinement condition constructed is
$\varphi := P \wedge \neg C$ ($P$ holds and $C$ fails). If $\varphi$
is unsatisfiable ($\forall \rho.\; \rho \not\models \varphi$), then $P
\models C$ on every concrete path execution.
In BCF, the kernel sends the refinement condition to user space, which
runs the cvc5 SMT solver to generate a refutation certificate $\pi$
and returns $\pi$ to the kernel.  In Proof-Carrying BPF, the compiler
runs a user-space verifier mirror ahead of time, generates all
certificates at compile time, and embeds them in the ELF binary.
The kernel checker checks $\pi$ against $\varphi$; if accepted, the
verifier refines its abstract state (pruning a branch or discharging
$C$) and continues. We will refer to the certificate as the BCF proof
certificate in the rest of the paper.

\textbf{Refinement conditions for the
  example}. Figure~\ref{fig:example}(b) shows $\varphi$ for
Figure~\ref{fig:example}(a).  Slicing backwards from \code{raw\_data +
  usize} and \code{max\_len - usize} reaches the first
\code{bpf\_get\_stack} call, whose 32-bit return value \code{usize}
becomes $\mathit{sym}$.  The branch condition $usize < 0$ induces the
path condition. Propagating forward from \code{raw\_data} at 64-bit
map offset $0^{64}$, eBPF's 64-bit pointer arithmetic sign-extends
$\mathit{sym}$ to 64 bits ($\mathrm{sext}_{64}(\mathit{sym})$), adds
$0^{64}$, and extracts bits $31..0$ for the 32-bit check, yielding $o
:= \mathrm{extract}_{31:0}(0^{64} +
\mathrm{sext}_{64}(\mathit{sym}))$.  The second write ends at $t := o
+ (800 - \mathit{sym})$ with $\neg C$ emitted as $t >_s 800$; refuting
$\varphi = P \wedge \neg C$ proves safety.

\textbf{Proof certificates and refutation}.  A BCF certificate $\pi =
(\Pi_\pi, s_0, \dots, s_{n-1})$ encodes both $\varphi$'s terms and the
proof steps as a flat array of 32-bit words (\code{u32}): (1)~an
\emph{expression pool} $\Pi_\pi$ storing each DAG term once in
topological order (a 32-bit header packing an 8-bit operator, 8-bit
arity, and 16-bit width/parameter field, followed by one \code{u32}
pool index per child); and (2)~\emph{proof steps} $s_0, \dots,
s_{n-1}$ from the bit-vector and Boolean fragment of cvc5's
Cooperating Proof Calculus (CPC).  Each step $s_i = (R, \vec{j},
\vec{u})$ names one of \NumRules{} rules $R$, cites earlier premises $j_\ell <
i$, supplies parameters $\vec{u}$, and derives $\psi_i$ (encoding
$\neg\psi$ equationally as $\psi = \bot$).  A valid certificate has a
unique $\Rule{assume}$ step matching $\varphi$ (up to variable
renaming) and concludes $\psi_{n-1} = \bot$.  The proof rules are
designed to balance proof size and the checker's complexity. They can
be divided into three classes.
The first class consists of \RulesSimple{} structural rules for
equality ($\Rule{refl}$, $\Rule{symm}$, $\Rule{trans}$,
$\Rule{cong}$), Booleans ($\Rule{and\_elim}$, $\Rule{and\_intro}$,
$\Rule{not\_and}$), and clausal resolution ($\Rule{resolution}$,
$\Rule{factoring}$, $\Rule{eq\_resolve}$, CNF clausification).
The second class of proof rules consists of \RulesComputation{}
computation rules.
They execute normalizers inside the checker: $\Rule{evaluate}$
(constant folding), $\Rule{distinct\_values}$, $\Rule{poly\_norm}$ and
$\Rule{poly\_norm\_eq}$ (reducing linear forms $\sum_i k_i \cdot
\mathit{atom}_i + c \bmod 2^w$ over $\mathbb{Z}/2^w\mathbb{Z}$),
$\Rule{aci\_norm}$ and $\Rule{absorb}$ (flattening/sorting ACI
operators), and $\Rule{bitblast}$ (word-to-bit circuits). The third
proof rule is a table-driven rewrite rule for mathematical identities.
The proof rule, $\Rule{rewrite}$, instantiates schematic equations
from a static table of cvc5's RARE rules discharging side conditions
via cited premises.

Figure~\ref{fig:example}(c) traces the \ExSteps{}-step proof, deriving
equational lemmas before assuming $\varphi$ at step 30. There are
three main components for this refutation proof. The first part strips
64-bit pointer wrappers to prove $o = \mathit{sym}$ (steps 3--22).
The second part recovers the relational invariant $t = 800$.  Step 2
uses $\Rule{poly\_norm}$ to rewrite subtraction in bit-vector
arithmetic as $800 - \mathit{sym} = 800 + -\mathit{sym}$, and step 23
($\Rule{cong}$) substitutes $o = \mathit{sym}$ (step 22) and step 2
into $t := o + (800 - \mathit{sym})$ to get $t = \mathit{sym} + (800 +
-\mathit{sym})$.  Step 1 applies $\Rule{poly\_norm}$ to cancel
$+\mathit{sym}$ and $-\mathit{sym}$ ($1\cdot\mathit{sym} +
(-1)\cdot\mathit{sym} + 800 \equiv 800 \pmod{2^{32}}$), proving
$\mathit{sym} + (800 + -\mathit{sym}) = 800$, which step 24
($\Rule{trans}$ with 23) chains to get $t = 800$.

The third part refutes the refinement condition $\neg C$. Steps 25--29
use $t = 800$ to derive $(t >_s 800) = \bot$. Steps 30
($\Rule{assume}\ \varphi$), 31 ($\Rule{and\_elim}: t >_s 800$), and 32
($\Rule{eq\_resolve}$) derive $\bot$, the contradiction.

\textbf{Motivation for \tool}.  Moving proof search to user space
shifts all trust to BCF's approximately five-thousand-line C checker,
where a single bug can compromise the kernel.  When formalizing BCF's
checker, we uncovered 13 bugs such as silently unchecked steps,
unsound rules, and memory-safety errors (see \S\ref{sec:eval}),
which motivates us to build a formally verified in-kernel checker.

\section{High-level Overview of \tool}
\label{sec:overview}

\begin{wrapfigure}{R}{0.7\textwidth}
\centering
\resizebox{\linewidth}{!}{%
\begin{tikzpicture}[x=1cm, y=1cm,
  font=\scriptsize,
  box/.style={draw, rounded corners=2pt, align=center, minimum height=1.5em, inner sep=2pt, fill=white},
  trusted/.style={box, fill=black!12},
  keel/.style={box, draw=blue!55!black, fill=blue!10, thick},
  part/.style={draw=black!40, rounded corners=1.5pt, align=center, inner sep=1.5pt, fill=white, font=\tiny},
  proof/.style={-{Stealth[length=3pt]}, thick, black!85},
  flow/.style={-{Stealth[length=2.5pt]}, black!60},
  lab/.style={font=\tiny, align=center, inner sep=1pt, text=black},
  region/.style={draw=black!50, dashed, rounded corners=3pt},
  regionlab/.style={font=\tiny, fill=white, inner sep=1.5pt, text=black!70}]
  \node (leant) at (3.50,0.38) {Lean checker {\tiny\color{black!55}\S\ref{sec:checker}}};
  \node[part, below=1pt of leant] (replay) {check the proof steps};
  \node[part, below=1pt of replay] (rules) {\NumRules{} proof rules};
  \node (embct) at (7.15,0.38) {EmbC program {\tiny\color{black!55}\S\ref{sec:refine}, \S\ref{sec:cost}}};
  \node[part, below=1pt of embct] (recl) {memory reclamation};
  \node[part, below=1pt of recl] (budget) {work + memory budgets};
  \node[trusted] (sem) at (0.75,0 |- replay) {bit-vector\\ semantics};
  \node[trusted, anchor=west] (esem) at (sem.west |- 0,-0.95) {EmbC semantics {\tiny\color{black!55}\S\ref{sec:embc}}};
  \node[box] (tbl) at (5.32,1.22) {declarative rewrite table {\tiny\color{black!55}\S\ref{sec:checker}}\\[-1pt]
    {\tiny rules are mathematical identities}};
  \node[box] (c) at (9.95,0 |- replay) {C file,\\ \PrintedCLines{} lines};
  \begin{scope}[on background layer]
    \node[draw, rounded corners=2pt, fill=black!4, fit={(leant)(replay)(rules)}, inner sep=2pt] (lean) {};
    \node[draw, rounded corners=2pt, fill=black!4, fit={(embct)(recl)(budget)}, inner sep=2pt] (embc) {};
  \end{scope}
  \draw[proof] (lean.west |- sem) -- node[lab, above] {Thm.~\ref{thm:sound:formal}:\\ soundness} (sem);
  \draw[proof] (embc.west |- replay) -- node[lab, above] {Thm.~\ref{thm:refine:formal}:\\ refinement} (lean.east |- replay);
  \draw[proof] (tbl.west) -| node[lab, above, pos=0.3] {one lemma per entry} (sem.north);
  \draw[proof] (embc.south) |- node[lab, above, pos=0.75] {Thm.~\ref{thm:cost}: cost bound} (esem.east);
  \draw[flow] ([xshift=-1.15cm]tbl.south) -- ([xshift=-1.15cm]tbl.south |- lean.north);
  \draw[flow] ([xshift=1.15cm]tbl.south) -- node[lab, right] {encoded for C} ([xshift=1.15cm]tbl.south |- embc.north);
  \draw[flow] (embc.east |- replay) -- node[lab, above, pos=0.55] {printer} (c.west |- replay);
  \node[box, anchor=west] (us) at (sem.west |- 0,-2.05) {compiler + proof-carrying BPF backend};
  \node[box] (ver) at (6.75,-2.05) {eBPF verifier};
  \node[keel, minimum width=1.4cm] (keel) at (9.95,-2.05) {\textbf{\Keel}};
  \draw[flow] (us) -- node[lab, above] {program +\\ certificates} (ver);
  \draw[flow] ([yshift=2pt]ver.east) -- node[lab, above] {$\varphi$ + certificate} ([yshift=2pt]keel.west);
  \draw[flow] ([yshift=-2pt]keel.west) -- node[lab, below] {accept / reject} ([yshift=-2pt]ver.east);
  \draw[flow] (c) -- node[lab, left] {gcc / clang} (keel);
  \begin{scope}[on background layer]
    \node[region, fit={(sem)(esem)(lean)(embc)(tbl)}, inner xsep=2mm, inner ysep=1.5mm] (ld) {};
    \node[region, fit={(ver)(keel)}, inner xsep=2mm, inner ysep=1.5mm] (k) {};
    \node[region, fit={(us)}, inner xsep=2mm, inner ysep=1.5mm] (u) {};
  \end{scope}
  \node[regionlab, anchor=west] at ([xshift=5pt]ld.north west) {Lean development};
  \node[regionlab, anchor=west] at ([xshift=5pt]k.south west) {kernel};
  \node[regionlab, anchor=west] at ([xshift=5pt]u.south west) {user space};
\end{tikzpicture}%
}
\caption{\small We show how \tool{} is built using Lean (top) and where
  it runs (bottom). A proof-carrying-code backend provides the
  proof certificates. The blue box is \tool, the
  compiled C code that runs in the kernel.}
\label{fig:overview}
\end{wrapfigure}
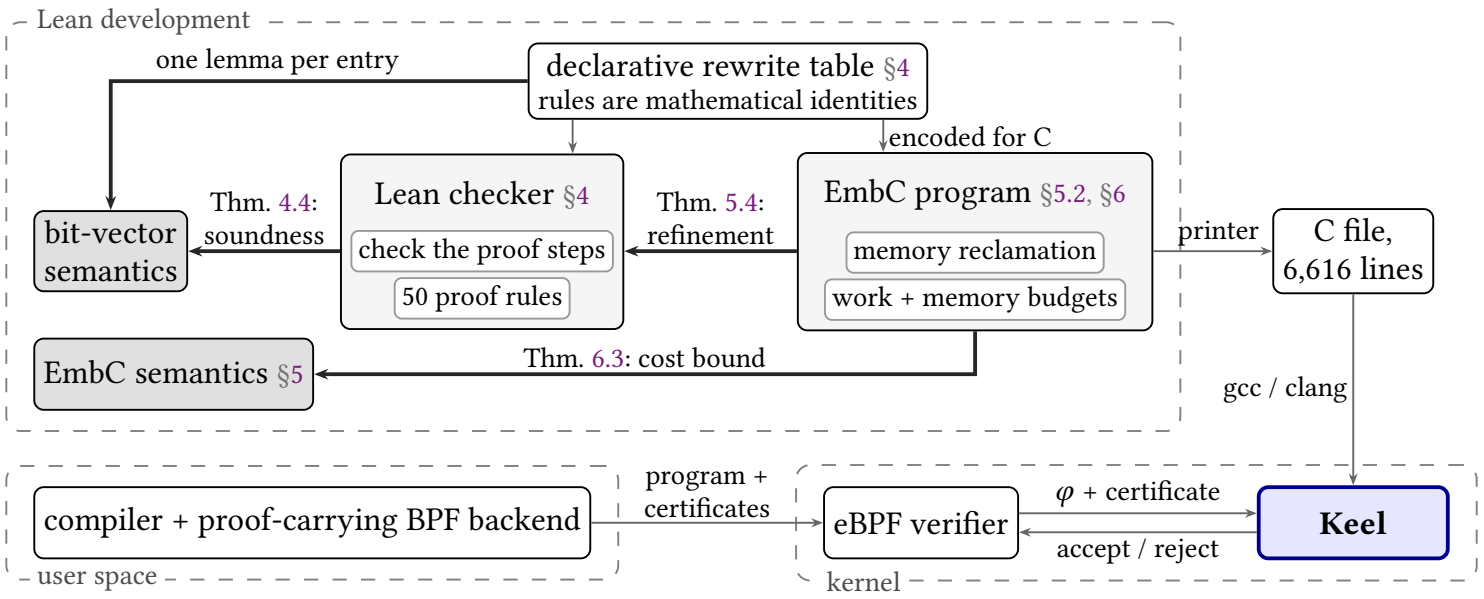

\tool replaces BCF's unverified checker while accepting the same proof
certificates and using the same hooks to attach to the Linux kernel.
When the safety check $C$ fails under path condition $P$, the eBPF
verifier passes the refinement condition $\varphi := P \wedge \neg C$
and an untrusted certificate $\pi$ to \tool.
Unlike BCF's checker, the C implementation generated by \tool comes
with machine-checked proofs that it is sound, memory-safe and has
bounded execution time and memory usage.

Figure~\ref{fig:overview} shows the design of \tool.  Rather than
combining reasoning about refutation proofs on bit-vectors with the
details of the C implementation such as pointer manipulation, hashing,
and in-place compaction of expression pools, we separate them into two
tasks.
First, we develop a pure functional program in Lean over the set of
well-typed terms and derived formulas and show that if
the functional proof checker accepts the proof certificate, then
the refinement condition $\varphi$ is unsatisfiable.
It serves as a stable specification of the proof checker that does not
rely on the C semantics. It also allows us to prove implementations in
different languages beyond C (\eg, Rust).
We create a deep embedding of a small subset of C in Lean, which we
call \embc, and develop the implementation of the proof checker using
the \embc abstract syntax tree in Lean.  We validate the \embc
semantics against CompCert's Clight. The \embc implementation of the
proof certificate checker handles pointer operations, reasons about
undefined behavior, and performs in-place compaction of expressions.
We show that the \embc implementation refines the functional
specification.  We use a trusted printer to create a C implementation
that executes in the Linux kernel (see right side of
Figure~\ref{fig:overview}).

\textbf{Threat model}. The adversary controls the user-space
certificate $\pi$ and may submit arbitrary, malformed certificates.
Our trusted computing base consists of the eBPF verifier that
constructs the refinement condition, our formalization of bit-vector
semantics, the \embc semantics, the specification of the kernel APIs,
our \embc-to-C printer, and the C compiler.

\textbf{\tool's guarantees}.  For \emph{every} refinement condition
$\varphi$, every user buffer $\pi$ of input size $B = |\pi| +
|\varphi|$ bytes, and \emph{every} valid kernel API behavior
(including arbitrary allocation or copy failures), \tool's in-kernel
checker guarantees three properties. (1) (\textbf{Logical
  soundness}). \tool accepts the certificate $\pi$ only when the
refinement condition $\varphi$ is unsatisfiable under bit-vector
semantics. (2) (\textbf{Memory and undefined behavior
  safety}). \tool's C implementation triggers no undefined behavior
and frees every allocated block. (3) (\textbf{Linear resource
  bounds}). \tool's C implementation terminates in linear time with
bounded kernel memory usage (rejection sacrifices precision, never
safety).

\textbf{The functional checker}.  The first component of \tool is a
checker for the BCF format, written in
Lean~\cite{moura:lean4:cade:2021} as an ordinary functional program
over arrays and hash maps.  The functional Lean checker serves as
\tool's logical specification, isolating pure quantifier-free
bit-vector validity from low-level C memory layout and resource
reclamation.  It operates over a proof certificate consisting of an
initial pool of well-typed Boolean and fixed-width bit-vector
expressions followed by a sequence of deduction steps.

\textbf{Handling table-driven rewrite rules}. Among the \NumRules{} rules, the
proof checker includes a rewrite proof step (see proof step 28 in
Figure~\ref{fig:example}(c)), which is driven by a table of
mathematical equalities. The functional checker decouples the
verification of its generic rewrite proof step from the algebraic
soundness of individual rewrite rules. This allows the addition of new
rewrite rules to the table without reverifying the proof checker's
stateful certificate verification machinery.

\tool stores all the rewrite rules as data in a lookup table rather
than hard-coding them into the checker~(see declarative rewrite table
in Figure~\ref{fig:overview}). Each table entry is written in a small
pattern language that specifies a left-hand pattern, a right-hand
pattern, typed parameters, and an optional side condition.
As a result, adding a new rewrite rule requires no changes to the
checker or its soundness proof. It only needs a stand-alone proof of
the new rewrite entry's mathematical identity.

\textbf{Deep embedding of C with \embc}. The Linux kernel requires
stand-alone C code with explicit memory management and direct kernel
API calls. Automatically extracting C code from Lean will link against
the Lean runtime, which makes it infeasible for use in the kernel.
We develop \embc, a deep embedding of a fragment of C in Lean inspired
by the Bedrock2 and Fiat-Crypto
projects~\cite{erbsen:bedrock2:pldi:2021,erbsen:fiatcrypto:sp:2019}.
\embc models memory as blocks of typed cells rather than raw bytes,
allowing our proofs to reason about array elements and struct fields
directly without repeatedly handling byte offsets and
alignment.
Programs execute on a small-step operational machine whose transition
rules check the exact conditions under which C defines each
operation—such as in-bounds array accesses, initialized reads,
non-zero divisors, and valid bit shifts—so that any C undefined
behavior causes the machine to get stuck.
Calls to Linux kernel APIs (for memory allocation, user-space copying,
and pending-signal checks) are specified by contracts and modeled by a
non-deterministic oracle that can return any permitted outcome,
forcing the proofs to handle allocation failures, copy errors, and
pending signals at every call.
To guarantee that an \embc program proven never to get stuck cannot
trigger undefined behavior in C, we prove forward and backward
simulations between \embc and CompCert's Clight semantics (ported to
Lean as Clean).

\textbf{Proving that the \embc implementation refines the functional
  specification}.  Using \embc, we implement the in-kernel checker. It
is approximately \PrintedCLinesK{}K lines of C across \PrintedCFunctions{} functions. We prove that it
safely refines the functional Lean checker. The two checkers hold the 
same state, but in very different data structures.
The functional checker's arrays and hash maps are Lean values that 
never fail to grow, never need to be freed, and cannot be indexed out 
of bounds. The C checker instead stores expressions in one flat 
\code{u32} array with a fixed capacity, whose allocations can fail.

Although the C checker may reject a certificate early if a kernel
allocation fails, a signal is pending, or a resource budget runs out,
our refinement theorem guarantees that it never performs undefined
behavior, leaks no memory, leaves pre-existing kernel memory
unchanged, and accepts a certificate only if the functional Lean
checker accepts it~(see Theorem~\ref{thm:refine:formal}).

\textbf{Proving linear time and memory bounds}.  Even a sound checker
is a denial-of-service vulnerability in the kernel if an attacker can
craft a small certificate with shared sub-terms that triggers
exponential traversals or gigabyte-scale allocations, which is allowed
by our threat model. To rule out such attacks, we prove that \tool
always terminates in time and memory linear in the length of its
inputs. We equip \embc's operational semantics with a cost metric that
measures time in ticks (one tick per basic transition) and memory as
the cumulative bytes allocated.
At entry, \tool initializes two unsigned counters, a work budget and a
memory budget, to linear functions of the input byte size. Before any
operation whose cost depends on the input (such as running an
input-dependent loop, adding a term to the pool, probing the hash
table, or growing an array), \tool decrements the corresponding
counter by that operation's size and rejects the certificate if the
counter is smaller than the required amount. 

To prove that these budgets cannot be bypassed across \tool's C
functions, we treat the two runtime counters as potential functions in
an amortized analysis~\cite{tarjan:amortized:1985}. \tool decrements
the work counter only before input-dependent operations. Our proof
therefore splits the ticks of each function into two parts:
input-dependent ticks, which are paid for by the preceding counter
decrement, and code-bounded ticks (for straight-line statements and
loops over the term's operands), which are bounded by a static
constant for each function. By induction over \embc statements and up
the function call graph, we prove that every function terminates with
a cost bounded by the potential it consumes without needing to reason
about \tool's data structures. Combined with our soundness and
refinement theorems on the same operational semantics, this
establishes \tool's total correctness: \emph{under every permitted
kernel API behavior, \tool terminates in linear time and memory,
performs no undefined behavior, leaks no memory, and accepts a
certificate only when the refinement condition is unsatisfiable}.

\section{A Sound and Extensible Functional Proof Certificate Checker}
\label{sec:checker}

\tool's logical specification is the functional Lean
checker~(Figure~\ref{fig:overview}), which is a pure functional
checker for proof certificates that isolates quantifier-free
bit-vector reasoning from C memory management. We formally define the
proof certificates and the monotonically growing state maintained
during the certificate checking process. We subsequently use the
semantic consequence invariant under bit-vector reasoning to prove
that acceptance means that the refinement condition is
unsatisfiable. Finally, we show how \tool separates its stateful
certificate checking engine from a declarative table of stateless
rewrite rules, which ensures that adding a new rewrite rule only
requires a proof of its validity.

\textbf{Certificates and checking the proof certificate}.  A decoded
proof certificate $\cert$ consists of an initial term pool
$\pool_\cert$ and a sequence of steps $s_0,\ldots,s_{n-1}$.  A \emph{term}
is a constant, a variable, or a bit-vector operator applied to other
terms of type $\mathsf{Bool}$ or $\mathsf{BV}_w$ ($w \in \{1, \dots,
\RwMaxWidth\}$). A \emph{formula} is a Boolean term.  The certificate stores
terms in a pool and refers to them by indices. Mathematically, $\pool$
denotes the set of available terms.  In Figure~\ref{fig:example}(b),
$\sym$ and $o$ have type $\mathsf{BV}_{32}$, $t=o+(800-\sym)$ is a
term, and $\varphi=(0\le\sym\le800)\wedge\neg(t\le800)$ is a formula.

Each step $s_i = (R,\, (j_1, \dots, j_m),\, \vec{u})$ names a rule
$R$, indices of earlier steps $0 \le j_\ell < i$ whose conclusions
serve as \emph{premises}, and rule-specific \emph{parameters}
$\vec{u}$ (resolved term references and, for \Rule{rewrite}, a table
entry index).
The certificate has a single \Rule{assume} step whose
parameter is the formula $\varphi$ to refute.

\textbf{Checker state}. After $k$ steps, the checker holds the current
pool and conclusions $\psi_0,\ldots,\psi_{k-1}$ indexed by the step
number. A later step naming $j$ uses $\psi_j$ as a premise.  For the
soundness proof, let $\concl=\{\psi_0,\ldots,\psi_{k-1}\}$ be the set of
conclusions derived so far, and write $\cfg{\pool}{\concl}$ for the
configuration (initially $\concl = \emptyset$ and $\Pi = \Pi_\pi
\cup\{\top, \bot\}$). In Figure~\ref{fig:example}(c) step~32 names
\textsc{eq\_resolve} with premises~31 ($t > 800$) and~29 ($(t > 800) =
\bot$). The rule \textsc{eq\_resolve} requires the second premise to
equate the first premise to some right-hand side, concluding that
right-hand side ($\bot$).

\begin{definition}[Rule check]\label{def:rulecheck}
  To check step $s_i=(R,(j_1,\ldots,j_m),\vec u)$, the checker runs
  rule $R$'s procedure on the current pool $\pool$, the earlier
  conclusions $\psi_{j_1},\ldots,\psi_{j_m}$, and the parameters $\vec
  u$.
  The procedure either fails (rejecting $\pi$) or returns a conclusion
  $\psi_i$ and a pool $\pool' \supseteq \pool$.  We write a successful
  check as
\[
  \mathit{check}_R(\pool;\,\psi_{j_1},\ldots,\psi_{j_m};\,\vec u)
    = (\pool',\psi_i).
\]
\end{definition}

The checker records $\psi_i$ as step $i$'s conclusion and extends
$\Pi$ when constructing $\psi_i$. In Figure~\ref{fig:example}(c),
step~28 (\textsc{rewrite} with entry~89 and terms $t, 800$) builds $(t
> 800) = (800 < t)$, adding terms to $\Pi'$.

\begin{definition}[Step of a certificate checker]
  \label{def:step}
A step of the certificate takes $\cfg{\pool}{\concl}$ to
$\cfg{\pool'}{\concl \cup \{\psi'\}}$, written $\cfg{\pool}{\concl}
\;\stepto\; \cfg{\pool'}{\concl \cup \{\psi'\}}$, when the step's rule
check succeeds and returns $(\pool',\psi')$.  The checking procedure
of a certificate with $n$ steps is the chain $\cfg{\pool_0}{\emptyset}
\stepto \cfg{\pool_1}{\concl_1} \stepto \cdots \stepto
\cfg{\pool_n}{\concl_n}$, written $\cfg{\pool_0}{\emptyset} \replayto
\cfg{\pool_n}{\concl_n}$, where
$\pool_0=\pool_\cert\cup\{\top,\bot\}$.
\end{definition}

\begin{wrapfigure}{R}{0.45\textwidth}
  \vspace{-20pt}
\footnotesize
\begin{algorithmic}[1]
\Require a certificate $\cert$: a pool $\pool_\cert$ and steps $s_0, \ldots, s_{n-1}$
\Ensure $\accept$ or $\reject$
\Function{$\checkL$}{$\cert$}
  \IfThen{$\neg \mathit{single\_assume\_step}(\cert)$}{\textbf{$\reject$}}
  \State $\varphi \gets$ the parameter of the \textsc{assume} step
  \IfThen{$\neg \mathit{well\_typed}(\pool_\cert)$}{\textbf{$\reject$}}
  \State $\pool \gets \pool_\cert \cup \{\top, \bot\}$
  \For{$i = 0, \ldots, n-1$}
    \State $(R,\ (j_1, \ldots, j_m),\ \vec u) \gets s_i$
    \IfThen{$\exists_{j_\ell} \neg(0 \le j_\ell < i)$}{\textbf{$\reject$}}
    \State $\mathit{result} \gets \mathit{check}_R(\pool;\, \psi_{j_1}, \ldots, \psi_{j_m};\, \vec u)$
    \IfThen{$\mathit{result}$ is failure}{\textbf{$\reject$}}
    \State $(\pool, \psi_i) \gets \mathit{result}$
  \EndFor
  \IfThen{$n \ge 1$ and $\psi_{n-1} = \bot$}{\textbf{$\accept$} \textbf{else $\reject$}}
\EndFunction
\end{algorithmic}
\caption{\small The algorithm employed by the functional checker to
  check a proof certificate $\cert$, which consists of an expression
  pool of terms $\pool_\cert$ and a sequence of steps $s_0, \ldots,
  s_{n-1}$.}
\label{fig:check}
\vspace{-20pt}
\end{wrapfigure}

\begin{definition}[Acceptance]
\label{def:accept}
$\checkL(\cert) = \accept$ when $\pi$ has a single well-formed
\textsc{assume} step, well-typed initial terms, all premises come from
earlier steps, the certificate checking procedure succeeds, and
$\psi_{n-1} = \bot$ (with each new term type-checked on being added to
$\Pi$). Otherwise $\checkL(\cert) = \reject$.
\end{definition}

Figure~\ref{fig:check} sketches our certificate checking procedure.
For \RulesSimple{} of the \NumRules{} rules in the BCF checker, the
check inspects the shapes of its inputs and assembles a conclusion
from their pieces. The other \RulesNonSimple{} compute a constant value, a normal
form, an operator's circuit, or an instance of a rewrite-table
entry. We describe the rewrite table and how we develop a generic
engine to check the \textsc{rewrite} step later in the section.

\textbf{Soundness of the certificate checker}.  Soundness means that
accepting a certificate establishes that its assumed formula $\varphi$
is unsatisfiable under the semantics of Boolean and fixed-width
bit-vector terms.  An assignment $\asg$ gives each variable a value of
its type, and $\semr{u}{\asg}$ is the resulting value of term $u$. We
write $\asg \models \psi$ when $\semr{\psi}{\asg} = \mathrm{true}$. A
formula is unsatisfiable if no assignment satisfies it.

\vspace{12pt}
\begin{theorem}[Soundness]
  \label{thm:sound:formal}
$\checkL(\cert) = \accept \;\Longrightarrow\; \forall \asg.\ \asg \not\models \varphi$.
\end{theorem}

\begin{definition}[Certificate Checker's Invariant]
  \label{def:inv}
  A formula $\psi$ is a \emph{semantic consequence} of $\varphi$ if
  $\psi$ holds under every type-correct assignment $\rho$ satisfying
  $\varphi$:
\[
  \Cn(\varphi) =
  \{\,\psi \mid \forall\asg.\;\asg\models\varphi
      \Longrightarrow \asg\models\psi\,\}.
\]
 The certificate checker's invariant is $\concl \subseteq
 \Cn(\varphi)$: every conclusion derived so far is a semantic
 consequence of $\varphi$.
\end{definition}

\begin{lemma}[Every step preserves the certificate checker's invariant]\label{lem:step}
If $\concl \subseteq \Cn(\varphi)$ and $\cfg{\pool}{\concl} \stepto \cfg{\pool'}{\concl \cup \{\psi'\}}$,
then $\psi' \in \Cn(\varphi)$, and hence $\concl \cup \{\psi'\} \subseteq \Cn(\varphi)$.
\end{lemma}

We prove Lemma~\ref{lem:step} for each of the \NumRules{} rules.  The
\textsc{assume} rule returns $\varphi$ itself, which is in
$\Cn(\varphi)$. For every other rule, we show that its conclusion
follows from its premises.  Computational rules (\Rule{poly\_norm},
\Rule{bitblast}) rely on correctness lemmas showing normalization and
bit-blasting preserve the checker's invariant. Rule \textsc{rewrite}
reuses one proof for a table of mathematical identities.

\begin{proof}[Proof of Theorem~\ref{thm:sound:formal}]
Initially $\Delta_0 = \emptyset \subseteq
\Cn(\varphi)$. Lemma~\ref{lem:step} preserves the invariant at each
step, so by induction along $\langle \Pi_0 ; \emptyset \rangle
\longrightarrow^* \langle \Pi_n ; \Delta_n \rangle$, $\Delta_n
\subseteq \Cn(\varphi)$.  On acceptance, $\psi_{n-1} = \bot$
(Definition~\ref{def:accept}), so $\bot \in \Cn(\varphi)$; since no
assignment satisfies $\bot$, $\varphi$ is unsatisfiable.
\end{proof}

\begin{wrapfigure}[9]{r}{0.56\textwidth}
\centering
\scriptsize
\renewcommand{\arraystretch}{0.85}
\setlength{\tabcolsep}{2.5pt}
\begin{tabular}{@{}p{0.56\textwidth}@{}}
$\begin{aligned}
  \tau &\;::=\; \Any \mid \Bool \mid \BV_{32} \mid \BV_* \mid \List(\Bool) \mid \List(\BV) \\
  p &\;::=\; x \mid \top \mid \bot \mid n \mid n{:}w \mid \mathit{op}(p_1, \dots, p_k) \mid \mathrm{ite}(p, p, p) \\
    &\;\phantom{::=}\; \mid \mathrm{extract}(p_h, p_l, p) \mid \mathrm{zext}(p_e, p) \mid \mathrm{sext}(p_e, p) \mid \mathrm{repeat}(p_n, p) \\
  \mathit{entry} &\;::=\; \forall x_1 : \tau_1, \dots, x_k : \tau_k.\; \mathit{lhs} \to \mathit{rhs}\; [\,\text{if }\mathit{cond}\,]
\end{aligned}$ \\[1pt]
\quad $\text{entry 89:}\;\; \forall x, y : \BV_*.\;\; x >_s y \;\to\; y <_s x$ \\
\quad $\text{entry 53:}\;\; \forall x : \BV_*,\; n : \BV_{32}.\;\; \mathrm{extract}(n, 0, x) \to x \;\;\text{if } n \ge_s \mathrm{size}(x) - 1$ \\
\end{tabular}
\caption{The rewrite pattern language and two sample rewrite rules in the pattern language.}
\label{fig:pattern-lang}
\end{wrapfigure}

\textbf{An extensible rewrite library}. To enable kernel developers to
add new rewrite rules without re-proving the checker, \tool stores all
rewrite rules as data in a declarative lookup table rather than
hard-coding them into the checker (Figure~\ref{fig:overview}).  While
the proof certificate checker is stateful in that it reads $\Pi$ and extends
it to $\Pi'$, and checks premises in $\Delta$, each rewrite rule is a
stateless mathematical identity.  \tool therefore separates the
soundness of an entry from the task of verifying the \textsc{rewrite}
step, which instantiates an entry inside the checker state.

\textbf{A pattern language to express rewrites}.  Each rewrite table
entry is written in a small pattern language
(Figure~\ref{fig:pattern-lang}) with typed parameters $x_1 : \tau_1,
\dots, x_k : \tau_k$, a left-hand pattern $\mathit{lhs}$, a right-hand
pattern $\mathit{rhs}$, and an optional Boolean side condition
$\mathit{cond}$ (default $\top$)~\cite{kundu:pec:pldi:2009,
  lopes:alive:pldi:2015, lopes:alive2:pldi:2021,
  notzli:rare:fmcad:2022}.
Patterns use the same bit-vector
operators as terms, generalized so one entry covers a family of
rewrites.
$\BV_*$ parameters match bit-vectors of any width $w \le \RwMaxWidth$, and
list parameters ($\List(\Bool)$, $\List(\BV)$) supply multiple
operands to operators like $\wedge$ or $+$.
Given a valuation $\zeta$ mapping each $x_i$ to a value of type
$\tau_i$, a pattern $p$ evaluates under the standard SMT-LIB rules to
a value $\semr{p}{\zeta}$.  Because this meaning is defined over the
valuation $\zeta$ alone, and not over the pool or the conclusions, it
is possible to define and prove soundness for each rewrite entry on
its own, without mentioning $\Pi$ or $\Delta$. Pattern evaluation is
\emph{partial}: ill-typed operations (\eg, comparing bit-vectors of
different widths, which $\BV_*$ parameters allow) or out-of-bounds bit
extractions (\eg, when an extraction index depends on a parameter) are
undefined.

\textbf{Soundness of a rewrite entry}. An entry $\mathit{lhs} \to
\mathit{rhs}\text{ if }\mathit{cond}$ is \emph{sound} if for every
valuation $\zeta$ where $\semr{\mathit{cond}}{\zeta} = \mathrm{true}$
and both $\semr{\mathit{lhs}}{\zeta}$ and $\semr{\mathit{rhs}}{\zeta}$
are defined, the two sides evaluate to the same value
($\semr{\mathit{lhs}}{\zeta} = \semr{\mathit{rhs}}{\zeta}$).
Requiring both sides to be defined supplies width- and
bounds-compatibility preconditions automatically: in entry~89
(step~28), $x >_s y$ is defined only when $x$ and $y$ have the same
width, where $x >_s y$ and $y <_s x$ always agree; in entry~53
(step~8), $\mathrm{extract}(n, 0, x)$ is defined only when $n <
\mathrm{size}(x)$, which with $n \ge_s \mathrm{size}(x) - 1$ forces $n
= \mathrm{size}(x) - 1$ and returns $x$.

\textbf{Checking a rewrite step}. Instead of writing a separate
procedure for each of the entries in the rewrite table, \tool checks
every \Rule{rewrite} step with a single generic routine,
$\mathit{check}_{\Rule{rewrite}}$.  Given the current pool $\Pi$, the
step's premise conclusions $\psi_1, \dots, \psi_m$, a table entry
index $e$, and term arguments $\vec{a} = (a_1, \dots, a_k)$ from
$\Pi$, $\mathit{check}_{\Rule{rewrite}}$ performs four checks: (1)~it
looks up entry $e$'s parameter types $\tau_1, \dots, \tau_k$, patterns
$\mathit{lhs}$ and $\mathit{rhs}$, and side condition $\mathit{cond}$
(defaulting to $\top$), and checks that each argument $a_i$ matches
type $\tau_i$; (2)~it substitutes $\vec{a}$ for $\vec{x}$ to construct
the instantiated terms $\mathit{lhs}[\vec{a}/\vec{x}]$,
$\mathit{rhs}[\vec{a}/\vec{x}]$, and $\mathit{cond}[\vec{a}/\vec{x}]$
in $\Pi$, rejecting the step if any constructed term is ill-typed or
out of bounds; (3)~for each conjunct $c$ of
$\mathit{cond}[\vec{a}/\vec{x}]$, it verifies that $c$ (or $c = \top$)
appears among the premises $\psi_1, \dots, \psi_m$ or that $c$ is a
constant expression evaluating to $\top$; and (4)~if all checks
succeed, it returns the extended pool $\Pi'$ and records
$\mathit{lhs}[\vec{a}/\vec{x}] = \mathit{rhs}[\vec{a}/\vec{x}]$ as the
step's conclusion.

\begin{lemma}[Sound entries make sound steps]
  \label{lem:rwstep}
Let $T$ be a rewrite table in which every entry is sound. Then every
\textsc{rewrite} step over $T$ preserves the invariant of
Definition~\ref{def:inv}: if $\concl \subseteq \Cn(\varphi)$ and the
step takes $\cfg{\pool}{\concl}$ to $\cfg{\pool'}{\concl \cup
  \{\psi'\}}$, then $\psi' \in \Cn(\varphi)$.
\end{lemma}

\begin{proof}[Proof of Lemma~\ref{lem:rwstep}]
Fix $\rho \models \varphi$ and define the parameter valuation
$\zeta(x_i) = \semr{a_i}{\rho}$.  By structural induction on patterns,
evaluating an instantiated pattern $p[\vec{a}/\vec{x}]$ under $\rho$
agrees with evaluating $p$ under $\zeta$:
$\semr{p[\vec{a}/\vec{x}]}{\rho} = \semr{p}{\zeta}$.
Since $\Delta \subseteq \Cn(\varphi)$ and
$\mathit{check}_{\Rule{rewrite}}$ justifies each conjunct of
$\mathit{cond}[\vec{a}/\vec{x}]$ by a premise in $\Delta$ or constant
evaluation, $\semr{\mathit{cond}}{\zeta} = \mathrm{true}$.
Successfully building $\mathit{lhs}[\vec{a}/\vec{x}]$ and
$\mathit{rhs}[\vec{a}/\vec{x}]$ in $\Pi'$ ensures that both
$\semr{\mathit{lhs}}{\zeta}$ and $\semr{\mathit{rhs}}{\zeta}$ are
defined.
Entry soundness then gives $\semr{\mathit{lhs}[\vec{a}/\vec{x}]}{\rho}
= \semr{\mathit{lhs}}{\zeta} = \semr{\mathit{rhs}}{\zeta} =
\semr{\mathit{rhs}[\vec{a}/\vec{x}]}{\rho}$, so $\psi' =
(\mathit{lhs}[\vec{a}/\vec{x}] = \mathit{rhs}[\vec{a}/\vec{x}]) \in
\Cn(\varphi)$.
\end{proof}

\begin{lemma}[The library is sound]
  \label{lem:table}
Each of the \NumRewritesKeel{} entries of \tool's rewrite library is
sound.
\end{lemma}

Lemmas~\ref{lem:rwstep} and~\ref{lem:table} discharge the
\textsc{rewrite} case of Lemma~\ref{lem:step}. Because
$\mathit{check}_{\textsc{rewrite}}$ and Lemma~\ref{lem:rwstep} are
proved once for any sound table, adding a rewrite in the pattern
language requires no changes to the checker or
Theorem~\ref{thm:sound:formal}. A script transcribes the entry into
Lean and generates its stand-alone obligation for
Lemma~\ref{lem:table}.

\section{The Verified Checker in \embc and Refinement}
\label{sec:embc}

The functional Lean checker is sound over $\langle \Pi ; \Delta
\rangle$, but the Linux kernel requires stand-alone C code (\ie, no
external dependencies) with explicit kernel API calls and bounded
memory.
We address this requirement in two steps. First, we define \embc
(\S\ref{sec:embc:semantics}), a first-order C fragment in Lean whose
typed cell blocks avoid CompCert
Clight's~\cite{leroy:compcert:cacm:2009} per-access byte-offset and
alignment obligations while preserving Clight behavior and
undefined-behavior reflection (Theorem~\ref{thm:embc-clight}).
Second, we implement \tool in \embc with a flat \code{u32} pool,
hash-table based insertion, and in-place compaction (\S\ref{sec:refine}), and prove
end-to-end refinement and memory
safety~(Theorem~\ref{thm:refine:formal}).

\subsection{Syntax and Semantics of \embc}
\label{sec:embc:semantics}

\begin{wrapfigure}{R}{0.45\textwidth}
\small
\centering
\small
\renewcommand{\arraystretch}{0.85}
\setlength{\tabcolsep}{2.5pt}
\begin{tabular}{@{}l@{}}
$\begin{aligned}
 \tau &::= \mathsf{u32}\mid\mathsf{u64}\mid\tau^*
                  \mid\mathsf{struct}(S)\mid\mathsf{bpfptr},\\
 e &::= x\mid a_{32}\mid a_{64}\mid\mathsf{NULL}
         \mid\mathsf{table}(g,e)\\
   &\quad\mid e\mathbin\diamond e\mid !e\mid\mathord\sim e
         \mid e\mathbin{\&\&}e\mid e\mathbin{||}e\mid(\tau)e\\
   &\quad\mid e[e]\mid e\mathbin{\to}f\mid e+_{\rm ptr}e,\\
 s &::= \mathsf{skip}\mid x:=e\mid e[e]:=e\mid e\mathbin{\to}f:=e
         \mid s;s\\
   &\quad\mid\mathsf{if}\ e\ \mathsf{then}\ s\ \mathsf{else}\ s
         \mid\mathsf{while}\ e\ \mathsf{do}\ s\\
   &\quad\mid d:=F(\bar e)\mid d:=\kappa(\bar e)
         \mid\mathsf{return}\ e?\mid\mathsf{break}.
\end{aligned}$ \\
\end{tabular}
\caption{\small \embc program syntax.}
\label{fig:embc:syntax}
\end{wrapfigure}

Figure~\ref{fig:embc:syntax} defines \embc.  A program $P$ declares
structs $S$, read-only \code{u32} tables $g$ (used for rewrite
bytecode), and first-order functions $F$. Types $\tau$ comprise 32-
and 64-bit unsigned integers (\code{u32}, \code{u64}), pointers
$\tau^*$, structs $\code{struct}(S)$, and user-space handles
$\code{bpfptr}$.  To keep weakest-precondition reasoning
compositional, expressions $e$ are pure: constants $a_w$, table
lookups $\code{table}(g, e)$, casts $(\tau)e$, arithmetic/bitwise
operations $e_1 \diamond e_2$ (with short-circuit $\code{\&\&}$,
$\code{||}$), array loads $e_1[e_2]$, field reads $e \to f$, and
pointer offsets $e_1 +_{\code{ptr}} e_2$.  All state mutations,
internal calls $d := F(\bar{e})$, and kernel calls $d :=
\kappa(\bar{e})$ are statements $s$; locals live in a register
environment $\eta$ with no address-of operator.

\begin{wrapfigure}{r}{0.3\textwidth}
\small
\centering
\small
\renewcommand{\arraystretch}{0.85}
\setlength{\tabcolsep}{2.5pt}
\begin{tabular}{@{}l@{}}
$\begin{aligned}
    v \;::=\;{}&  \mathsf{u32}(a) \mid \mathsf{u64}(a) \mid \vptr{b}{i} \\
             &  \llap{$\mid\;$}\vnull \mid \vuptr \mid \vundef \\
    B \;::=\;{}&  \mathsf{dead} \mid (\tau,[v_0,\ldots,v_{\ell-1}]) \\
    \mu \;=\;{}& [B_0,\ldots,B_{k-1}] .
  \end{aligned}$ \\
\end{tabular}
\caption{\small Memory in \embc.}
\label{fig:embc:memory}
\end{wrapfigure}

\textbf{Memory}.  A value $v$ is a 32-bit or 64-bit unsigned integer,
a pointer $\vptr{b}{i}$, null, the input handle, or $\vundef$, which
marks a value that has not been written. The memory $\mu$ is a list of
blocks. Pointers follow CompCert's block-and-offset
approach~\cite{leroy:backend:jar:2009}, with offsets measured in typed
cells as shown in Figure~\ref{fig:embc:memory}. A live block pairs a
type with a list of values. We call position $i$ of the list
\emph{cell} $i$ of the block. In an array of $\tau$, each cell holds
one element of type $\tau$. In a struct, the cells are its fields, in
order. A pointer $\vptr{b}{i}$ names cell $i$ of block $b$, so offsets
count cells, not bytes. Freeing a block makes it $\mathsf{dead}$, and
allocation appends a new block.

\textbf{Operational semantics}. We define small-step operational
semantics for \embc.  The local environment $\lframe$ maps each local
variable to its declared type and current value. A callee $D$ is
either an internal function $F$ or a kernel function $\kappa$. The
machine carries an oracle $o$ and a continuation $K$ describing
pending work.  Its states $c$ have three forms:
\[
  \begin{aligned}
  c &\;::=\; \langle s,\lframe,\mu,o,K\rangle
      \;\mid\; \mathsf{call}(D,\bar v,\mu,o,K)
      \;\mid\; \mathsf{ret}(\hat v,\mu,o,K) ,\\
  K &\;::=\; \mathsf{stop} \mid \mathsf{seq}(s,K) \mid \mathsf{loop}(e,s,K)
      \mid \mathsf{fun}(\tau?,K) \mid \mathsf{caller}(d,\lframe,K) .
  \end{aligned}
\]
A regular state $\langle s,\lframe,\mu,o,K\rangle$ runs the statement
$s$ with environment $\lframe$, memory $\mu$, oracle $o$ and
continuation $K$.  The \emph{continuation} $K$ describes what the
machine does after $s$ finishes. $K$ is a stack of pending tasks, with
the next task on top: the statements left in the current sequence
($\mathsf{seq}$), the loops around $s$ ($\mathsf{loop}$), the end of
the current function ($\mathsf{fun}$, which records its result type),
and the callers waiting for its result ($\mathsf{caller}$, which saves
the caller's variables).

\begin{wrapfigure}{R}{0.615\textwidth}
\centering
\scriptsize
\setlength{\tabcolsep}{0pt}
\begin{tabular}{@{}c@{\hspace{3pt}}l@{}}
$\dfrac{\sem{p}=\vptr{b}{j}\quad \mathsf{idx}(\sem{i})=a\quad \sem{e}=v
       \quad\mathsf{write}_{\rm arr}(\mu,b,j+a,v)=\mu'}
      {\stepj{\langle p[i]:=e,\lframe,\mu,o,K\rangle}{\langle\mathsf{skip},\lframe,\mu',o,K\rangle}}$
  & \rulename{Store} \\[11pt]
$\dfrac{\sem{\bar e}=\bar v}
      {\stepj{\langle d:=D(\bar e),\lframe,\mu,o,K\rangle}{\mathsf{call}(D,\bar v,\mu,o,\mathsf{caller}(d,\lframe,K))}}$
  & \rulename{Call} \\[11pt]
$\dfrac{\mathit{Pre}_\kappa(\bar v,\mu)
       \quad \operatorname{ask}(o,\kappa,\bar v,\mu)=(\hat v,\mu',o')}
      {\stepj{\mathsf{call}(\kappa,\bar v,\mu,o,K)}{\mathsf{ret}(\hat v,\mu',o',K)}}$
  & \phantomsection\label{rule:embc:kernel}\rulename{Kernel}
\end{tabular}
\caption{\small Three of the transition rules of \embc. The
  supplementary material gives all of them.}
\label{fig:embc:rules}
\end{wrapfigure}

A call state $\mathsf{call}(D,\bar v,\mu,o,K)$ is about to enter the
function or kernel function $D$ with the arguments $\bar v$, and a
return state $\mathsf{ret}(\hat v,\mu,o,K)$ holds the optional result
$\hat v$ of a call. The empty stack $\mathsf{stop}$ marks the
outermost call, so a run of $F$ on $\bar v$ goes from
$\mathsf{call}(F,\bar v,\mu,o,\mathsf{stop})$ to $\mathsf{ret}(\hat v,\mu',o',\mathsf{stop})$.
A transition $\stepj{c}{c'}$ is one move of the
machine. Figure~\ref{fig:embc:rules} shows three rules where $\sem{e}$
is the value of $e$ under $\lframe$ and $\mu$, and $\mathsf{idx}(v)$
is the number that an integer value $v$ holds. Both are partial: $\sem{e}$ is undefined, for
example, when $e$ reads a cell that holds $\vundef$. In rule
\rulename{Store}, $\mathsf{write}_{\rm arr}(\mu,b,j+a,v)$ is defined
only if $b$ is live, $j+a$ is a cell of $b$, and $v$ has the type of
its cells. Rule \rulename{Call} saves the destination, the caller's
variables and its continuation in $\mathsf{caller}(d,\lframe,K)$. A
later transition assigns the result to $d$. The supplementary material
provides detailed semantics of \embc.

\textbf{Kernel oracle}. \tool interacts with the Linux kernel solely
through the seven primitives $\kappa$: three allocators
($\code{kvcalloc}$, $\code{kzalloc}$, $\code{kvrealloc}$), free
($\code{kvfree}$), user copy ($\code{copy\_from\_bpfptr}$), signal
check ($\code{signal\_pending}$), and preemption
($\code{cond\_resched}$) (see supplementary material for a detailed
description).  Each primitive has a \emph{precondition}
$\mathit{Pre}_\kappa(\bar{v}, \mu)$ and \emph{postcondition}
$\mathit{Post}_\kappa(\bar{v}, \mu, \hat{v}, \mu')$. To model
nondeterministic kernel outcomes (allocation failure, user bytes
$\bar{a}$, signals) deterministically, \embc uses a stateful
\emph{oracle}:

\begin{definition}[Kernel Oracle]
\label{def:oracle}
An \emph{oracle} $o = (\xi, \mathit{ans})$ pairs an internal state
$\xi$ with a function $\mathit{ans}$ that, for every $\kappa, \bar{v},
\mu$ satisfying $\mathit{Pre}_\kappa(\bar{v}, \mu)$, returns
$(\hat{v}, \mu', \xi') = \mathit{ans}(\xi, \kappa, \bar{v}, \mu)$
satisfying $\mathit{Post}_\kappa(\bar{v}, \mu, \hat{v}, \mu')$.  We
write $\mathrm{ask}(o, \kappa, \bar{v}, \mu) = (\hat{v}, \mu', (\xi',
\mathit{ans}))$.
\end{definition}

Rule \rulename{Kernel} (Figure~\ref{fig:embc:rules}) requires the
precondition and uses $\operatorname{ask}$ to obtain the reply and
updated oracle. A call that violates the precondition has no successor
and is \emph{stuck}. Every valid call receives a reply satisfying the
postcondition; in Lean, this guarantee is part of the oracle's type.
The checker specifications quantify over every oracle. The proofs
establish each call's precondition and use its postcondition without
inspecting the oracle's state.

\textbf{Stuck states}.  A state $c$ is \emph{stuck} if $c \neq
\mathsf{ret}(\hat{v}, \mu, o, \mathsf{stop})$ and no $c'$ satisfies $c
\leadsto c'$.  Every \embc rule enforces the validity conditions of C
and the kernel functions as premises: (1)~reads/writes access a cell
$0 \le i < \ell$ of a live block $B_b \neq \mathsf{dead}$ with $\ell$
cells, at matching type, and never read $\vundef$; (2)~divisors are
nonzero and shift counts are below the operand width $w$; (3)~pointer
offsets stay in $[0, \ell]$ of an array block, and pointer comparisons
compare pointers in the \emph{same} live block (or $\vnull$);
(4)~non-void functions exit via $\code{return}$; and
(5)~\rulename{Kernel} requires $\mathit{Pre}_\kappa(\bar{v}, \mu)$.

For example, consider $\code{x := kvcalloc}_{\code{u32}}(4)$ in $\mu$
($k = |\mu|$).  Since $4 > 0$, $\mathit{Pre}_{\code{kvcalloc}}$ holds;
$o$ may either fail ($\hat{v} = \vnull, \mu' = \mu$) or succeed
($\hat{v} = \vptr{k}{0}$ with 4 zeroed cells).
Writing $\code{x[0] := 1}$ without checking $\code{x != NULL}$ steps
under the success reply but \emph{gets stuck} under the failure reply;
since safety quantifies over \emph{all} oracles $o$, every fallible
call must be checked.  Conversely, calling
$\code{kvcalloc}_{\code{u32}}(0)$, freeing $\vptr{b}{i}$ with $i
> 0$, or freeing a block that is already $\mathsf{dead}$ violates $\mathit{Pre}_\kappa$
and gets stuck for \emph{every} oracle.

\textbf{Correspondence with CompCert Clight}.  A \PrinterLines-line printer
emits C source from an \embc AST $P$.  To further gain confidence
about \embc semantics, we validate \embc by embedding $P$ as $P_{\mathrm{Cl}}$
in CompCert Clight (in Lean via Clean~\cite{certora:clean:2026}). We
relate $\mu$ to Clight byte memories $m$ via a memory representation
relation $R(\mu, m)$ under an injective block map $\iota$: each live
$B_b = (\tau, [v_0, \dots, v_{\ell-1}])$ in $\mu$ maps to a disjoint
live Clight block $\iota(b)$ in $m$ of byte size $\ell \cdot
\mathrm{sizeof}(\tau) < 2^{64}$ (LP64), storing each $v_i \neq
\vundef$ at byte offset $i \cdot \mathrm{sizeof}(\tau)$ (or field
$f$ at $\delta(S, f)$) and mapping $\vptr{b}{i}$ to
$\code{Vptr}(\iota(b), i \cdot \mathrm{sizeof}(\tau))$.

It is worth noting that \embc is \textbf{stricter} than
Clight. Comparing pointers $\code{p == q}$ from distinct live blocks
$b_1 \neq b_2$ is stuck in \embc (simplifying checker proofs, which
only compare pointers within one array), whereas Clight evaluates
$\code{p == q}$ to $0$.  Because \embc has more stuck states, we prove
a two-part correspondence—\emph{forward behavior equivalence} for safe
programs and \emph{backward UB reflection}:

\begin{theorem}[Clight Equivalence and Reflection of Target UB]
\label{thm:embc-clight}
For any well-formed \embc program $P$ whose translation
$P_{\mathrm{Cl}}$ exists, and for initial memories with $R(\mu_0,
m_0)$, assuming that Clight's built-in kernel functions have exactly
the outcomes of the contracts and that every reachable \embc state
fits Clight's 64-bit representation (the supplementary material states
these assumptions exactly):
\begin{enumerate}

  \item \emph{(Behavior equivalence on safe inputs)} If calling
    $F(\bar{v})$ in \embc from $\mu_0$ never gets stuck, whatever
    replies the kernel contracts allow, all its behaviors and final
    memories (up to $R(\mu, m)$) coincide with $P_{\mathrm{Cl}}$ in
    Clight from $m_0$.

  \item \emph{(Reflection of target UB)} If a Clight run of
    $P_{\mathrm{Cl}}$ from $m_0$ gets stuck, then for some replies that
    the kernel contracts allow, the \embc run from $\mu_0$ gets stuck
    after a prefix of the Clight trace. This part does not assume that
    the \embc run is safe.

\end{enumerate}
\end{theorem}

\begin{proof}[Proof sketch]
Both parts follow from a forward and a backward simulation in the
style of CompCert~\cite{leroy:backend:jar:2009}, which relate memories
by $R$, local variables, and continuations, and allow finitely many
extra steps on either side. The forward simulation maps each non-stuck
\embc step to Clight steps; it uses $\ell \cdot \mathrm{sizeof}(\tau)
< 2^{64}$ so that cell offsets never overflow 64-bit byte offsets. The
backward simulation maps each Clight step back to \embc. At a kernel
call it picks the \embc reply that matches the Clight reply. Part~(1) uses both
simulations. Part~(2) uses the backward one alone.
\end{proof}

\subsection{\embc Checker Refines the Functional Checker}
\label{sec:refine}

\tool implements the in-kernel checker in \embc. Implementing the
functional checker described in \S\ref{sec:checker} in the
kernel requires us to address two challenges. First, we need to avoid
unnecessary allocation of nodes and recursive tree walks.  Second,
monotonically growing terms in the pool can cause significant memory
overheads. \tool's in-kernel checker overcomes both by representing
$\cfg{\pool}{\concl}$ in five disjoint flat arrays of 32-bit units.
Each array is one block of the memory $\mu$, and a \emph{state struct}
in another block points to all five.
(1) A contiguous array \cid{pool} stores $\Pi$
as a linearized directed acyclic graph (DAG).  The \cid{pool} is built
bottom-up. Every compound term satisfies the topological invariant,
which ensures that the operand indices $i_1, \ldots, i_m$ of the
compound term are less than the compound term's index $i$.
It ensures that \cid{pool} is cycle-free and well-founded.
We write $\poolat{i}$ as the term that \cid{pool} holds at $i$,
$\poolat[\mu]{i}$ when the memory matters, and $\mathrm{dom}(\cid{pool}_\mu)$ for the
indices that hold a term.
Further, any backward pass over \cid{pool} visits every parent before
its children, and any forward pass visits every child before its
parents, which enables reclamation of dead terms.
(2) To store each term of $\Pi$ at most once in \cid{pool}, the
open-addressed table \cid{htab} holds $0$ in empty slots and $i + 1$
for a term at index $i$.  Before adding a new term to the pool,
insertion hashes the expression and probes \cid{htab}. It reuses the
expression on a hit or adds a new expression to the pool. By induction
on construction, each stored term has a unique canonical index,
reducing deep DAG equality to $O(1)$ index equality (\ie, $i_1 =
i_2$).
(3) To avoid searching for premises or re-parsing terms at each step,
the conclusions array \cid{facts[j]} stores the canonical \cid{pool}
index of step $j$'s conclusion $\psi_j \in \Delta$.
(4) Initialization copies the user-space certificate $\pi$ via
$\code{copy\_from\_bpfptr}$ into a private buffer \cid{cert}
consisting of a term section (serializing $\Pi_\pi$) followed by a
step section (serializing $s_0, \dots, s_{n-1}$), and inserts $\Pi_0 =
\Pi_\pi \cup \{\top, \bot\}$ into the pinned prefix $\cid{pool}[0..\cid{pin})$.
The fields \cid{tt} and \cid{ff} hold the indices of $\top$ and $\bot$.
The loader overwrites each certificate term in \cid{cert} with its pool
index plus one, so the term section of \cid{cert} also serves as a
lookup table from certificate positions to pool indices.
(5) Before checking steps, a single linear scan over the step section
records in \cid{last\_cit[j]} the largest step index $i > j$ that
cites step $j$ as a premise (or $0$ if step $j$ is never cited).
After $k$ steps ($s_0, \dots, s_{k-1}$) have been checked, a term in
$\Pi$ is \emph{live} if it belongs to the initial pool $\Pi_0$ ($i <
\cid{pin}$), is a derived conclusion $\psi_j$ ($j < k$) still
needed by a future step ($\cid{last\_cit}[j] \ge k$), or is a
subterm of a live term; all other dynamically generated terms are
\emph{dead}. Once the pool has outgrown its live terms by a set margin,
compaction ($\code{drop}$) reclaims dead terms in place by
sliding each live index, rewriting operands, updating live
\cid{facts}, and rebuilding the hash table \cid{htab}.

\textbf{Step execution in $O(1)$ lookups}.  These five arrays make
step execution fast.
To check step $s_i = (R, (j_1, \dots, j_m), (p_1, \dots, p_r))$,
$\checkC$ resolves each premise index $j_\ell < i$ to $q_\ell =
\cid{facts}[j_\ell]$ (decoding to $\psi_{j_\ell}$) and each
certificate parameter $p_\ell$ that names a term to $a_\ell = \cid{cert}[p_\ell] - 1$
(decoding to $u_\ell$), invokes rule $R$'s checker
$\mathit{check}_{R,\mathrm{C}}$ on $(q_1, \dots, q_m)$ and $(a_1,
\dots, a_r)$, and writes the resulting pool index $c$ to
$\cid{facts}[i]$.

\textbf{State correspondence}.
A key verification challenge is that $\checkL$ is
monotonic---$\Pi$ and $\Delta$ only grow---whereas
$\checkC$ periodically deletes dead terms and
renumbers surviving dynamic terms from $i \ge \cid{pin}$ to
$\sigma(i) \le i$.  If our simulation relation required all of $\Pi$
to stay in $\cid{pool}$, or fixed the concrete pool indices of
dynamic terms, each of the proofs for various inference rules would
have to reason about compaction.
To avoid this issue, we define \emph{state correspondence} $\mu
\crep \cfg{\pool}{\concl}$ over \emph{live} terms, leaving
the concrete indices $i \ge \cid{pin}$ of dynamically generated
terms unconstrained while pinning $\Pi_0$ at $i < \cid{pin}$:

\begin{definition}[State Correspondence]
\label{def:correspondence}

Let $\cfg{\pool}{\concl}$ be the abstract configuration of
$\checkL$ after $k$ steps ($0 \le k \le n$), with
$\Delta = \{\psi_0, \dots, \psi_{k-1}\}$. A concrete memory $\mu$
\emph{corresponds} to $\cfg{\pool}{\concl}$, written $\mu
\crep \cfg{\pool}{\concl}$, if:
\begin{enumerate}
\item[(D1)] \emph{(Pool soundness + uniqueness)} Every
  $\poolat[\mu]{i} \in \Pi$, and $\poolat[\mu]{i_1} =
  \poolat[\mu]{i_2} \implies i_1 = i_2$.

\item[(D2)] \emph{(Live completeness)} Every live term is decoded by
  some $i$ in $\cid{pool}$,\newline
  and $\poolat[\mu]{\cid{tt}} = \top$, $\poolat[\mu]{\cid{ff}} = \bot$.

\item[(D3)] \emph{(Hash table consistency)} $\cid{htab}$ stores $i
  + 1$ for each term $i$ in \cid{pool} ($0$ elsewhere), and each term is
  reachable from the slot of its hash through occupied slots.

\item[(D4)] \emph{(Live conclusions)} For every completed step $j < k$
  that is cited by a remaining step ($\cid{last\_cit}[j] \ge k$),
  $\poolat[\mu]{\cid{facts}[j]} = \psi_j$.

\item[(D5)] \emph{(Certificate and citations)} The step section of
  $\cid{cert}$ encodes $s_0, \dots, s_{n-1}$; for each certificate
  term $u_p \in \Pi_\pi$, $\cid{cert}[p] = r + 1$ with $r <
  \cid{pin}$ and $\poolat[\mu]{r} = u_p$; and
  $\cid{last\_cit}[j]$ holds the last step index $> j$ citing step
  $j$ ($0$ if uncited).
\end{enumerate}
\end{definition}

Definition~\ref{def:correspondence} makes in-place compaction
completely transparent to the inference rules of the certificate
checker, because $\mu \crep \cfg{\pool}{\concl}$ specifies
\emph{which} live terms are represented in $\mu$ rather than
\emph{where} dynamic terms sit.
Further, compaction transitions $\mu$ to a compacted memory $\mu'$
satisfying $\mu' \crep \cfg{\pool}{\concl}$ for the
\emph{exact same} abstract state $\cfg{\pool}{\concl}$
(Lemma~\ref{lem:compaction}).

\subsection{The Refinement and Safety Theorem}
\label{sec:c-proof}

We now state and prove our main result for $\checkC$: for every input
formula $\varphi$, user-space certificate pointer $\code{ubuf}$, and
kernel oracle $o$, $\checkC$ executes safely without getting stuck,
preserves all caller memory while freeing all of its own allocations,
and accepts only when the functional checker $\checkL$ accepts.

\begin{theorem}[Refinement and Safety]
\label{thm:refine:formal}
Let $\checkC$ be called in an initial memory $\mu_{\mathrm{in}}$
encoding a valid formula $\varphi$.  For every $\varphi$, user buffer
$\code{ubuf}$, and oracle $o$, $\checkC$ never
gets stuck, and if it returns in memory $\mu'$:
\begin{enumerate}
  \item \emph{(Memory safety and cleanup)} Every block $b$ in
    $\mu_{\mathrm{in}}$ is unchanged in $\mu'$ ($\mu'(b) =
    \mu_{\mathrm{in}}(b)$), and every block allocated during
    $\checkC$ is $\mathsf{dead}$ in $\mu'$;
  \item \emph{(Soundness by refinement)} On $\accept$, the bytes
    copied from $\code{ubuf}$ decode to a certificate $\pi$ whose
    \textsc{assume} step has the formula $\varphi$ (up to variable
    renaming), and $\checkL(\pi) = \accept$.
\end{enumerate}
\end{theorem}

By Theorem~\ref{thm:embc-clight} (Clight equivalence and UB
reflection), Theorem~\ref{thm:refine:formal} transfers to
the Clight program $P_{\mathrm{Cl}}$: for any initial Clight
memory $m_{\mathrm{in}}$ with $R(\mu_{\mathrm{in}}, m_{\mathrm{in}})$,
$P_{\mathrm{Cl}}$ is free of Clight undefined behavior, leaves caller
memory untouched, frees all allocated blocks, and returns
$\accept$ only when $\checkL(\pi) =
\accept$---which by Theorem~\ref{thm:sound:formal} guarantees that
$\varphi$ is unsatisfiable.

\textbf{Overview of the four-layer proof architecture}.
We write $\{P\}\; F(\bar{v})\; \{Q\}$ for total-safety Hoare triples
in \embc: starting from any state $(\mu, o)$ satisfying $P$,
calling $F(\bar{v})$ never gets stuck for any oracle replies, and if
$F(\bar{v})$ returns value $\hat{v}$ in memory $\mu'$ with oracle
$o'$, then $(\hat{v}, \mu', o')$ satisfies $Q$.  We decompose the
proof of Theorem~\ref{thm:refine:formal} into four modular
layers.

\textbf{Layer 1: Footprint invariant}.  In \embc, every allocation
($\code{kvcalloc}$, $\code{kzalloc}$, $\code{kvrealloc}$) appends a
fresh block at index $b \ge |\mu_{\mathrm{in}}|$, and $\code{kvfree}$
marks a block $\mathsf{dead}$ in place without reusing its block ID.
Thus, any block allocated by $\checkC$
automatically has block ID $b \ge |\mu_{\mathrm{in}}|$, structurally
separating \tool's state from all caller blocks $b <
|\mu_{\mathrm{in}}|$:

\begin{definition}[Footprint Invariant]
\label{def:footprint}
A memory $\mu$ satisfies the \emph{footprint invariant} $I$ if:
\begin{enumerate}
  \item[(I1)] \emph{(Disjoint checker state)} The state struct, each
    allocated array it points to (\cid{pool}, \cid{htab}, \cid{facts},
    \cid{last\_cit}, and the scratch arrays \cid{prem}, for the current
    step's premises, and \cid{stk}, a stack that rules and compaction
    use) and the copy \cid{cert}, once allocated, are
    distinct live blocks at $b \ge |\mu_{\mathrm{in}}|$. Each array sits
    at offset~$0$ of a $\code{u32}$ block whose length
    matches its capacity bound.
  \item[(I2)] \emph{(No leaked blocks)} All other blocks at $b \ge
    |\mu_{\mathrm{in}}|$ are $\mathsf{dead}$.
  \item[(I3)] \emph{(Caller preservation)} Every caller block $b <
    |\mu_{\mathrm{in}}|$ is unchanged from its initial state ($\mu(b) =
    \mu_{\mathrm{in}}(b)$).
\end{enumerate}
\end{definition}

\begin{lemma}[Footprint Preservation]
\label{lem:footprint}
Every helper and rule function $F$ invoked during certificate checking
satisfies $\{I\}\; F(\bar{v})\; \{I\}$.
\end{lemma}

\begin{proof}
Each helper and rule function reads caller memory only during
$\code{compare}$ (the read-only check that the assumed formula is
$\varphi$) and mutates only the
checker state at $b \ge |\mu_{\mathrm{in}}|$ or temporary buffers
(the forwarding table $\code{fw}$ in $\code{drop}$) allocated via $\code{kvcalloc}$ at $b
\ge |\mu_{\mathrm{in}}|$ and freed via $\code{kvfree}$ before returning.
When term construction grows $\cid{pool}$ via $\code{kvrealloc}$ or
$\cid{htab}$ via $\code{kvcalloc}$ and $\code{kvfree}$, the old block at $b \ge |\mu_{\mathrm{in}}|$ becomes
$\mathsf{dead}$ and a fresh block at $|\mu| \ge |\mu_{\mathrm{in}}|$
replaces it (or the allocation fails and leaves $\mu$ unchanged).
So $I$ holds on exit.
\end{proof}

Because $I$ is preserved across every call, all specifications below
implicitly conjoin $I$ to their preconditions and postconditions.

\textbf{Layer 2: Compositional rule refinement}.  Both
term construction and every proof rule
$R$ of \S\ref{sec:checker} can be given the functional signature $f(\Pi, \vec{u}) = (\Pi', u')$,
taking the current term pool $\Pi$ and tuple of input terms $\vec{u} =
(u_1, \dots, u_m)$ and either rejecting or returning an extended pool
$\Pi' \supseteq \Pi$ and output term $u' \in \Pi'$.  In
$\checkC$, the corresponding \embc function
$F$ receives a tuple of pool indices $\vec{\imath} = (i_1, \dots, i_m)$ and
returns either $\code{NOID}$ ($2^{32}-1$, signaling rejection or
resource exhaustion) or the canonical pool index $\hat{v}$ of $u'$:

\begin{definition}[Refinement]
\label{def:refinement}
An \embc function $F$ \emph{refines} a functional operation
$f(\Pi, \vec{u}) = (\Pi', u')$ if for every abstract configuration
$\cfg{\pool}{\concl}$, term tuple $\vec{u} = (u_1, \dots,
u_m)$, and index tuple $\vec{\imath} = (i_1, \dots, i_m)$, calling
$F(\vec{\imath})$ from $\mu \crep \cfg{\pool}{\concl}$ with
$\forall \ell \in [1, m].\, \poolat[\mu]{i_\ell} = u_\ell$ satisfies
the postcondition
\[
\begin{aligned}
  \hat{v} = \code{NOID} \;\vee\; \bigl(&f(\Pi, \vec{u}) = (\Pi', u') \;\wedge\; \poolat[\mu']{\hat{v}} = u' \;\wedge\; \mu' \crep \cfg{\pool'}{\concl} \\
  &{}\wedge\; \forall i \in \mathrm{dom}(\cid{pool}_\mu).\, \poolat[\mu']{i} = \poolat[\mu]{i}\bigr).
\end{aligned}
\]
\end{definition}

Notice the monotonicity clause $\forall i \in \mathrm{dom}(\cid{pool}_\mu).\, \poolat[\mu']{i} =
\poolat[\mu]{i}$ in Definition~\ref{def:refinement}: when a rule
constructs multiple intermediate subterms in sequence, each hash-table
insertion adds or reuses a node without invalidating any previously
resolved pool index.  This makes multi-allocation rule proofs
completely compositional.

\begin{lemma}[Checking a Rule]
\label{lem:check-rule}
Hash-table term insertion refines functional term construction
$((\Pi, \vec{u}) \mapsto (\Pi \cup \{\mathrm{op}(\vec{u})\},
\mathrm{op}(\vec{u})))$, and for each of the \NumRules{} proof inference rules
$R$, $\mathit{check}_{R,\mathrm{C}}$ refines $\mathit{check}_R$
(Definition~\ref{def:rulecheck}).
\end{lemma}

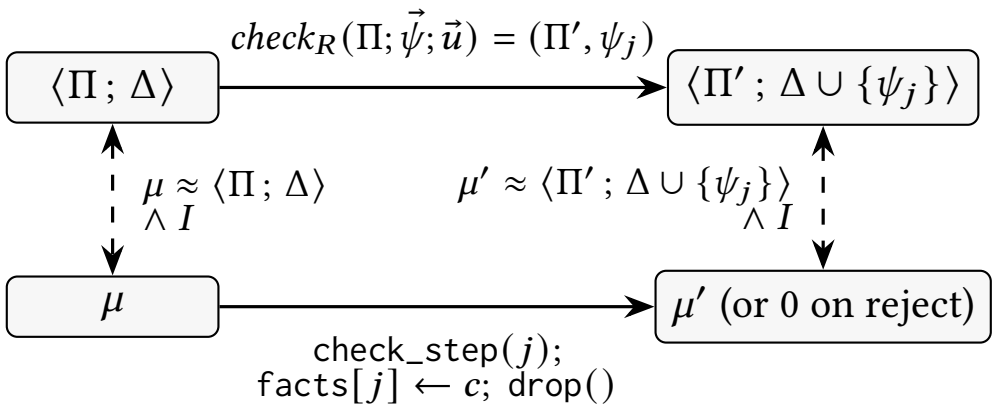
\begin{wrapfigure}[9]{r}{0.5\textwidth}
  \centering
  \vspace{-12pt}
\begin{tikzpicture}[
  >=Stealth,
  font=\footnotesize,
  box/.style={draw, rounded corners=2pt, inner sep=3pt, minimum width=1.4cm, fill=gray!6}
]
  \node[box] (abs1) at (0, 1.45) {$\cfg{\pool}{\concl}$};
  \node[box, minimum width=1.85cm] (abs2) at (4.7, 1.45) {$\cfg{\pool'}{\concl \cup \{\psi_j\}}$};
  \node[box] (conc1) at (0, 0) {$\mu$};
  \node[box, minimum width=1.85cm] (conc2) at (4.7, 0) {$\mu'$ (or $0$ on reject)};

  \draw[->, semithick] (abs1) -- node[above=1.5pt, font=\scriptsize] {$\mathit{check}_R(\Pi; \vec{\psi}; \vec{u}) = (\Pi', \psi_j)$} (abs2);
  \draw[->, semithick] (conc1) -- node[below=1.5pt, font=\scriptsize, align=center] {$\code{check\_step}(j);$\\[-1.5pt]$\cid{facts}[j] \gets c;\; \code{drop}()$} (conc2);
  \draw[<->, dashed, semithick] (abs1) -- node[right=2pt, font=\scriptsize, align=left] {$\mu \crep \cfg{\pool}{\concl}$\\[-1.5pt]$\wedge\; I$} (conc1);
  \draw[<->, dashed, semithick] (abs2) -- node[left=2pt, font=\scriptsize, align=right] {$\mu' \crep \cfg{\pool'}{\concl \cup \{\psi_j\}}$\\[-1.5pt]$\wedge\; I$} (conc2);
\end{tikzpicture}
\caption{Simulation diagram for step $j \in \{0, \dots,
  n-1\}$ of $\checkC$
  (Lemma~\ref{lem:simulation}).}
\vspace{12pt}
\label{fig:sim-overview}
\end{wrapfigure}

\textbf{Layer 3: Soundness of in-place compaction}.  Between steps,
$\code{drop}$ reclaims dead dynamic terms while preserving $\mu
\crep \cfg{\pool}{\concl}$:

\begin{lemma}[Compaction]
\label{lem:compaction}
After $k < n$ steps, $\{\mu \crep \cfg{\pool}{\concl}\}\; \code{drop}()\; \{\hat{v} = 0 \vee \mu' \crep \cfg{\pool}{\concl}\}$.
\end{lemma}

We proved Lemmas~\ref{lem:check-rule} and~\ref{lem:compaction} in Lean, and the
supplementary material gives the ideas of the proofs.

\textbf{Layer 4: Single-step simulation}.  Combining
Lemmas~\ref{lem:footprint} to \ref{lem:compaction} yields the
invariant of the step-checking loop (Figure~\ref{fig:sim-overview}).

\begin{lemma}[Simulation]
\label{lem:simulation}
If $\mu \crep \cfg{\pool}{\concl}$ holds before step $j \in
\{0, \dots, n-1\}$ (where $\Delta = \{\psi_0, \dots, \psi_{j-1}\}$),
checking step $j$ never gets stuck and either returns $0$ (rejecting)
or transitions $\cfg{\pool}{\concl} \stepto \cfg{\pool'}{\concl \cup \{\psi_j\}}$ (Definition~\ref{def:step}) in memory
$\mu'$ with $\mu' \crep \cfg{\pool'}{\concl \cup \{\psi_j\}}$ (and $\poolat[\mu']{\cid{facts}[n-1]} = \psi_{n-1}$ when
$j = n - 1$).
\end{lemma}

See the supplementary material for the proof of Lemma~\ref{lem:simulation} and Theorem~\ref{thm:refine:formal}.

\section{Time and Memory Bounds with Potentials}
\label{sec:cost}

Given that the kernel invokes $\checkC$ on untrusted user inputs when
an eBPF program is loaded, we want to ensure that a malicious user
does not cause denial of service by stalling it or exhausting kernel
memory. To address this issue, \tool follows the eBPF verifier (\eg,
bounded programs) in maintaining two runtime counters, $\code{work}$
and $\code{mem}$, initialized linearly in $B$ (where $B =|\pi| +
|\varphi|$) and decremented via a helper $\code{spend}(k)$ before
every input-dependent loop, term insertion, hash probe, and
allocation.  We turn them into the potential functions of a
machine-checked amortized analysis on \embc (\S\ref{sec:embc:semantics}), proving
that \emph{every} execution terminates in $O(B)$ time and $O(B)$
memory.

\textbf{Measuring time and memory usage}. We measure time $T$ (in
ticks) and memory $M$ (in bytes) directly on each small-step
transition $c \leadsto c'$ of \embc (\S\ref{sec:embc:semantics}).  Each transition is
charged according to the work it performs.  First, every internal
statement, branch, function call, preemption check
($\code{signal\_pending}$, $\code{cond\_resched}$), and
$\code{kvfree}$ costs $1$ tick and $0$ bytes.  Second, copying $n$
bytes from user space ($\code{copy\_from\_bpfptr}$) touches each byte
once, costing $1 + n$ ticks and $0$ bytes.  Third, allocating or
resizing $n$ bytes ($\code{kvcalloc}$, $\code{kzalloc}$,
$\code{kvrealloc}$) costs $n$ bytes and $1 + 2n$ ticks.  This time
charge covers $n$ ticks to zero or copy the block upon allocation,
plus $n$ ticks prepaid for the eventual $\code{kvfree}$ so that
cleanup takes $1$ tick regardless of block size.

\begin{definition}[Execution and Cost]
\label{def:run-cost}
An \emph{execution} of $F(\bar{v})$ from $\sigma = (\mu, o)$ to
$\sigma' = (\mu', o')$, written $\sigma \vdash F(\bar{v}) \Downarrow
\hat{v}, \sigma' \mid (T, M)$, is a finite sequence $\mathsf{call}(F,
\bar{v}, \mu, o, \mathsf{stop}) \leadsto^* \mathsf{ret}(\hat{v}, \mu', o',
\mathsf{stop})$ whose transitions sum to $T$ ticks and $M$ bytes (where
$M$ never subtracts freed bytes, bounding both cumulative and peak
memory).
\end{definition}

\textbf{Budgets as potential}. Before any operation whose cost depends
on the input or dynamic state, $\checkC$ invokes $\code{spend}(k)$,
which checks whether $\code{work} \ge k$, deducts $k$ from
$\code{work}$, and otherwise aborts.  A loop of $k$ input-dependent
iterations calls $\code{spend}(k)$ up front before entering the
loop. Appending an $m$-operand term to \cid{pool} (\S\ref{sec:refine}) calls
$\code{spend}(1 + m)$, paying $1$ unit per $\code{u32}$ cell
written. Each collision probe in \cid{htab} calls
$\code{spend}$ before advancing to the next slot. Growing
\cid{pool} via $\code{kvrealloc}$ or \cid{htab} via $\code{kvcalloc}$ to a new
capacity $n$ deducts $n$ from both $\code{work}$ and $\code{mem}$.

\begin{definition}[Budgets]
\label{def:budgets}
For an input of $B = |\pi| + |\varphi|$ bytes, the \emph{time budget}
$W_T(B) = O(B)$ is the initial value of $\code{work}$ set by
$\code{init}$, and the \emph{memory budget} $W_M(B) = O(B)$ is the
initial value of $\code{mem}$ plus the allocations made before
$\code{mem}$ is set (the $|\pi|$-byte certificate buffer \cid{cert} and
fixed-size state tables).
\end{definition}

\textbf{Why metering eliminates data-structure invariants}.  Proving
worst-case bounds for DAG and hash-table algorithms normally requires
bounding hash-table load factors and DAG expansion.  Because every
collision probe in \cid{htab} and every variable-length loop
calls $\code{spend}$ before executing, our proof needs \emph{no}
hash-distribution or DAG-shape invariants: even if every probe
collided, each probe consumes at least $1$ unit of $\code{work}$ until
$\code{spend}$ aborts.  In Tarjan's amortized framework, $\code{work}$
and $\code{mem}$ act directly as \emph{potential functions}
$\Phi_T(\sigma) = R \cdot \sigma.\code{work} \ge 0$ and
$\Phi_M(\sigma) = \sigma.\code{mem} \ge 0$, where $R \in \mathbb{N}$
is the maximum ticks per unit of $\code{work}$ spent, so $T +
\Phi_T(\sigma') \le \Phi_T(\sigma_0)$ immediately implies $T \le R
\cdot \sigma_0.\code{work} = O(B)$.

\begin{theorem}[Cost Bound]
\label{thm:cost}
There is a static constant $R \in \mathbb{N}$ such that for every
formula $\varphi$ and certificate $\pi$ of $B = |\pi| + |\varphi|$
bytes, initial memory $\mu_{\mathrm{in}}$ encoding $\varphi$, and kernel
oracle $o$, calling $\mathit{check}_{\mathrm{C}}(\varphi,
\code{ubuf})$ terminates in an execution $\sigma_{\mathrm{in}} \vdash
\mathit{check}_{\mathrm{C}}(\varphi, \code{ubuf}) \Downarrow \hat{v},
\sigma' \mid (T, M)$ satisfying $T \le R \cdot W_T(B) = O(B)$ and $M
\le W_M(B) = O(B)$.
\end{theorem}

\begin{lemma}[Total Correctness of \tool]
\label{cor:total}
Combining Theorems~\ref{thm:sound:formal}, \ref{thm:embc-clight}, \ref{thm:refine:formal}, and~\ref{thm:cost}, for every input
$(\varphi, \pi)$ of $B$ bytes and every kernel oracle $o$,
$\mathit{check}_{\mathrm{C}}$ terminates within $O(B)$ time and $O(B)$ memory
without undefined behavior, leaves caller memory $\mu_{\mathrm{in}}$
unchanged, frees all allocated blocks, and returns $\accept$
only if $\varphi$ is unsatisfiable.
\end{lemma}

\textbf{Proving the cost bound}. To prove Theorem~\ref{thm:cost}
compositionally across all functions of $\checkC$, we must account for
the code that runs \emph{between} $\code{spend}$ calls.  We partition
the ticks of any call $F(\bar{v})$ into: (1)~\emph{metered ticks}
(variable-bound loops, term insertions, \cid{htab} probes, and
$n$-byte kernel copies/allocations), which follow $\code{spend}(k)$
and cost at most $R \cdot k$ ticks; and (2)~\emph{unmetered overhead}
(straight-line code, branches, and loops over a term's operands,
bounded by the arity field, \S\ref{sec:background}), which we bound by a static
per-function allowance $A_F^T$.

\begin{definition}[Bounded Function]
\label{def:bounded-fn}
An \embc function $F$ is \emph{bounded} with static allowances
$(A_F^T, A_F^M)$ if every call of $F(\bar{v})$ from $\sigma = (\mu,
o)$ terminates (in a stuck state or an execution $\sigma \vdash
F(\bar{v}) \Downarrow \hat{v}, \sigma' \mid (T, M)$ with
$\sigma'.\code{work} \le \sigma.\code{work}$ and $\sigma'.\code{mem}
\le \sigma.\code{mem}$) and every execution satisfies $T \le R \cdot
(\sigma.\code{work} - \sigma'.\code{work}) + A_F^T$ and $M \le
(\sigma.\code{mem} - \sigma'.\code{mem}) + A_F^M$.
\end{definition}

\begin{lemma}[Cost of a Statement]
\label{lem:stmt-cost}
If every callee in $F$ is bounded, then every statement $s$ in $F$ terminates with static allowances $(A_s^T, A_s^M)$ satisfying $T + R \cdot \sigma'.\code{work} \le R \cdot \sigma.\code{work} + A_s^T$ and $M + \sigma'.\code{mem} \le \sigma.\code{mem} + A_s^M$.
\end{lemma}

\begin{lemma}[Cost of a Function]
\label{lem:fn-cost}
Every function of $\checkC$ is bounded.
\end{lemma}

Lemma~\ref{lem:stmt-cost} follows by induction on statements and
Lemma~\ref{lem:fn-cost} by induction up the call graph. The supplementary
material gives these proofs and that of Theorem~\ref{thm:cost}.

\section{Experimental Evaluation}
\label{sec:eval}
We describe our evaluation of \tool focusing on four dimensions: (1)
how the formalization with \tool helped us discover the bugs in the
BCF checker, (2) \tool's performance and memory overheads compared to
BCF, (3) how \tool's optimizations such as compaction and cost
metering affect performance, and (4) additional effort needed to extend
and verify with new rewrite rules.

\textbf{Prototype and methodology}. \tool is open source and publicly
available~\cite{keel-repo}. \tool is specifically built for Linux kernel~6.19, which is
the version that the BCF checker targets.  \tool takes as input a binary
proof certificate $\pi$ and, in the kernel, the refinement condition
$\varphi$ from the eBPF verifier, which it checks against the formula
that $\pi$ assumes. The certificate reaches \tool through a user-space
buffer, supplied by the loader in the kernel or read from a stand-alone
certificate file in our experiments.
We perform the evaluation using object files from BCF's
repository~\cite{bcf-repo}. It contains two kinds of object files: (1)
\IdxObjects{} object files from real-world projects including Calico,
Cilium, Inspektor Gadget, BCC and a few others and (2)
\CorpusShipped{} proof certificate files generated by BCF for their
evaluation. During the evaluation, we classified the BCF certificates
with the expected outcome: accept or reject. In the process of building
\tool, we used fuzzing to generate a corpus of \FuzzInputsM{} million proof
certificates through mutation of existing proof certificates. We use
them to differentially test \tool and the BCF checkers. We perform all
our experiments on one server with an \XCPU{}. The loading tests run in
QEMU 6.2 virtual machines running Linux 6.19-rc1 with BCF's patches
applied, built twice, once with BCF's checker and once with \tool.

\begin{wraptable}{r}{0.5\textwidth}
\vspace{-10pt}
\caption{The bugs in BCF's checker.}
\label{tab:eval:bugs}
\vspace{-8pt}
\scriptsize
\setlength{\tabcolsep}{3pt}
\renewcommand{\arraystretch}{0.92}
\begin{tabular}{@{}r@{\;\;}l@{\;\;}l@{}}
\toprule
\multicolumn{3}{@{}l}{\textbf{unchecked proof steps}} \\
1 & \Rule{rewrite} 0, \Rule{poly\_norm} & any equation is accepted \\
\multicolumn{3}{@{}l}{\textbf{unsound proof rules}} \\
2 & \Rule{rewrite} 73, 74 & products can wrap around \\
3 & \Rule{rewrite} 80 & constant overflows 32 bits \\
4 & \Rule{aci\_norm} & sorts a concatenation \\
5 & \Rule{aci\_norm} & changes a shared term \\
6 & XOR rules & ignore later operands \\
7 & \Rule{evaluate} & wrong shifts by 64 or more \\
\multicolumn{3}{@{}l}{\textbf{unsoundness due to missing checks}} \\
8 & equivalence rules & accept bit-vectors \\
9 & \Rule{distinct\_values} & accepts lists \\
10 & \Rule{cong} & null-pointer read \\
11 & \Rule{rewrite} & null-pointer read \\
\multicolumn{3}{@{}l}{\textbf{memory safety errors}} \\
12 & \Rule{absorb} & writes past a stack array \\
13 & \Rule{bitblast} & null-pointer read in the log \\
\bottomrule
\end{tabular}
\vspace{-10pt}
\end{wraptable}

\textbf{Discovery of bugs in the BCF checker}. Our goal in building
\tool was to get rich guarantees for a critical infrastructure of the
Linux kernel.  During the development of \tool, we surprisingly
discovered \BCFDefects{} bugs in BCF's checker, including unchecked
proof steps, unsound rule semantics, and memory safety faults (see
Table~\ref{tab:eval:bugs}). Surprisingly, the proof steps used in
Bug\#1 in Table~\ref{tab:eval:bugs} are frequently used by other
correct proof certificates. On the \IdxBcfEval{} object files of BCF's evaluation,
\ProgSetTrustedPct\% of the proof certificate steps in the in-kernel checker
are unchecked proof steps. Our supplementary material describes each
of the distinct bugs that we discovered in the BCF checker.

\textbf{Unsound rewrite rule discovered in the BCF checker}.  We
describe a specific case study of unsound rewrite rules in the BCF
checker, which are named rewrite 73, 74 in Table~\ref{tab:eval:bugs}.
The rewrite rule 73 replaces a signed comparison of two products with
a formula that contains no multiplication because multiplication is
much harder for the solver to reason about than comparison.  With
$\mathrm{sext}_W(u)$ the sign extension of $u$ to $W$ bits, it states
\[
  \mathrm{sext}_W(y) \cdot \mathrm{sext}_W(a) <_s \mathrm{sext}_W(x) \cdot \mathrm{sext}_W(a)
  \;\to\;
  (y \ne x) \wedge (a \ne 0) \wedge \big((y <_s x) \leftrightarrow (a >_s 0)\big) ,
\]
where $\leftrightarrow$ says that both formulas are true or both are
false.

Over mathematical integers, this rewrite rule is correct. Multiplying
by a positive $a$ keeps the order of $y$ and $x$, multiplying by a
negative $a$ reverses it, and the products are equal when $a = 0$ or
$y = x$. However, with bit-vectors, we need to ensure that products do
not wrap around.
The multiplication of a $w$-bit $y$ and a $w'$-bit $a$ can produce a
product that may need $w + w'$ bits.  Hence, the entry needs $W \ge w
+ w'$. We discovered that BCF's checker did not have this
condition. We found this bug when the proof of Lemma~\ref{lem:table}
failed on the rewrite rule of BCF while building \tool.  Here is a
concrete example with 8-bit bit-vectors, $y = 64$, $x = 0$, $a = 2$,
and $W = 8$. The product $64 \cdot 2$ wraps around to $-128$, and
$-128 <_s 0$ makes the left side true. On the right, $64 <_s 0$ is
false and $2 >_s 0$ is true, so the right side is false.  By the
definition in \S\ref{sec:checker}, the entry is not sound. It
can be used to derive false.
In a refutation proof setting, one false fact is enough to refute any
formula. We built a certificate of \ForgeTableSteps{} steps whose
\Rule{rewrite} step concludes the false equation above. Its other
steps then use sound rules to derive $\bot$ from that equation,
without using $\varphi$. BCF's checker therefore accepts it for every
$\varphi$, even $\varphi = \top$.  We also tested forged certificates
in the kernel: a kernel with BCF's checker accepts them, and a kernel
with \tool rejects them. \tool's entry 73 has one more conjunct in its
side condition, which gives $W \ge w + w'$. With it,
Lemma~\ref{lem:table} holds and \tool rejects the forged certificate
at the \Rule{rewrite} step.

In summary, \tool avoids all the distinct bugs that we discovered with
the BCF checker.  \tool did not crash or hang or leak memory with the
\FuzzInputsM{} million proof certificates from our mutation testing
experiment, which demonstrates the value of formal verification.

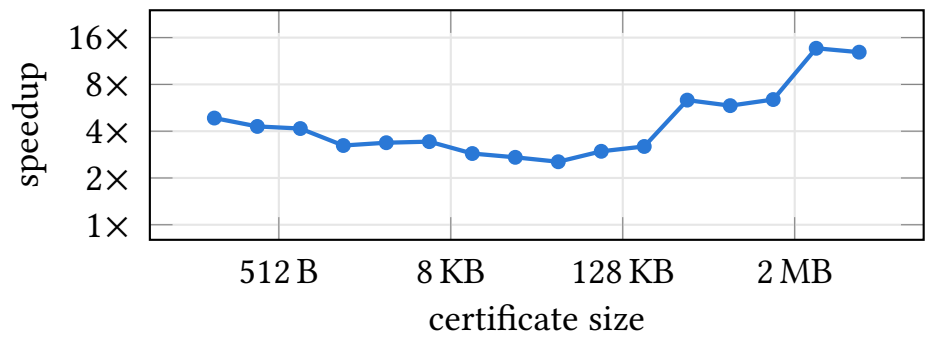
\begin{wrapfigure}{r}{0.47\textwidth}
\vspace{-12pt}
\centering
\pgfplotstableread[row sep=\\]{
  bytes median p10 p90 n \\
  181 4.859 4.167 5.400 364 \\
  362 4.292 3.667 4.583 62 \\
  724 4.160 3.562 4.455 30 \\
  1448 3.235 2.972 3.500 323 \\
  2896 3.372 3.138 4.028 116 \\
  5793 3.426 3.271 3.648 216 \\
  11585 2.873 2.520 3.273 111 \\
  23170 2.716 2.526 2.873 159 \\
  46341 2.543 2.413 2.859 61 \\
  92682 2.969 2.448 7.505 36 \\
  185364 3.190 2.508 7.108 15 \\
  370728 6.338 2.688 7.790 4 \\
  741455 5.840 5.799 5.953 18 \\
  1482910 6.393 6.393 6.393 1 \\
  2965821 13.663 7.355 15.635 4 \\
  5931642 12.896 8.142 17.650 2 \\
}\whereuser
\begin{tikzpicture}
\begin{axis}[
  width=6.7cm, height=3.1cm,
  xmode=log, ymode=log, log basis x=2, log basis y=2,
  xlabel={certificate size}, ylabel={speedup},
  xtick={512,8192,131072,2097152},
  xticklabels={512\,B,8\,KB,128\,KB,2\,MB},
  ytick={1,2,4,8,16}, yticklabels={$1\times$,$2\times$,$4\times$,$8\times$,$16\times$},
  ymin=0.8, ymax=24, grid=major, grid style={black!10},
  tick label style={font=\scriptsize}, label style={font=\scriptsize},
  xlabel shift=-3pt, ylabel shift=-3pt]
\addplot[keelblue, thick, mark=*, mark size=1pt] table[x=bytes, y=median] {\whereuser};
\end{axis}
\end{tikzpicture}
\vspace{-8pt}
\caption{\small The median speedup over BCF's checker for proof
  certificates of a specific size.}
\label{fig:eval:where}
\Description{Median speedup of Keel over BCF's checker on the corpus certificates, by size. It falls from about five times on the smallest certificates to about two and a half times near 46 KB, then rises to about thirteen times on the largest.}
\vspace{-10pt}
\end{wrapfigure}

\textbf{Ability to verify proofs of real-world proof
  certificates}. The Linux kernel with \tool and the kernel with the BCF
checker agree on the expected outcomes for the valid proof
certificates. On every one of the \IdxObjects{} object files, both
kernels reach the same verdict: both load \IdxLoaded{} and both reject
\IdxRejected{}. Outside the kernel, both checkers accept the same
\CorpusShared{} of the \CorpusShipped{} certificates. No certificate
is rejected for exhausting a budget. \tool accepts \CorpusOnlyKeel{}
more valid certificates that BCF's checker rejects because its
constant evaluator is not defined for every operator. In summary,
\tool rejects all unsound proof certificates and accepts more valid
proof certificates compared to the BCF checker.

\textbf{Proof checking with \tool is faster than BCF}. Inside the
kernel, \tool checks the certificates \IdxCheckSpeedup$\times$ faster
in total than BCF's checker. Most of them are small: \IdxFastPct\%
take about \IdxFastMedUs{}\,$\mu$s with either checker. The other
\IdxSlowPct\% are larger, with a median of \IdxSlowMedKB{}\,KB. On
these, BCF's checker takes a median of \IdxSlowMedBCF{}\,$\mu$s and
\tool \IdxSlowMedKeel{}\,$\mu$s. The larger the certificate, the
larger the speedup. At \PfKernelSizedLo{}--\PfKernelSizedHi{}\,KB, on the \IdxBcfEval{}
object files of BCF's evaluation, \tool is \WhyKernelLargeSpeedup$\times$ faster.  Outside the kernel
(Figure~\ref{fig:eval:where}), \tool is \CorpusSpeedupGeo$\times$
faster per certificate on average and \BigSpeedup$\times$ faster on
the largest one that both accept, of \BigCertMB{}\,MB.

\textbf{Why}. Most of the difference comes from two decisions about how
to store terms. First, BCF's
checker stores every new term that a rule builds at a new index, even
when an equal term is already stored. To check whether two terms are
equal, BCF's checker must compare their structure.
\tool never stores a term twice.
Two terms are therefore equal exactly when their indices are equal, and
\tool checks equality with one comparison of two integers. Second, BCF's
checker allocates every term it builds separately and keeps track of it
in the \texttt{xarray}. \tool
keeps all its terms in one array.

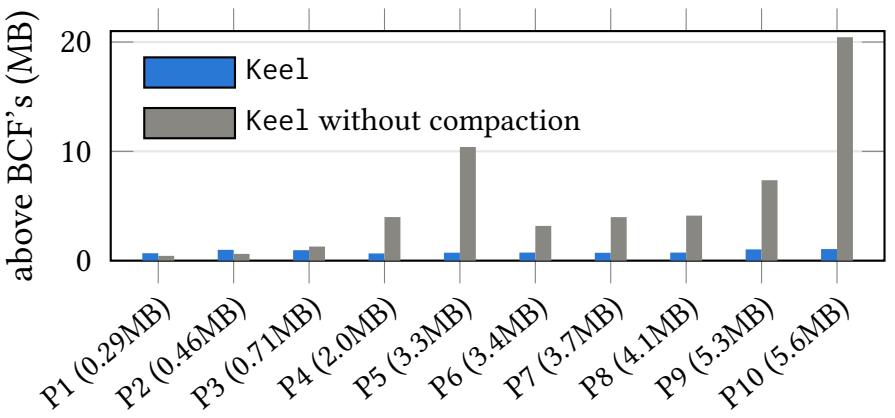
\begin{wrapfigure}{r}{0.47\textwidth}
\vspace{-12pt}
\centering
\pgfplotstableread[row sep=\\]{
  label keel nocompact \\
  {0.29} 0.68 0.44 \\
  {0.46} 0.99 0.62 \\
  {0.7--0.8} 0.96 1.29 \\
  {2.0} 0.66 3.99 \\
  {3.3} 0.73 10.40 \\
  {3.4} 0.74 3.18 \\
  {3.7} 0.72 3.98 \\
  {4.1} 0.74 4.12 \\
  {5.3} 1.03 7.36 \\
  {5.6} 1.06 20.44 \\
}\compactionextra
\begin{tikzpicture}
\begin{axis}[
  width=6.7cm, height=3.1cm, ybar=0pt, bar width=3pt,
  symbolic x coords={0.29,0.46,0.7--0.8,2.0,3.3,3.4,3.7,4.1,5.3,5.6},
  xtick=data, enlarge x limits=0.07,
  xticklabels={P1 (0.29MB),P2 (0.46MB),P3 (0.71MB),P4 (2.0MB),P5 (3.3MB),P6 (3.4MB),P7 (3.7MB),P8 (4.1MB),P9 (5.3MB),P10 (5.6MB)},
  x tick label style={rotate=40, anchor=north east, inner sep=1pt},
  ymin=0, ymax=21, ytick={0,10,20},
  ylabel={above BCF's (MB)},
  tick label style={font=\tiny}, label style={font=\scriptsize},
  ylabel shift=-3pt,
  ymajorgrids, grid style={black!10},
  legend style={font=\tiny, draw=none, at={(0.02,0.98)}, anchor=north west, fill=none}, legend cell align=left]
\addplot[fill=keelblue, draw=none, area legend] table[x=label, y=keel] {\compactionextra};
\addlegendentry{\tool}
\addplot[fill=offgray, draw=none, area legend] table[x=label, y=nocompact] {\compactionextra};
\addlegendentry{\tool without compaction}
\end{axis}
\end{tikzpicture}
\vspace{-8pt}
\caption{Peak memory overhead (in MB) of \tool with and without
  compaction relative to BCF's checker, on the \CompRuns{} corpus
  certificates on which compaction runs. Each P$i$ is a separate
  certificate, shown with its size. The \CompSharedBarN{} certificates
  of 0.7--0.8\,MB share one bar (P3), their median.}
\label{fig:eval:compaction}
\Description{With compaction Keel stays about one megabyte above BCF's peak at every size. Without it the excess reaches twenty megabytes.}
\vspace{-10pt}
\end{wrapfigure}

\textbf{\tool uses a bit more memory than BCF}. \tool uses slightly
more memory than BCF's checker: at the median certificate its peak is
\MemMedKeel{}\,KB against \MemMedBCF{}\,KB, and on the largest
certificate \BigMemKeel{}\,MB against \BigMemBCF{}\,MB
(Figure~\ref{fig:eval:compaction}).

\textbf{Sensitivity to design decisions}. Two decisions in \tool's design
serve the kernel rather than the checking itself: \tool reclaims memory,
and its two counters enforce the budgets of
\S\ref{sec:cost}. To see what each costs, we turned it off
in the printed C. The variants return the same verdicts as \tool on all
\CorpusShipped{} certificates. Without compaction, the peak
grows to \CompMemOff{}\,MB against \BigMemBCF{}\,MB on the largest.
Overall, compaction actually saves time by putting live terms together tightly
and therefore reducing cache misses. The \CompRuns{} certificates
take \CompNetSavePct\% less time with it. The counters cost very little.
Removing them makes checking the ten largest certificates
\CountersCostHi\% faster in total.

\textbf{Extending the checker and proof effort}.  To highlight the
ease of adding additional proof rules, we added \NumRewritesAdded{}
rewrites from cvc5's rewrite rules that BCF was lacking.  The
\NumRewritesAdded{} rules took \AddedRuleLeanLines{} lines of Lean in
total, a median of \MedianLinesPerRule{} per rule for the entry and
its lemma. No C function was changed: the checker C code emitted
differs only in the table.  The refinement proof needed no new lemmas,
only two lines to prove the mathematical identity. The two proofs that
connect the C code to the Lean checker are by far the largest part:
the refinement proof is \SoundnessProofLines{} lines and the cost
proof \CostProofLines{}. We anticipate that this extensibility will
encourage kernel developers to adopt \tool.

\section{Related Work}
\label{sec:related}
\tool builds on a body of work related to proof-carrying
code~\cite{necula:pcc:osdi:1996,necula:pcc:popl:1997}, foundational
verification~\cite{appel:felty:popl:2000,appel:fpcc:lics:2001,feng:openfpcc:tldi:2007},
eBPF
verifiers~\cite{tnums-cgo22,agni-cav23,gershuni:prevail:pldi:2019},
and deep embedding of C in interactive theorem
provers~\cite{erbsen:bedrock2:pldi:2021,erbsen:fiatcrypto:sp:2019}.

\textbf{Proof-carrying code}.  A seminal work on proof-carrying
code~\cite{necula:pcc:osdi:1996,necula:pcc:popl:1997} relies on the
idea that checking the proof is much easier than generating
it. Indeed, PCC was employed in the old BPF context. In PCC, the proof
checker and the rules of the safety logic are trusted.  Foundational
proof-carrying code removes these components from the trusted
computing
base~\cite{appel:felty:popl:2000,appel:fpcc:lics:2001,feng:openfpcc:tldi:2007}. ExoBPF~\cite{nelson:exobpf:lpc:2021}
revisits proof-carrying code for eBPF, where proofs turned out to be
very large and the project was disbanded.  BCF~\cite{sun:bcf:sosp:2025}
and proof-carrying BPF~\cite{cai:pcb:ebpf:2026} generate proofs with
cvc5 in user space, but trust both the proof rules and the in-kernel
checker. \tool is designed as a foundational checker which removes
both from the trusted base.

\textbf{Verifying the eBPF components}.  There are numerous attempts
to verify the existing eBPF verifier. Among them,
Agni~\cite{tnums-cgo22,agni-cav23} proves the soundness of the range
analysis by translating the kernel's C code into first-order logic
formulas, whose satisfiability is checked with SMT solvers. It is now
part of the verifier continuous integration
pipeline~\cite{chaignon:agni:lpc:2024}. Verifiers and fuzzers have
also been explored to show the correctness of just-in-time compilation
engines~\cite{nelson:jitterbug:osdi:2020}. Fuzzers find
bugs in the verifier and in the runtime~\cite{haosun-bvf,brf}. BCF
adds a third component that the kernel relies on, the proof
checker. \tool is a verified replacement for that component.

\textbf{Improving precision of the eBPF verifier}.  \tool's goal is to
make the verifier accept safe programs with proof-carrying code when
its default analysis fails. Other tools add language runtimes and
verifiers to the trusted computing base and move the verification
outside the kernel. Prevail is an abstract interpreter for eBPF which
employs stronger abstract domains that track the relationship between
registers~\cite{gershuni:prevail:pldi:2019}.  VEP verifies annotated C
programs in user space, compiles the annotations and proofs to
bytecode together with the code, and checks them in the
kernel~\cite{wu:vep:nsdi:2025}. Rex and KFlex move part of the safety
argument out of the verifier and into the source language or into
run-time
checks~\cite{jiaRexClosingLanguageverifier2025,dwivedi:kflex:sosp:2024}.

\textbf{Checking SAT and SMT certificates}.  \tool builds a verified
in-kernel proof checker for refutation proofs. There is a large body
of work on building refutation proofs and proof checkers in the SAT
and SMT
community~\cite{cruzfilipe:lrat:cade:2017,ekici:smtcoq:cav:2017,
  lammich:unsat-check:ijcar:2024, tan:cake-lpr:tacas:2021,
  bodingbauer:vampire-lean:arxiv:2026,
  reeves:smt-skeletons:ijcar:2026}. For SMT, SMTCoq checks
certificates from SMT solvers with a checker written and proved in
Coq~\cite{ekici:smtcoq:cav:2017}, and lean-smt turns cvc5's proofs
into Lean proofs~\cite{mohamed:leansmt:cav:2025}. These checkers run
inside a proof assistant or as code extracted from one.
The Vampire project~\cite{bartek:vampire:cav:2025} also explores
a Lean-based checker for a class of refutation proofs but it does not
have a verified C checker
yet~\cite{bodingbauer:vampire-lean:arxiv:2026}, which makes it unusable
in the Linux kernel.
cvc5's Ethos~\cite{reynolds:ethos:ijcar:2026} checks CPC
proofs~\cite{reynolds:cpc:cav:2026} against trusted Eunoia rule files, and
IsaRare~\cite{lachnitt:isarare:tacas:2024} proves \code{cvc5}'s RARE
rewrites~\cite{notzli:rare:fmcad:2022} in Isabelle.  \tool proves each
RARE entry in Lean and checks all entries via one verified routine
(\S\ref{sec:checker}). Unlike prior checkers, it runs in the
kernel as runtime-free C with linear time and memory bounds.

\textbf{Verified low-level code}. Verifying low-level C code is a widely
studied topic.  seL4 proves that the C code of a microkernel refines
an abstract specification~\cite{klein:sel4:sosp:2009}. bedrock2 and
fiat-crypto write low-level programs, do a deep embedding as syntax
trees inside a proof assistant, prove them there, and print C
code~\cite{erbsen:bedrock2:pldi:2021,erbsen:fiatcrypto:sp:2019}.
\tool follows the same plan on a smaller scale: a specification in
Lean, an \embc program proved to refine it, and a printer that turns
the program into C (\S\ref{sec:overview}).
To model kernel API calls without full module
composition~\cite{stewart:compcomp:popl:2015}, our Clight simulation
(Theorem~\ref{thm:embc-clight}) is parameterized by forward/backward
oracle contracts for the seven kernel primitives in a single
translated Clight program.

\textbf{Resource bounds}. Automatic amortized resource analysis
derives such bounds of a program from the code. It assigns the input a
potential, for example a fixed amount per list element, chosen so that
the potential never drops below zero when the cost of each step is
subtracted from it. The bound is then the potential of the
input~\cite{hofmann:jost:popl:2003,hoffmann:multivariate:toplas:2012,kahn:linearmaps:arxiv:2025}.
C4B applies this analysis to C programs and produces Coq proofs of the
bounds it finds~\cite{carbonneaux:c4b:pldi:2015}. In \tool, the
potential is not found by an analysis but written into the
program. Before it reads the certificate, the checker sets two
budgets, one for time and one for memory, each a fixed amount per byte
of input plus a constant. It decrements a budget before any code whose
cost depends on the input, and rejects the certificate when a budget
would drop below zero (\S\ref{sec:cost}).

\section{Conclusion}
\label{sec:conclusion}

Selective proof-carrying code with an in-kernel certificate checker
can increase the expressivity of the eBPF verifier. \tool is a
drop-in replacement for an in-kernel certificate checker that comes
equipped with machine-checked proofs of logical soundness, absence of
undefined behavior and memory leaks, and linear time and memory
bounds. We made \tool extensible by separating the certificate checker
from the mathematical rewrite rules, which we believe will enable
adoption by the Linux kernel developers. This is a first step in our
journey to make the core infrastructure formally verified with limited
trusted assumptions.

\section{AI Use}
The authors claim responsibility for the ideas, experimental design,
research methodology, findings, and exposition in this paper. We used
Anthropic's Claude Code and Google's Gemini 4 Argon to develop Lean
proofs. The authors set the goals and the design of \tool and drew out
the theorems and lemmas to prove. Under their direction, the AI tools
assisted in developing the Lean checker and its soundness proof, \embc
and its printer, the checker as an \embc program (and therefore the
printed C code), and the refinement and cost proofs.
After the authors designed the fuzzing experiment, the AI tools also
helped develop the fuzzer and the experiment scripts. The AI tools
also performed the audit that compared BCF's C code with our
specification.
We do not rely on the AI tools for the correctness of our proofs. Lean 
checks every proof. The theorems about the checker depend only on
Lean's three standard axioms, and the correspondence with Clight also
uses the two axioms of the Clight model for external functions and inline
assembly. We confirmed each bug with a failed proof,
a certificate, a fuzzer input, or by reading BCF's code.

\begin{acks}
We thank Harishankar Vishwanathan, Srinivas Narayana, and and members
of the Rutgers Architecture and Programming Languages (RAPL) lab for
their feedback on this project.  This material is based upon work
supported in part by the Google GEMINI research award from Google Inc.
\end{acks}

\bibliographystyle{ACM-Reference-Format}
\bibliography{references}

@inproceedings{lopes:alive:pldi:2015,
 author = "Nuno P. Lopes and David Menendez and Santosh Nagarakatte and John Regehr",
 title = "Provably Correct Peephole Optimizations with {Alive}",
 booktitle = "Proceedings of the 36th ACM SIGPLAN Conference on Programming Language Design and Implementation",
 pages={22--32},
 series={PLDI '15},
 year = {2015},
 publisher = {Association for Computing Machinery},
 address = {New York, NY, USA},
 url = {https://doi.org/10.1145/2737924.2737965},
 doi = {10.1145/2737924.2737965},
}

@inproceedings{kundu:pec:pldi:2009,
author = {Kundu, Sudipta and Tatlock, Zachary and Lerner, Sorin},
title = {Proving Optimizations Correct Using Parameterized Program Equivalence},
year = {2009},
publisher = {Association for Computing Machinery},
address = {New York, NY, USA},
url = {https://doi.org/10.1145/1542476.1542513},
doi = {10.1145/1542476.1542513},
booktitle = {Proceedings of the 30th ACM SIGPLAN Conference on Programming Language Design and Implementation},
pages = {327--337},
numpages = {11},
location = {Dublin, Ireland},
series = {PLDI '09}
}

@inproceedings{lopes:alive2:pldi:2021,
author = {Lopes, Nuno P. and Lee, Juneyoung and Hur, Chung-Kil and Liu, Zhengyang and Regehr, John},
title = {{Alive2}: Bounded Translation Validation for {LLVM}},
year = {2021},
publisher = {Association for Computing Machinery},
address = {New York, NY, USA},
url = {https://doi.org/10.1145/3453483.3454030},
doi = {10.1145/3453483.3454030},
booktitle = {Proceedings of the 42nd ACM SIGPLAN International Conference on Programming Language Design and Implementation},
pages = {65--79},
numpages = {15},
location = {Virtual, Canada},
series = {PLDI '21}
}

@inproceedings{necula:pcc:popl:1997,
author = {Necula, George C.},
title = {Proof-carrying code},
year = {1997},
isbn = {0897918533},
publisher = {Association for Computing Machinery},
address = {New York, NY, USA},
url = {https://doi.org/10.1145/263699.263712},
doi = {10.1145/263699.263712},
booktitle = {Proceedings of the 24th ACM SIGPLAN-SIGACT Symposium on Principles of Programming Languages},
pages = {106–119},
numpages = {14},
location = {Paris, France},
series = {POPL '97}
}

@inproceedings{necula:pcc:osdi:1996,
author = {Necula, George C. and Lee, Peter},
title = {Safe kernel extensions without run-time checking},
year = {1996},
isbn = {1880446820},
publisher = {Association for Computing Machinery},
address = {New York, NY, USA},
url = {https://doi.org/10.1145/238721.238781},
doi = {10.1145/238721.238781},
booktitle = {Proceedings of the Second USENIX Symposium on Operating Systems Design and Implementation},
pages = {229–243},
numpages = {15},
location = {Seattle, Washington, USA},
series = {OSDI '96}
}

@inproceedings{agni-cav23,
author = {Vishwanathan, Harishankar and Shachnai, Matan and Narayana, Srinivas and Nagarakatte, Santosh},
title = {Verifying the Verifier: eBPF Range Analysis Verification},
year = {2023},
isbn = {978-3-031-37708-2},
publisher = {Springer Nature Switzerland},
address = {Cham},
series = {Lecture Notes in Computer Science},
volume = {13966},
doi = {10.1007/978-3-031-37709-9_12},
booktitle = {Computer Aided Verification: 35th International Conference, CAV 2023, Paris, France, July 17--22, 2023, Proceedings, Part III},
pages = {226--251},
numpages = {26}
}

@inproceedings{agni-sas24,
  author = {Shachnai, Matan and Vishwanathan, Harishankar and Narayana, Srinivas and Nagarakatte, Santosh},
  title = {Fixing Latent Unsound Abstract Operators in the eBPF Verifier of the Linux Kernel},
  year = {2024},
  isbn = {978-3-031-74776-2},
  publisher = {Springer Nature Switzerland},
  address = {Cham},
  booktitle= {Static Analysis: 31st International Symposium, SAS 2024, Pasadena, CA, USA, October 20--22, 2024, Proceedings},
  series = {Lecture Notes in Computer Science},
  volume = {14995},
  pages={386--406},
  doi = {10.1007/978-3-031-74776-2_15},
}

@inproceedings{tnums-cgo22,
  author    = {Vishwanathan, Harishankar and Shachnai, Matan and Narayana, Srinivas and Nagarakatte, Santosh},
  title     = {Sound, Precise, and Fast Abstract Interpretation with Tristate Numbers},
  year      = {2022},
  isbn      = {9781665405843},
  publisher = {IEEE Press},
  doi       = {10.1109/CGO53902.2022.9741267},
  booktitle = {Proceedings of the 20th IEEE/ACM International Symposium on Code Generation and Optimization},
  pages     = {254--265},
  numpages  = {12},
  location  = {Virtual Event, Republic of Korea},
  series    = {CGO '22}
}

@article{gange:zones-octagons-fresh-look:toplas:2021,
author = {Gange, Graeme and Ma, Zequn and Navas, Jorge A. and Schachte, Peter and S\o{}ndergaard, Harald and Stuckey, Peter J.},
title = {A Fresh Look at Zones and Octagons},
year = {2021},
issue_date = {September 2021},
publisher = {Association for Computing Machinery},
address = {New York, NY, USA},
volume = {43},
number = {3},
issn = {0164-0925},
url = {https://doi.org/10.1145/3457885},
doi = {10.1145/3457885},
journal = {ACM Trans. Program. Lang. Syst.},
month = sep,
articleno = {11},
numpages = {51}
}

@article{brf,
author = {Hung, Hsin-Wei and Amiri Sani, Ardalan},
title = {BRF: Fuzzing the eBPF Runtime},
year = {2024},
issue_date = {July 2024},
publisher = {Association for Computing Machinery},
address = {New York, NY, USA},
volume = {1},
number = {FSE},
url = {https://doi.org/10.1145/3643778},
doi = {10.1145/3643778},
journal = {Proc. ACM Softw. Eng.},
month = jul,
articleno = {52},
numpages = {20}
}

@inproceedings{haosun-bvf,
author = {Sun, Hao and Xu, Yiru and Liu, Jianzhong and Shen, Yuheng and Guan, Nan and Jiang, Yu},
title = {Finding Correctness Bugs in eBPF Verifier with Structured and Sanitized Program},
year = {2024},
isbn = {9798400704376},
publisher = {Association for Computing Machinery},
address = {New York, NY, USA},
url = {https://doi.org/10.1145/3627703.3629562},
doi = {10.1145/3627703.3629562},
booktitle = {Proceedings of the Nineteenth European Conference on Computer Systems},
pages = {689--703},
numpages = {15},
location = {Athens, Greece},
series = {EuroSys '24}
}

@inproceedings{sun:bcf:sosp:2025,
author = {Sun, Hao and Su, Zhendong},
title = {Prove It to the Kernel: Precise Extension Analysis via Proof-Guided Abstraction Refinement},
year = {2025},
isbn = {9798400718700},
publisher = {Association for Computing Machinery},
address = {New York, NY, USA},
url = {https://doi.org/10.1145/3731569.3764796},
doi = {10.1145/3731569.3764796},
booktitle = {Proceedings of the ACM SIGOPS 31st Symposium on Operating Systems Principles},
pages = {736–751},
numpages = {16},
location = {Lotte Hotel World, Seoul, Republic of Korea},
series = {SOSP '25}
}

@inproceedings{gershuni:prevail:pldi:2019,
author = {Gershuni, Elazar and Amit, Nadav and Gurfinkel, Arie and Narodytska, Nina and Navas, Jorge A. and Rinetzky, Noam and Ryzhyk, Leonid and Sagiv, Mooly},
title = {Simple and precise static analysis of untrusted Linux kernel extensions},
year = {2019},
isbn = {9781450367127},
publisher = {Association for Computing Machinery},
address = {New York, NY, USA},
url = {https://doi.org/10.1145/3314221.3314590},
doi = {10.1145/3314221.3314590},
booktitle = {Proceedings of the 40th ACM SIGPLAN Conference on Programming Language Design and Implementation},
pages = {1069–1084},
numpages = {16},
location = {Phoenix, AZ, USA},
series = {PLDI '19}
}

@misc{bcf-repo,
  author       = {Hao Sun and Zhendong Su},
  title        = {{BCF}: {eBPF} Certificate Framework},
  howpublished = {\url{https://github.com/SunHao-0/BCF}},
  note         = {Accessed: 2026-06-14},
  year         = {2025}
}

@inproceedings{jiaRexClosingLanguageverifier2025,
  title = {Rex: Closing the Language-Verifier Gap with Safe and Usable Kernel Extensions},
  shorttitle = {Rex},
  booktitle = {2025 {USENIX} Annual Technical Conference},
  author = {Jia, Jinghao and Qin, Ruowen and Craun, Milo and Lukiyanov, Egor and Bansal, Ayush and Phan, Minh and Le, Michael V. and Franke, Hubertus and Jamjoom, Hani and Xu, Tianyin and Williams, Dan},
  year = {2025},
  month = jul,
  series = {{{USENIX ATC}} '25},
  pages = {325--342},
  publisher = {USENIX Association},
  location = {USA},
  isbn = {978-1-939133-48-9}
}

@misc{nelson:exobpf:lpc:2021,
  author       = {Luke Nelson and Xi Wang and Emina Torlak},
  title        = {A Proof-Carrying Approach to Building Correct and Flexible {BPF} Verifiers},
  howpublished = {Linux Plumbers Conference (LPC)},
  year         = {2021},
  note         = {Accessed: 2026-06-17. \url{https://lpc.events/event/11/contributions/944/}}
}

@inproceedings{appel:fpcc:lics:2001,
  author    = {Appel, Andrew W.},
  title     = {Foundational Proof-Carrying Code},
  booktitle = {Proceedings of the 16th Annual IEEE Symposium on Logic in Computer Science (LICS '01)},
  year      = {2001},
  pages     = {247--256},
  publisher = {IEEE Computer Society},
  address   = {Boston, MA, USA},
  doi       = {10.1109/LICS.2001.932501}
}

@inproceedings{appel:felty:popl:2000,
  author    = {Appel, Andrew W. and Felty, Amy P.},
  title     = {A Semantic Model of Types and Machine Instructions for Proof-Carrying Code},
  booktitle = {Proceedings of the 27th ACM SIGPLAN-SIGACT Symposium on Principles of Programming Languages (POPL '00)},
  year      = {2000},
  pages     = {243--253},
  publisher = {Association for Computing Machinery},
  doi       = {10.1145/325694.325727}
}

@inproceedings{moura:lean4:cade:2021,
  author    = {de Moura, Leonardo and Ullrich, Sebastian},
  title     = {The {Lean} 4 Theorem Prover and Programming Language},
  booktitle = {Automated Deduction -- CADE 28},
  year      = {2021},
  pages     = {625--635},
  series    = {Lecture Notes in Computer Science},
  volume    = {12699},
  publisher = {Springer},
  address   = {Cham},
  doi       = {10.1007/978-3-030-79876-5\_37}
}

@inproceedings{barbosa:cvc5:tacas:2022,
  author    = {Barbosa, Haniel and Barrett, Clark and Brain, Martin and Kremer, Gereon
               and Lachnitt, Hanna and Mann, Makai and Mohamed, Abdalrhman and Mohamed, Mudathir
               and Niemetz, Aina and N{\"o}tzli, Andres and Ozdemir, Alex and Preiner, Mathias
               and Reynolds, Andrew and Sheng, Ying and Tinelli, Cesare and Zohar, Yoni},
  title     = {cvc5: A Versatile and Industrial-Strength {SMT} Solver},
  booktitle = {Tools and Algorithms for the Construction and Analysis of Systems (TACAS 2022)},
  year      = {2022},
  pages     = {415--442},
  series    = {Lecture Notes in Computer Science},
  volume    = {13243},
  publisher = {Springer},
  doi       = {10.1007/978-3-030-99524-9\_24}
}

@inproceedings{cruzfilipe:lrat:cade:2017,
  author    = {Cruz-Filipe, Lu{\'i}s and Heule, Marijn J. H. and Hunt, Warren A. and
               Kaufmann, Matt and Schneider-Kamp, Peter},
  title     = {Efficient Certified {RAT} Verification},
  booktitle = {Automated Deduction -- CADE 26},
  year      = {2017},
  pages     = {220--236},
  series    = {Lecture Notes in Computer Science},
  volume    = {10395},
  publisher = {Springer},
  address   = {Cham},
  doi       = {10.1007/978-3-319-63046-5\_14}
}

@inproceedings{lammich:unsat-check:ijcar:2024,
  author    = {Lammich, Peter},
  title     = {Fast and Verified {UNSAT} Certificate Checking},
  booktitle = {Automated Reasoning -- IJCAR 2024},
  year      = {2024},
  pages     = {439--457},
  series    = {Lecture Notes in Computer Science},
  volume    = {14739},
  publisher = {Springer},
  address   = {Cham},
  doi       = {10.1007/978-3-031-63498-7\_26}
}

@inproceedings{tan:cake-lpr:tacas:2021,
  author    = {Tan, Yong Kiam and Heule, Marijn J. H. and Myreen, Magnus O.},
  title     = {\texttt{cake\_lpr}: Verified Propagation Redundancy Checking in {CakeML}},
  booktitle = {Tools and Algorithms for the Construction and Analysis of Systems (TACAS 2021)},
  year      = {2021},
  pages     = {223--241},
  series    = {Lecture Notes in Computer Science},
  volume    = {12652},
  publisher = {Springer},
  address   = {Cham},
  doi       = {10.1007/978-3-030-72013-1\_12}
}

@misc{bodingbauer:vampire-lean:arxiv:2026,
  author        = {Bodingbauer, Jonas and Hajdu, M{\'a}rton and Kov{\'a}cs, Laura and
                   Polaczek, Axel and Rawson, Michael},
  title         = {Lean on {Vampire} Proofs (Short Paper)},
  year          = {2026},
  eprint        = {2603.26342},
  archivePrefix = {arXiv},
  primaryClass  = {cs.LO},
  doi           = {10.48550/arXiv.2603.26342}
}

@inproceedings{bartek:vampire:cav:2025,
  author    = {B{\'a}rtek, Filip and Bhayat, Ahmed and Coutelier, Robin and Hajdu, M{\'a}rton and
               Hetzenberger, Matthias and Hozzov{\'a}, Petra and Kov{\'a}cs, Laura and Rath, Jakob and
               Rawson, Michael and Reger, Giles and Suda, Martin and Schoisswohl, Johannes and
               Voronkov, Andrei},
  title     = {The {Vampire} Diary},
  booktitle = {Computer Aided Verification -- CAV 2025},
  year      = {2025},
  pages     = {57--71},
  series    = {Lecture Notes in Computer Science},
  volume    = {15933},
  publisher = {Springer},
  address   = {Cham},
  doi       = {10.1007/978-3-031-98682-6\_4}
}

@inproceedings{reeves:smt-skeletons:ijcar:2026,
  author    = {Reeves, Joseph E. and Barbosa, Haniel and Reynolds, Andrew and Heule, Marijn J. H.},
  title     = {A General Approach for {SMT} Proof Skeletons},
  booktitle = {Automated Reasoning -- IJCAR 2026},
  year      = {2026},
  pages     = {156--174},
  series    = {Lecture Notes in Computer Science},
  volume    = {16688},
  publisher = {Springer},
  address   = {Cham},
  doi       = {10.1007/978-3-032-32589-1\_10}
}

@inproceedings{ekici:smtcoq:cav:2017,
  author    = {Ekici, Burak and Mebsout, Alain and Tinelli, Cesare and Keller, Chantal
               and Katz, Guy and Reynolds, Andrew and Barrett, Clark},
  title     = {{SMTCoq}: A Plug-In for Integrating {SMT} Solvers into {Coq}},
  booktitle = {Computer Aided Verification (CAV 2017)},
  year      = {2017},
  pages     = {126--133},
  series    = {Lecture Notes in Computer Science},
  volume    = {10427},
  publisher = {Springer},
  doi       = {10.1007/978-3-319-63390-9\_7}
}

@article{leroy:compcert:cacm:2009,
  author  = {Leroy, Xavier},
  title   = {Formal Verification of a Realistic Compiler},
  journal = {Communications of the ACM},
  volume  = {52},
  number  = {7},
  pages   = {107--115},
  year    = {2009},
  doi     = {10.1145/1538788.1538814}
}

@article{leroy:backend:jar:2009,
  author  = {Leroy, Xavier},
  title   = {A Formally Verified Compiler Back-end},
  journal = {Journal of Automated Reasoning},
  volume  = {43},
  number  = {4},
  pages   = {363--446},
  year    = {2009},
  doi     = {10.1007/s10817-009-9155-4},
  url     = {https://xavierleroy.org/publi/compcert-backend.pdf}
}

@inproceedings{stewart:compcomp:popl:2015,
  author    = {Stewart, Gordon and Beringer, Lennart and Cuellar, Santiago
               and Appel, Andrew W.},
  title     = {Compositional {CompCert}},
  booktitle = {Proceedings of the 42nd Annual ACM SIGPLAN-SIGACT Symposium
               on Principles of Programming Languages (POPL '15)},
  year      = {2015},
  pages     = {275--287},
  publisher = {Association for Computing Machinery},
  address   = {New York, NY, USA},
  doi       = {10.1145/2676726.2676985},
  url       = {https://www.cs.princeton.edu/~appel/papers/compcomp.pdf}
}

@inproceedings{klein:sel4:sosp:2009,
  author    = {Klein, Gerwin and Elphinstone, Kevin and Heiser, Gernot and Andronick, June
               and Cock, David and Derrin, Philip and Elkaduwe, Dhammika and Engelhardt, Kai
               and Kolanski, Rafal and Norrish, Michael and Sewell, Thomas and Tuch, Harvey
               and Winwood, Simon},
  title     = {{seL4}: Formal Verification of an {OS} Kernel},
  booktitle = {Proceedings of the ACM SIGOPS 22nd Symposium on Operating Systems Principles (SOSP '09)},
  year      = {2009},
  pages     = {207--220},
  publisher = {Association for Computing Machinery},
  doi       = {10.1145/1629575.1629596}
}

@inproceedings{nelson:jitterbug:osdi:2020,
  author    = {Nelson, Luke and Van Geffen, Jacob and Torlak, Emina and Wang, Xi},
  title     = {Specification and Verification in the Field: Applying Formal Methods to
               {BPF} Just-in-Time Compilers in the {Linux} Kernel},
  booktitle = {14th USENIX Symposium on Operating Systems Design and Implementation (OSDI '20)},
  year      = {2020},
  pages     = {41--61},
  publisher = {USENIX Association}
}

@inproceedings{reynolds:cpc:cav:2026,
  author    = {Reynolds, Andrew and Schurr, Hans-J{\"o}rg and Barbosa, Haniel and Israel, Ofec
               and Jakpor, Jibiana Zoe and Lachnitt, Hanna and Mohamed, Abdalrhman and Niemetz, Aina
               and Preiner, Mathias and Zohar, Yoni and Jones, Robert and Barrett, Clark and Tinelli, Cesare},
  title     = {The Cooperating Proof Calculus: Comprehensive Proofs for an {SMT} Solver},
  booktitle = {Computer Aided Verification (CAV 2026)},
  series    = {Lecture Notes in Computer Science},
  pages     = {188--212},
  publisher = {Springer},
  address   = {Cham},
  year      = {2026},
  doi       = {10.1007/978-3-032-32526-6_9}
}

@inproceedings{reynolds:ethos:ijcar:2026,
  author    = {Reynolds, Andrew and Schurr, Hans-J{\"o}rg and Soldevila, Mallku and Barbosa, Haniel
               and Barrett, Clark and Tinelli, Cesare},
  title     = {Ethos: A Fast Proof Checker for the {Eunoia} Logical Framework},
  booktitle = {Automated Reasoning -- 13th International Joint Conference, IJCAR 2026},
  series    = {Lecture Notes in Computer Science},
  volume    = {16688},
  pages     = {305--314},
  publisher = {Springer},
  address   = {Cham},
  year      = {2026},
  doi       = {10.1007/978-3-032-32589-1_19}
}

@inproceedings{mohamed:leansmt:cav:2025,
  author    = {Mohamed, Abdalrhman and Mascarenhas, Tomaz and Khan, Harun and Barbosa, Haniel
               and Reynolds, Andrew and Qian, Yicheng and Tinelli, Cesare and Barrett, Clark},
  title     = {lean-smt: An {SMT} Tactic for Discharging Proof Goals in {Lean}},
  booktitle = {Computer Aided Verification (CAV 2025)},
  series    = {Lecture Notes in Computer Science},
  volume    = {15933},
  pages     = {197--212},
  publisher = {Springer},
  address   = {Cham},
  year      = {2025},
  doi       = {10.1007/978-3-031-98682-6\_11}
}

@inproceedings{lachnitt:isarare:tacas:2024,
  author    = {Lachnitt, Hanna and Fleury, Mathias and Aniva, Leni and Reynolds, Andrew
               and Barbosa, Haniel and N{\"o}tzli, Andres and Barrett, Clark and Tinelli, Cesare},
  title     = {{IsaRare}: Automatic Verification of {SMT} Rewrites in {Isabelle/HOL}},
  booktitle = {Tools and Algorithms for the Construction and Analysis of Systems (TACAS 2024)},
  series    = {Lecture Notes in Computer Science},
  volume    = {14570},
  pages     = {311--330},
  publisher = {Springer},
  address   = {Cham},
  year      = {2024},
  doi       = {10.1007/978-3-031-57246-3\_17}
}

@inproceedings{feng:openfpcc:tldi:2007,
author = {Feng, Xinyu and Ni, Zhaozhong and Shao, Zhong and Guo, Yu},
title = {An Open Framework for Foundational Proof-Carrying Code},
year = {2007},
publisher = {Association for Computing Machinery},
address = {New York, NY, USA},
url = {https://doi.org/10.1145/1190315.1190325},
doi = {10.1145/1190315.1190325},
booktitle = {Proceedings of the 2007 ACM SIGPLAN International Workshop on Types in Language Design and Implementation},
pages = {67--78},
numpages = {12},
location = {Nice, France},
series = {TLDI '07}
}

@inproceedings{dwivedi:kflex:sosp:2024,
author = {Dwivedi, Kumar Kartikeya and Iyer, Rishabh and Kashyap, Sanidhya},
title = {Fast, Flexible, and Practical Kernel Extensions},
year = {2024},
publisher = {Association for Computing Machinery},
address = {New York, NY, USA},
url = {https://doi.org/10.1145/3694715.3695950},
doi = {10.1145/3694715.3695950},
booktitle = {Proceedings of the ACM SIGOPS 30th Symposium on Operating Systems Principles},
pages = {249--264},
numpages = {16},
location = {Austin, TX, USA},
series = {SOSP '24}
}

@inproceedings{wu:vep:nsdi:2025,
author = {Xiwei Wu and Yueyang Feng and Tianyi Huang and Xiaoyang Lu and Shengkai Lin and Lihan Xie and Shizhen Zhao and Qinxiang Cao},
title = {{VEP}: A Two-stage Verification Toolchain for Full {eBPF} Programmability},
booktitle = {22nd USENIX Symposium on Networked Systems Design and Implementation (NSDI 25)},
year = {2025},
isbn = {978-1-939133-46-5},
address = {Philadelphia, PA},
pages = {277--299},
url = {https://www.usenix.org/conference/nsdi25/presentation/wu-xiwei},
publisher = {USENIX Association},
month = apr
}

@book{rice:learning-ebpf:book:2023,
  author    = {Rice, Liz},
  title     = {Learning {eBPF}: Programming the {Linux} Kernel for Enhanced Observability, Networking, and Security},
  year      = {2023},
  publisher = {O'Reilly Media},
  isbn      = {9781098135126}
}

@misc{chaignon:agni:lpc:2024,
  author       = {Harishankar Vishwanathan and Matan Shachnai and Paul Chaignon and Santosh Nagarakatte and Srinivas Narayana},
  title        = {Agni: Fast Formal Verification of the Verifier's Range Analysis},
  howpublished = {Linux Plumbers Conference (LPC), eBPF Track},
  year         = {2024},
  note         = {Presented by Paul Chaignon; covers Agni's use in {BPF} {CI}. Accessed: 2026-09-01. \url{https://lpc.events/event/18/contributions/1937/}}
}

@inproceedings{notzli:rare:fmcad:2022,
  author    = {N{\"o}tzli, Andres and Barbosa, Haniel and Niemetz, Aina and Preiner, Mathias and Reynolds, Andrew and Barrett, Clark and Tinelli, Cesare},
  title     = {Reconstructing Fine-Grained Proofs of Rewrites Using a Domain-Specific Language},
  booktitle = {Formal Methods in Computer-Aided Design (FMCAD 2022)},
  pages     = {65--74},
  year      = {2022},
  publisher = {TU Wien Academic Press},
  doi       = {10.34727/2022/isbn.978-3-85448-053-2_12}
}

@inproceedings{erbsen:bedrock2:pldi:2021,
  author    = {Erbsen, Andres and Gruetter, Samuel and Choi, Joonwon and Wood, Clark and Chlipala, Adam},
  title     = {Integration Verification across Software and Hardware for a Simple Embedded System},
  booktitle = {Proceedings of the 42nd ACM SIGPLAN International Conference on Programming Language Design and Implementation (PLDI '21)},
  pages     = {604--619},
  year      = {2021},
  publisher = {Association for Computing Machinery},
  address   = {New York, NY, USA},
  doi       = {10.1145/3453483.3454065}
}

@inproceedings{erbsen:fiatcrypto:sp:2019,
  author    = {Erbsen, Andres and Philipoom, Jade and Gross, Jason and Sloan, Robert and Chlipala, Adam},
  title     = {Simple High-Level Code for Cryptographic Arithmetic -- With Proofs, Without Compromises},
  booktitle = {2019 IEEE Symposium on Security and Privacy (SP)},
  pages     = {1202--1219},
  year      = {2019},
  publisher = {IEEE},
  doi       = {10.1109/SP.2019.00005}
}

@inproceedings{cai:pcb:ebpf:2026,
  author    = {Cai, Yalu and Nagarakatte, Santosh},
  title     = {Proof-Carrying-{BPF} Programs for the {eBPF} Verifier},
  booktitle = {Proceedings of the 4th Workshop on eBPF and Kernel Extensions (eBPF '26)},
  publisher = {ACM},
  pages     = {9--17},
  year      = {2026},
  month     = sep,
  doi       = {10.1145/3837779.3838158}
}

@inproceedings{hofmann:jost:popl:2003,
  author    = {Hofmann, Martin and Jost, Steffen},
  title     = {Static Prediction of Heap Space Usage for First-Order Functional Programs},
  booktitle = {Proceedings of the 30th ACM SIGPLAN-SIGACT Symposium on Principles of Programming Languages (POPL '03)},
  pages     = {185--197},
  year      = {2003},
  publisher = {Association for Computing Machinery},
  address   = {New York, NY, USA},
  doi       = {10.1145/604131.604148}
}

@article{hoffmann:multivariate:toplas:2012,
  author    = {Hoffmann, Jan and Aehlig, Klaus and Hofmann, Martin},
  title     = {Multivariate Amortized Resource Analysis},
  journal   = {ACM Transactions on Programming Languages and Systems},
  volume    = {34},
  number    = {3},
  pages     = {1--62},
  year      = {2012},
  doi       = {10.1145/2362389.2362393}
}

@misc{kahn:linearmaps:arxiv:2025,
  author        = {Kahn, David M. and Hoffmann, Jan and Reps, Thomas and Grosen, Jessie},
  title         = {Efficient Cost Bounds with Linear Maps},
  year          = {2025},
  eprint        = {2509.22982},
  archivePrefix = {arXiv},
  primaryClass  = {cs.PL}
}

@article{tarjan:amortized:1985,
  author    = {Tarjan, Robert Endre},
  title     = {Amortized Computational Complexity},
  journal   = {SIAM Journal on Algebraic Discrete Methods},
  volume    = {6},
  number    = {2},
  pages     = {306--318},
  year      = {1985},
  doi       = {10.1137/0606031}
}

@inproceedings{carbonneaux:c4b:pldi:2015,
  author    = {Carbonneaux, Quentin and Hoffmann, Jan and Shao, Zhong},
  title     = {Compositional Certified Resource Bounds},
  booktitle = {Proceedings of the 36th ACM SIGPLAN Conference on Programming Language Design and Implementation (PLDI '15)},
  year      = {2015},
  pages     = {467--478},
  publisher = {Association for Computing Machinery},
  address   = {New York, NY, USA},
  doi       = {10.1145/2737924.2737955}
}

@misc{linux-bpf-uapi,
  title        = {{include/uapi/linux/bpf.h}: the {BPF\_PROG\_LOAD} arguments \texttt{log\_level}, \texttt{log\_size} and \texttt{log\_buf}},
  author       = {{The Linux kernel developers}},
  howpublished = {Linux v7.0-rc1. \url{https://git.kernel.org/pub/scm/linux/kernel/git/torvalds/linux.git/tree/include/uapi/linux/bpf.h?h=v7.0-rc1}},
  note         = {Accessed: 2026-10-19},
  year         = {2026}
}

@misc{certora:clean:2026,
  author       = {{Certora}},
  title        = {{Clean}: a {Lean}~4 port of {CompCert}'s {Clight} semantics},
  howpublished = {\url{https://github.com/Certora/clean}, directory \texttt{export/}, commit \texttt{123f7f7}},
  note         = {Experimental fork of {CompCert} 3.17. Accessed: 2026-11-02},
  year         = {2026}
}

@misc{keel-repo,
  author       = {Yalu Cai and Tung-Che Chang and Santosh Nagarakatte},
  title        = {{Keel}: A Verified In-Kernel Proof Checker for Proof-Carrying {BPF}},
  year         = 2026,
  howpublished = {\url{https://github.com/rutgers-apl/keel}}
}

\clearpage
\section*{Supplementary Material}
\pdfbookmark[1]{Supplementary Material}{supplementary-material}
\appendix
\section{EmbC}
\label{app:embc}

This appendix gives the full semantics of EmbC, its stuck states, and the proof of
Theorem~\ref{thm:embc-clight} of the paper.

\subsection{The Semantics of EmbC}
\label{sec:embc:supp-semantics}

\S\ref{sec:embc} of the main paper summarizes EmbC.  This subsection
gives the details of its syntax and its full small-step semantics.

Figure~\ref{fig:embc:syntax} of the main paper defines the program syntax.
Here \(S\) is a struct name and \(f\) a field name;
\(\mathsf{struct}(S)\) denotes the C type \cid{struct S}.
The literals \(a_{32}\) and \(a_{64}\) carry unsigned bit-vector payloads
of widths 32 and 64.  The expression \(\mathsf{table}(g,e)\) reads index
\(e\) of the global read-only table named \(g\), fixed by \(P\).
In \(e\mathbin{\to}f\), \(e\) must evaluate to a pointer to a struct
containing \(f\).  Structs reside in memory and are accessed through
pointers; EmbC does not support struct values or \(e.f\) access.
The operators \(\diamond\) are unsigned arithmetic, bitwise operations,
shifts, and comparisons.  Conversions are explicit.

The destination \(d\) and return expression \(e?\) are optional;
\(F\) names an internal function and \(\kappa\) a kernel operation.
Each function declaration
\((\bar x:\bar\tau,\tau_r?,\bar y:\bar\upsilon,s)\) gives parameters,
an optional return type, locals, and a body.  For a successful embedding,
parameters, locals, return values, and struct fields have scalar types,
and expressions pass the embedding's type checks.
So \cid{y = f(x) + 1} is written as
\cid{t = f(x); y = t + 1}.  This separation also maps directly to
Clight call statements~\cite{leroy:compcert:cacm:2009}.

Figure~\ref{supp-fig:embc:memory} defines environments, values, memory, and
runtime typing.  An environment \(\lframe\) is a finite map from variable
names to declared types and current values.
Lookup is written \(\lframe(x)=(\tau,v)\), and
\(\lframe[x\mapsto(\tau,v)]\) updates a binding.
For the well-formed programs considered here, parameter and local names
within a function are distinct.
Memory \(\mu\) is a sequence of \(k\) blocks \(B_0,\ldots,B_{k-1}\).
A live block contains an element type and a vector of values; it represents
an array of that type or one struct whose cells are its fields.
A dead entry records a freed block.  Block identifiers are natural-number
indices, and \(\mu(b)\) looks up a live block at index \(b\).
We write \(\bot\) for an undefined partial operation.

An array pointer \(\vptr{b}{i}\) identifies block \(b\) and cell offset
\(i\); struct field access requires offset zero.  Offsets count cells.
The value \(\vundef\) records uninitialized memory and cannot be loaded.
The runtime judgment \(\mu\vdash v:\tau\) checks value types;
for a non-null pointer it checks the element type of a live block.
Figure~\ref{supp-fig:embc:memory} also defines the zero-initialized value
\(0_\tau\) of each type, with integer width \(w\in\{32,64\}\).

\begin{embcfigure}
\hrule
\smallskip
\small
\[
\begin{gathered}
\begin{aligned}
 v &::=\mathsf{u32}(a)\mid\mathsf{u64}(a)\mid\vptr{b}{i}
        \mid\vnull\mid\vuptr\mid\vundef,\\
 \lframe &=\{x_1\mapsto(\tau_1,v_1),\ldots,x_n\mapsto(\tau_n,v_n)\},\\
 B &::=\mathsf{dead}\mid(\tau,[v_0,\ldots,v_{\ell-1}]),
       \qquad\mu=[B_0,\ldots,B_{k-1}].
\end{aligned}
\\[3pt]
 \mu(b)=
 \begin{cases}
 B_b & b<k\ \land\ B_b\ne\mathsf{dead},\\
 \bot & \text{otherwise}.
 \end{cases}
\\[3pt]
\begin{gathered}
 \mu\vdash\mathsf{u32}(a):\mathsf{u32},\quad
 \mu\vdash\mathsf{u64}(a):\mathsf{u64},\quad
 \mu\vdash\vnull:\tau^*,\\
 \mu\vdash\vuptr:\mathsf{bpfptr},\qquad
 \frac{\mu(b)=(\tau,\bar v)}{\mu\vdash\vptr{b}{i}:\tau^*}.
\end{gathered}
\\[3pt]
 0_{\mathsf{u}_w}=\mathsf{u}_w(0),\qquad 0_{\tau^*}=\vnull,\qquad
 0_{\mathsf{bpfptr}}=0_{\mathsf{struct}(S)}=\vundef.
\end{gathered}
\]
\par\smallskip
\hrule
\par\nopagebreak[4]
\caption{EmbC values, memory, and runtime typing.}
\Description{Value, environment, block, and heap definitions, partial
heap lookup, runtime value typing, and zero initialization.}
\label{supp-fig:embc:memory}
\end{embcfigure}

Bounds are checked when accessing
memory.  Stores check the declared cell type, and reads reject dead blocks,
invalid slots, and uninitialized cells.

The machine states and continuations are those of \S\ref{sec:embc:semantics} of the main paper.
A return state carries an optional result \(\hat v\), where \(\varnothing\) means no value; value
injections into optional results are left implicit.
Normal completion is represented by \(\mathsf{skip}\).
A break discards sequence continuations until it reaches a loop;
a return discards pending sequences and loops up to the function boundary.

Expression evaluation \(\sem{e}_{P,\lframe,\mu}\), abbreviated
\(\sem{e}\), is defined in Figure~\ref{fig:embc:expressions}.
Evaluation is partial and side-effect free: missing names, failed premises,
or undefined operands make it undefined.
Here \(w,w'\in\{32,64\}\), \([q]\in\{0,1\}\) is the indicator of \(q\),
and \(\mathsf{idx}(\mathsf{u}_w(a))=a\).
For a table name \(g\), \(P(g)\) is its declared sequence of words;
\(P(g)[i]\) selects its entry at index \(i\), starting from zero.
A missing table or an out-of-bounds index makes the lookup undefined.
Table lookup leaves the environment, heap, and oracle unchanged.

Heap reads require live, in-bounds, initialized cells.
Array reads reject struct blocks; field reads require the declared
struct and field.  Arithmetic and bitwise operations require equal-width
integer operands and compute modulo \(2^w\); unsigned comparisons return
\(\mathsf{u32}\) Booleans.  Division and remainder require a nonzero
divisor, and shifts require a count below \(w\).
Equality and inequality also accept null pointers and pointers into one
live block: null equals null, a live pointer differs from null, and
pointers in that block compare by their offsets.
All other pointer comparisons are undefined.
Short-circuit operators evaluate their right operand only when required.

\begin{embcfigure}
\small
\begin{align*}
 \sem{x}&=v\quad\text{if }\lframe(x)=(\tau,v),\\
 \sem{a_{32}}&=\mathsf{u32}(a),\quad \sem{a_{64}}=\mathsf{u64}(a),\\
 \sem{\mathsf{NULL}}&=\vnull,\\[2pt]
 \sem{\mathsf{table}(g,e)}&=\mathsf{u32}(P(g)[\mathsf{idx}(\sem{e})]),\\
 \sem{e_1\mathbin\diamond e_2}&=\mathsf{op}_\mu(\diamond,\sem{e_1},\sem{e_2}),\\
 \sem{!e}&=\mathsf{u32}([a=0])\quad\text{if }\sem{e}=\mathsf{u32}(a),\\
 \sem{\mathord\sim e}&=\mathsf{u}_w(\mathord\sim a)\quad\text{if }\sem{e}=\mathsf{u}_w(a),\\
 \sem{e_1\mathbin{\&\&}e_2}
   &=\begin{cases}
       \mathsf{u32}(0) & \sem{e_1}=\mathsf{u32}(0),\\
       \mathsf{u32}([b\ne0]) & \sem{e_1}=\mathsf{u32}(a),\ a\ne0,
                                  \ \sem{e_2}=\mathsf{u32}(b),
     \end{cases}\\
 \sem{e_1\mathbin{||}e_2}
   &=\begin{cases}
       \mathsf{u32}(1) & \sem{e_1}=\mathsf{u32}(a),\ a\ne0,\\
       \mathsf{u32}([b\ne0]) & \sem{e_1}=\mathsf{u32}(0),
                                 \ \sem{e_2}=\mathsf{u32}(b),
     \end{cases}\\
 \sem{(\mathsf{u}_w)e}&=\mathsf{u}_w(a\bmod 2^w)
                    \quad\text{if }\sem{e}=\mathsf{u}_{w'}(a),\\
 \sem{p[i]}&=\mathsf{read}_{\rm arr}(\mu,b,j+\mathsf{idx}(\sem{i}))
                    \quad\text{if }\sem{p}=\vptr{b}{j},\\
 \sem{p\mathbin{\to}f}&=\mathsf{read}_{\rm field}(P,\mu,b,f)
                    \quad\text{if }\sem{p}=\vptr{b}{0},\\
 \sem{p+_{\rm ptr}i}&=\vptr{b}{j+k}
       \quad\begin{array}{l}
         \text{if }\sem{p}=\vptr{b}{j},\ k=\mathsf{idx}(\sem{i}),\\
         \mu(b)=(\tau,\bar v),\ \tau\text{ non-struct},\ j+k\le|\bar v|.
       \end{array}
\end{align*}
\par\nopagebreak[4]
\caption{Expression evaluation in EmbC.}
\Description{A complete partial evaluation table for expressions,
including literals, read-only table lookup, arithmetic, casts, short-circuit logic, array and
field loads, and bounded pointer arithmetic.}
\label{fig:embc:expressions}
\end{embcfigure}

Figures~\ref{fig:embc:semantics} and~\ref{fig:embc:continuations}
define the small-step transition \(\stepj{c}{c'}\) for a fixed program
\(P\).  We write \(\sigma=(\mu,o)\) and \(\sigma'=(\mu',o')\) for heaps
paired with stateful oracles, and let \(\sem{\bar e}\) evaluate a list
elementwise.
The checked update \(\lframe[x\leftarrow_\mu v]\) equals
\(\lframe[x\mapsto(\tau,v)]\) when \(\lframe(x)=(\tau,v_0)\) and
\(\mu\vdash v:\tau\); it is undefined otherwise.
Array and field writes, denoted \(\mathsf{write}_{\rm arr}\) and
\(\mathsf{write}_{\rm field}\), require a live block, a valid slot, and
the declared cell type.  Array writes reject struct blocks; field writes
resolve the field through \(P\).
Successful assignments and stores complete as \(\mathsf{skip}\).
Kernel calls check the precondition \(\mathit{Pre}_\kappa\) and use
\(\operatorname{ask}(o,\kappa,\bar v,\mu)\), which
Definition~\ref{def:oracle} of the main paper defines.
In Lean the precondition check sits inside the oracle: the type of its
answer function makes it answer exactly the calls that meet
\(\mathit{Pre}_\kappa\), and the Lean rule for kernel calls has only the
oracle's answer as its premise.  The two rules allow the same transitions.

Internal and kernel calls share \rulename{Call}; \rulename{Kernel}
executes only a kernel callee and records
\((\kappa,\bar v,\sigma,\hat v,\sigma')\) for trace labeling.
The kernel action finishes before its result is assigned to the caller.
The return transitions check the declared return type and discard the
callee's environment.

\begin{embcfigure}
\small
\begin{gather*}
 \frac{\sem{e}=v\qquad \lframe[x\leftarrow_\mu v]=\lframe'}
      {\stepj{\langle x:=e,\lframe,\mu,o,K\rangle}{\langle\mathsf{skip},\lframe',\mu,o,K\rangle}}
    \quad\rulename{Assign}\\[5pt]
 \frac{\sem{p}=\vptr{b}{j}\quad \mathsf{idx}(\sem{i})=a\quad \sem{e}=v
       \quad\mathsf{write}_{\rm arr}(\mu,b,j+a,v)=\mu'}
      {\stepj{\langle p[i]:=e,\lframe,\mu,o,K\rangle}{\langle\mathsf{skip},\lframe,\mu',o,K\rangle}}
    \quad\rulename{Store}\\[5pt]
 \frac{\sem{p}=\vptr{b}{0}\quad \sem{e}=v
       \quad\mathsf{write}_{\rm field}(P,\mu,b,f,v)=\mu'}
      {\stepj{\langle p\mathbin{\to}f:=e,\lframe,\mu,o,K\rangle}{\langle\mathsf{skip},\lframe,\mu',o,K\rangle}}
    \quad\rulename{Field}\\[5pt]
 \stepj{\langle s_1;s_2,\lframe,\mu,o,K\rangle}{\langle s_1,\lframe,\mu,o,\mathsf{seq}(s_2,K)\rangle}
    \quad\rulename{Sequence}\\[3pt]
 \frac{\sem{e}=\mathsf{u32}(a)}
      {\stepj{\langle\mathsf{if}\ e\ \mathsf{then}\ s_1\ \mathsf{else}\ s_2,\lframe,\mu,o,K\rangle}{\langle s_{a\ne0\,?\,1:2},\lframe,\mu,o,K\rangle}}
    \quad\rulename{If}\\[5pt]
 \frac{\sem{e}=\mathsf{u32}(0)}
      {\stepj{\langle\mathsf{while}\ e\ \mathsf{do}\ s,\lframe,\mu,o,K\rangle}{\langle\mathsf{skip},\lframe,\mu,o,K\rangle}}
    \quad\rulename{While-False}\\[5pt]
 \frac{\sem{e}=\mathsf{u32}(a)\qquad a\ne0}
      {\stepj{\langle\mathsf{while}\ e\ \mathsf{do}\ s,\lframe,\mu,o,K\rangle}{\langle s,\lframe,\mu,o,\mathsf{loop}(e,s,K)\rangle}}
    \quad\rulename{While-True}\\[5pt]
 \frac{\sem{\bar e}=\bar v}
      {\stepj{\langle d:=D(\bar e),\lframe,\mu,o,K\rangle}{\mathsf{call}(D,\bar v,\mu,o,\mathsf{caller}(d,\lframe,K))}}
    \quad\rulename{Call}\\[5pt]
 \frac{\mathit{Pre}_\kappa(\bar v,\mu)
       \quad \operatorname{ask}(o,\kappa,\bar v,\mu)=(\hat v,\mu',o')}
      {\stepj{\mathsf{call}(\kappa,\bar v,\mu,o,K)}{\mathsf{ret}(\hat v,\mu',o',K)}}
    \quad\phantomsection\label{supp-rule:embc:kernel}\rulename{Kernel}\\[5pt]
 \stepj{\langle\mathsf{return},\lframe,\mu,o,\mathsf{fun}(\varnothing,K)\rangle}{\mathsf{ret}(\varnothing,\mu,o,K)}
    \quad\rulename{Return-Void}\\[3pt]
 \frac{\sem{e}=v\qquad\mu\vdash v:\tau}
      {\stepj{\langle\mathsf{return}\ e,\lframe,\mu,o,\mathsf{fun}(\tau,K)\rangle}{\mathsf{ret}(v,\mu,o,K)}}
    \quad\rulename{Return-Value}
\end{gather*}
\par\nopagebreak[4]
\caption{Small-step statement and kernel transitions of EmbC.}
\Description{Transitions for assignment, array and field stores,
sequencing, conditionals, while loops, calls, kernel outcomes, and returns.}
\label{fig:embc:semantics}
\end{embcfigure}

The return rules use the optional expression \(e?\) from the syntax.
Their recursive premises discard pending sequence and loop continuations
within a single return transition, stopping at the function boundary.
Function entry checks arguments and initializes locals; a returned value
is then discarded or assigned in the saved caller environment.
Function lookup \(P(F)\) returns the declaration of \(F\).
Vector typing \(\mu\vdash\bar v:\bar\tau\) requires matching lengths and
pointwise value typing.  The entry environment \(\lframe_F\) binds each
parameter \(x_i\) to \((\tau_i,v_i)\) and each local \(y_j\) to
\((\upsilon_j,0_{\upsilon_j})\), with no other bindings.

\begin{embcfigure}
\small
\begin{gather*}
 \stepj{\langle\mathsf{skip},\lframe,\mu,o,\mathsf{seq}(s,K)\rangle}{\langle s,\lframe,\mu,o,K\rangle}
    \quad\rulename{Next}\\[3pt]
 \stepj{\langle\mathsf{skip},\lframe,\mu,o,\mathsf{loop}(e,s,K)\rangle}{\langle\mathsf{while}\ e\ \mathsf{do}\ s,\lframe,\mu,o,K\rangle}
    \quad\rulename{Repeat}\\[3pt]
 \stepj{\langle\mathsf{break},\lframe,\mu,o,\mathsf{seq}(s,K)\rangle}{\langle\mathsf{break},\lframe,\mu,o,K\rangle}
    \quad\rulename{Break-Sequence}\\[3pt]
 \stepj{\langle\mathsf{break},\lframe,\mu,o,\mathsf{loop}(e,s,K)\rangle}{\langle\mathsf{skip},\lframe,\mu,o,K\rangle}
    \quad\rulename{Exit-Loop}\\[3pt]
 \frac{\stepj{\langle\mathsf{return}\ e?,\lframe,\mu,o,K\rangle}{\mathsf{ret}(\hat v,\mu,o,K')}}
      {\stepj{\langle\mathsf{return}\ e?,\lframe,\mu,o,\mathsf{seq}(s,K)\rangle}{\mathsf{ret}(\hat v,\mu,o,K')}}
    \quad\rulename{Return-Sequence}\\[5pt]
 \frac{\stepj{\langle\mathsf{return}\ e?,\lframe,\mu,o,K\rangle}{\mathsf{ret}(\hat v,\mu,o,K')}}
      {\stepj{\langle\mathsf{return}\ e?,\lframe,\mu,o,\mathsf{loop}(e_c,s,K)\rangle}{\mathsf{ret}(\hat v,\mu,o,K')}}
    \quad\rulename{Return-Loop}\\[5pt]
 \stepj{\langle\mathsf{skip},\lframe,\mu,o,\mathsf{fun}(\varnothing,K)\rangle}{\mathsf{ret}(\varnothing,\mu,o,K)}
    \quad\rulename{Fallthrough-Void}\\[5pt]
 \frac{P(F)=(\bar x:\bar\tau,\tau_r?,\bar y:\bar\upsilon,s)
       \qquad \mu\vdash\bar v:\bar\tau}
      {\stepj{\mathsf{call}(F,\bar v,\mu,o,K)}{\langle s,\lframe_F,\mu,o,\mathsf{fun}(\tau_r?,K)\rangle}}
    \quad\rulename{Enter-Function}\\[5pt]
 \stepj{\mathsf{ret}(\hat v,\mu,o,\mathsf{caller}(\varnothing,\lframe,K))}{\langle\mathsf{skip},\lframe,\mu,o,K\rangle}
    \quad\rulename{Resume-Discard}\\[3pt]
 \frac{\lframe[x\leftarrow_\mu v]=\lframe'}
      {\stepj{\mathsf{ret}(v,\mu,o,\mathsf{caller}(x,\lframe,K))}{\langle\mathsf{skip},\lframe',\mu,o,K\rangle}}
    \quad\rulename{Resume-Assign}
\end{gather*}
\par\nopagebreak[4]
\caption{Small-step continuation and function transitions of EmbC.}
\Description{Transitions for normal completion, break, return unwinding,
function entry, void fallthrough, and restoration of caller environments.}
\label{fig:embc:continuations}
\end{embcfigure}

A state \(\mathsf{ret}(\hat v,\mu,o,\mathsf{stop})\) is final and has no
successor.  All other states without a successor are stuck.
\S\ref{sec:embc:supp-stuck} describes these states.
The executable interpreter iterates the same transition function and
reports a failure when it encounters such a stuck state.
For fixed \(P\), the successor of a machine state is deterministic.

\paragraph{Kernel contracts.}
Table~\ref{tab:embc:supp-kernel} states the contract of each kernel API in full.
\S\ref{sec:embc:semantics} of the main paper names these APIs without their contracts.
Pending-signal checks and rescheduling are external calls in this sequential model.

\begin{longtable}{@{}p{.30\linewidth}p{.66\linewidth}@{}}
\caption{Logical kernel specifications used by EmbC. Allocation sizes are
cell counts.}
\label{tab:embc:supp-kernel}\\
\toprule
Operation & Preconditions and effects\\
\midrule
\endfirsthead
\multicolumn{2}{@{}l}{\tablename~\thetable\ (continued)}\\
\toprule
Operation & Preconditions and effects\\
\midrule
\endhead
\midrule
\multicolumn{2}{r@{}}{Continued on next page.}\\
\endfoot
\bottomrule
\endlastfoot
\(\cid{kvcalloc}_\tau(n)\) &
\(n>0\), non-struct \(\tau\). Returns null, or a fresh array of \(n\)
zero-initialized cells.\\
\(\cid{kzalloc}_S()\) &
\(S\) is declared. Returns null, or a fresh struct with zero-initialized
fields.\\
\(\cid{kvrealloc}_\tau(p,n)\) &
\(n>0\), non-struct \(\tau\); \(p\) is null or the base of a live array of type \(\tau\).
Failure preserves the old block. Success returns a fresh block, preserves
the retained prefix, leaves additional cells uninitialized, and frees the
old block if present.\\
\(\cid{kvfree}(p)\) &
\(p\) is null or a live block's base. Null leaves the heap unchanged;
otherwise the block becomes dead.\\
\(\cid{copy\_from\_bpfptr}(p,h,n)\) &
\(h\) is the opaque input handle; \(p\) addresses \(n\) cells of a live
\(\mathsf{u32}\) array. Success supplies arbitrary words and returns zero.
Failure marks the destination slice uninitialized and returns one.\\
\(\cid{signal\_pending}()\) &
Checks for a pending signal in the current task. Returns zero or one,
as selected by the environment, and leaves the heap unchanged.\\
\(\cid{cond\_resched}()\) &
Allows the scheduler to reschedule. Returns no value and leaves the
heap unchanged.\\
\end{longtable}

\subsection{Stuck States of EmbC}
\label{sec:embc:supp-stuck}

\S\ref{sec:embc:supp-semantics} gives the successful small-step transitions.  This subsection
describes the conditions under which no transition applies.
For a fixed program \(P\), write \(\stepj{c}{c'}\) for a transition and
\(\nostepj{c}\) when no successor exists.  A state returning to the stop
continuation is final; every other state with no successor is stuck.
The transition rules of Figures~\ref{fig:embc:semantics}
and~\ref{fig:embc:continuations} are exhaustive, so each stuck case negates
the premises of the rules for that state.
The state remains at the failing operation; there is no failure constructor
or transition to a separate error state.

For example, rule \rulename{Store} of Figure~\ref{fig:embc:semantics} gives
this stuck case:
\[
 \frac{\begin{gathered}
         \neg\exists b,j,a,v,\mu'.\
           \sem{p}=\vptr{b}{j}\ \land\ \mathsf{idx}(\sem{i})=a\\
         {}\land\ \sem{e}=v\ \land\
           \mathsf{write}_{\rm arr}(\mu,b,j+a,v)=\mu'
       \end{gathered}}
      {\nostepj{\langle p[i]:=e,\lframe,\mu,o,K\rangle}}
    \quad\rulename{Store-Stuck}
\]
The negated existential premise lists all values and operations required for
the store to succeed.

For statements and kernel calls, the stuck cases include undefined operands,
incompatible value forms, invalid indices, rejected memory updates, and
violated kernel preconditions.  A kernel operation whose specification is
undefined emits no observation.  A successful kernel operation retains its
observation and resulting kernel state even if the subsequent result
assignment is stuck.

For functions and continuations, a call is stuck on an undeclared name or on
arguments that do not match the parameters in arity and type.  Falling through
the end of a value-returning function is undefined.  A break cannot escape a
function boundary.  The return conditions reject a missing function boundary,
a mismatch between void and value returns, an undefined return expression,
or an ill-typed returned value.  A return state is also stuck when the caller
expects a result and none is returned, when the result cannot be assigned, or
when its continuation is neither a saved caller nor the stop continuation.
A return state with the stop continuation is final, rather than stuck.

\subsection{Formal EmbC--Clight Correspondence}
\label{sec:embc:supp}

This subsection gives the formal development behind the main paper's
correspondence results.  It states the separate forward and backward
simulations, their kernel contracts, and the derivation of
behavior equivalence and reflection of target undefined behavior.

\subsubsection{Prerequisites and Notation}
\label{sec:embc:supp-prerequisites}

The executable machine of \S\ref{sec:embc:supp-semantics}
carries the stateful oracle.
For correspondence, we omit that private component and let each kernel call
choose any outcome allowed by its specification.  The program is fixed, and
the resulting source machine states are those of \S\ref{sec:embc:semantics} of the main
paper without \(o\).
The control transitions are those of \S\ref{sec:embc:supp-semantics} with the oracle
component omitted.  A public kernel outcome is realizable when
\[
 \mathcal K_P(\kappa,\bar v,\mu,\hat v,\mu')
 \quad\Longleftrightarrow\quad
 \exists o,o'.\
 \operatorname{ask}(o,\kappa,\bar v,\mu)
   =(\hat v,\mu',o').
\]
The oracle's set of states is existentially quantified as part of realizability.
Unpacking the response certificate gives the effects in
Table~\ref{tab:embc:supp-kernel}.  Conversely, every outcome described there has a certified
oracle witness, so deriving this relation loses no permitted alternatives.
The executable interpreter follows this relation by invoking its stored oracle.

Kernel specifications are assumptions about the external
interface, rather than proofs of the Linux implementations.

\subsubsection{Embedding in Clight and Simulations}
\label{sec:embc:supp-simulation}

For an EmbC program \(P\), write \(P_{\mathrm{Cl}}\) for its translated Clight
program.  The translation constructs a Clight AST directly.  Arithmetic,
assignments, stores, calls, and returns have corresponding Clight constructs.
Array access becomes pointer addition and dereference; short-circuit
operators become conditionals with fresh temporaries; \cid{while} becomes
Clight's explicit loop-and-break form.  Parameters and locals become
temporaries, and kernel operations become Clight builtin statements
invoking abstract wrappers with the logical signatures of the kernel
interface specified in Table~\ref{tab:embc:supp-kernel}.
The results concern Clean's Lean Clight model; its fidelity to CompCert's
Rocq semantics is trusted.

\paragraph{Clight machine states.}
Clight also has regular, call, and return states.  A regular state retains
the current function \(f_C\), statement \(s_C\), continuation \(K_C\),
local-variable environment \(\gamma_C\), temporary-variable environment
\(\lframe_C\), and byte memory \(M_C\).  The local-variable environment maps
addressable locals to memory blocks and types; the temporary environment
maps temporaries to values.  A call state carries the callee \(D_C\),
evaluated arguments \(\bar v_C\), continuation, and memory; the callee
may be internal or external.  A return state carries the result \(v_C\),
continuation, and memory.  We use the following notation for the three forms:
\[
\begin{aligned}
 q ::= {}&\langle f_C,s_C,K_C,\gamma_C,\lframe_C,M_C\rangle_C\\
         &{}\mid\mathsf{call}_C(D_C,\bar v_C,M_C,K_C)
           \mid\mathsf{ret}_C(v_C,M_C,K_C).
\end{aligned}
\]
EmbC's \(\lframe\) corresponds to the target temporaries \(\lframe_C\);
\(\gamma_C\) is empty.  The relation below connects heaps \(\mu\) and
\(M_C\).  A kernel builtin has no Clight function body.  Its external-call
specification relates the arguments and memory before the call to allowed
traces, returned values, and memories after the call.  It may permit several
outcomes.  The relational EmbC semantics likewise permits every outcome of
its kernel specification; the contracts below require the two specifications
to agree.  Clight's kernel builtins
execute in regular states, which match the source kernel-call and return phases.

\paragraph{Corresponding memories and values.}
EmbC addresses cells; Clight addresses bytes.  To compare them, we use a
representation map \(j\) that records a target block and a positive byte
stride for each source block, together with the representation of the
opaque input handle.  Its block mapping is injective.  We write
\(j(b)\) for the target block assigned to source block \(b\),
\(\mu\rep{j}M_C\) when source heap \(\mu\) is represented by Clight memory
\(M_C\), and \(v\rep{j}v_C\) when source value \(v\) is represented by
Clight value \(v_C\).  The arguments distinguish the heap and value
relations.

Integer values retain their bit-vector contents.  For pointers, let \(b\)
be a source block, \(i\) its cell offset, and
\(\operatorname{stride}_j(b)\) the bytes per cell recorded by \(j\).
A Clight pointer, constructed by \(\mathsf{Vptr}\), pairs a target block
with a byte offset.  Writing \(\operatorname{repr}_{64}\) for
the encoding of a byte offset as a 64-bit word, pointer representation is
\[
 \vptr{b}{i}\rep{j}
 \mathsf{Vptr}\bigl(j(b),\operatorname{repr}_{64}
                       (i\cdot\operatorname{stride}_j(b))\bigr).
\]
The scaled offset must fit without wrapping.  Array cells are represented
by loads at these offsets; struct fields use Clight's computed field
offsets and padding.  Live blocks have the required permissions, and freed
blocks have no current permissions.  The heap relation also requires the
read-only table globals to retain their contents and to be disjoint from
the target blocks representing source allocations, including freed ones.

\paragraph{The initial call and its assumptions.}
Fix a source program whose translation succeeds and a declared entry
function \(F\).  Well-formedness requires unique declarations, scalar
value declarations, valid named struct types, bounded constant tables,
and a valid composite layout.  These checks establish structural and type
compatibility; program safety remains a separate proof obligation.

Fix initial source heap \(\mu_0\), Clight memory \(M_{C,0}\), and
representation map \(j_0\).  The initial heaps, table globals, and function
arguments must satisfy the representation relations.  Write
\(\bar v\) and \(\bar v_C\) for these source and target arguments, and
\(F_C\) for the translated entry function.  Both executions begin with a
pending call and the stop continuation.  Their initial machine states are
\[
 E_0=\mathsf{call}(F,\bar v,\mu_0,\mathsf{stop}),
 \qquad
 C_0=\mathsf{call}_C(F_C,\bar v_C,M_{C,0},\mathsf{stop}_C).
\]
Thus \(E_0\) is a state of \(P\), and \(C_0\) is a state of
\(P_{\mathrm{Cl}}\).

We call the initial source state \emph{admissible} when every machine state
reachable from it fits the target representation.  This includes the
block extents, type layouts, and pointers in the heap, environment, and
pending arguments or results.  The requirement covers every finite prefix
of an infinite execution as well.  It rules out representation overflow
and permits source failure.

\paragraph{Events and traces.}
Observations use Clight's event alphabet, denoted \(\mathit{Event}\).
Event values are integers, floating-point values, or pointers represented
by a global symbol and byte offset.
An external-call event records a name \(a\), argument event values
\(\bar d\), and returned event value \(d'\).  A volatile load or store
records a memory chunk \(\ell\), specifying the access size and value
format, a global symbol \(g\), byte offset \(i\), and transferred event
value \(d\).  An annotation records its name and
arguments.  An event \(\alpha\) has one of four forms:
\[
\begin{aligned}
 \alpha ::= {}&\mathsf{syscall}(a,\bar d,d')
    \mid\mathsf{vload}(\ell,g,i,d)\\
    &{}\mid\mathsf{vstore}(\ell,g,i,d)
    \mid\mathsf{annot}(a,\bar d).
\end{aligned}
\]
EmbC has no volatile-access constructs; we use the common alphabet to
compare its kernel observations with those of Clight.
A finite trace \(t\) is a finite sequence of events, and an infinite
trace \(T\) is an infinite sequence.  The empty trace is \(\epsilon\).
Thus \(t\in\mathit{Event}^{*}\) and
\(T\in\mathit{Event}^{\mathbb N}\).

An observation relation \(\mathcal O\) specifies the traces allowed for
a defined kernel outcome.  Its arguments are the operation \(\kappa\),
argument values \(\bar v\), initial heap \(\mu\), optional returned
value \(\hat v\), resulting heap \(\mu'\), and trace \(t\).
We write
\[
 \mathcal O(\kappa,\bar v,\mu,\hat v,\mu',t)
\]
when that outcome may emit \(t\).  A call may emit an empty trace or
several events.  We require every defined kernel outcome to admit at least
one trace, so adding observations does not remove executions.

We now label the small-step EmbC semantics of \S\ref{sec:embc:supp-semantics}
with traces.  The kernel-call transition
(rule \rulename{Kernel} in Figure~\ref{fig:embc:semantics})
emits any trace permitted by
\(\mathcal O\) for the recorded call.  Its return state retains the
resulting heap.  If the next step cannot assign the result,
execution is stuck after the kernel event has already been emitted.  Every other step emits \(\epsilon\).
Let \(c,c'\) range over EmbC machine states and \(q,q'\) over Clight
machine states; primed letters denote successors.  Subscripts \(E\) and
\(C\) identify the EmbC and Clight semantics.  Their labeled steps are
written
\[
 \lstepj{E}{c}{t}{c'}
 \qquad\text{and}\qquad
 \lstepj{C}{q}{t}{q'}.
\]
The kernel rule is therefore
\[
 \frac{\mathcal K_P(\kappa,\bar v,\mu,\hat v,\mu')
       \qquad\mathcal O(\kappa,\bar v,\mu,\hat v,\mu',t)}
      {\lstepj{E}{\mathsf{call}(\kappa,\bar v,\mu,K)}{t}
                    {\mathsf{ret}(\hat v,\mu',K)}}.
\]
For either semantics, a superscript \(*\) permits zero or more steps,
and \(+\) requires one or more.  The label on such a path is the
concatenation of its step traces.  A silent path has label \(\epsilon\),
even when it contains several computation steps.

\paragraph{Matching termination.}
A final EmbC machine state is
\(\mathsf{ret}(\hat v,\mu,\mathsf{stop})\), carrying the returned
optional value and final heap.  A final Clight state is
\(\mathsf{ret}_C(v_C,M_C,\mathsf{stop}_C)\).

Termination must preserve the returned value and heap representation.
Allocations may have extended the initial representation map.  For a
candidate final map \(j'\), write \(j'\succeq_{\mu_0}j_0\) when it
preserves the target block and stride of every block already allocated in
\(\mu_0\), including dead blocks, and preserves the opaque input handle.
We extend value representation to optional returns: an absent source
result corresponds to Clight's undefined value.  For source final heap
\(\mu\), the final-outcome relation \(\Phi_{j_0}\) requires such an extension
that relates the
returned values and heaps:
\[
 \Phi_{j_0}(\hat v,\mu,v_C,M_C)
 \;\equiv\;
 \exists j'\succeq_{\mu_0}j_0.\
     \hat v\rep{j'}v_C\ \land\ \mu\rep{j'}M_C.
\]
The heap relation includes the table-global requirements above.  The
state of an executable oracle is ignored when comparing outcomes.

\paragraph{Forward simulation.}
The kernel specifications must agree in the forward direction: each
observed source kernel outcome whose resulting heap fits the target
representation must have a Clight outcome with the same trace, a related
returned value, and related memory under an extended representation map.
We call this requirement the \emph{forward kernel contract}.

Fix an admissible initial source state \(E_0\).  We first assign a structural rank
\(r(c)\) to source states to bound bookkeeping steps.  For statements,
let \(w(s_1;s_2)=w(s_1)+w(s_2)+2\), and let \(w(s)=1\) otherwise.
For continuations, let
\(h(\mathsf{seq}(s,K))=w(s)+1+h(K)\),
\(h(\mathsf{loop}(e,s,K))=1+h(K)\),
\(h(\mathsf{fun}(\tau?,K))=2\), and \(h(K)=1\) for stop and caller forms.
A regular state executing \(s\) with continuation \(K\) has rank
\(r(c)=w(s)+h(K)\).  A return state with continuation
\(\mathsf{caller}(d,\lframe,K)\) has rank \(h(K)+3\); all other call
and return states have rank one.

We define \(c\,R_i\,q\), for a natural number \(i\), to require that
every source state reachable from \(c\) fits the target representation,
and that one of the following two cases holds.

\emph{Ordinary matching.}
The index is \(i=r(c)\).  There is a representation map extending the
initial map that relates the current heaps, preserves the read-only table
globals, and matches the control positions, local variables, pending
arguments or results, and continuations.  Control positions correspond
under the translation, allowing its bookkeeping phases: for example,
the target may have advanced over a translated skip or already bound a
builtin result that the source is about to bind.
The source heap extends the initial allocation history: block identities
are retained, blocks already dead initially stay dead, and retained live
blocks keep their element types.  Contents may change and live blocks
may be freed.

\emph{Source failure.}
The source reaches a non-final state with no successor within at most
\(i\) internal computation steps.  These steps perform no kernel calls,
even calls whose observable trace would be empty.  In this case, \(q\)
is unrestricted.  This clause accommodates EmbC's conservative failures.

The forward theorem proves the following properties of this relation,
starting with the initial match
\(E_0\,R_1\,C_0\).  From a related pair, each source step must
either be matched by one or more target steps with the same trace, or be
silent and leave the target unchanged while strictly decreasing the
index.  The successor index \(i'\) and target machine state \(q'\) may
depend on that step.  Formally,
\[
\begin{gathered}
 c\,R_i\,q\ \land\ \lstepj{E}{c}{t}{c'}
 \quad\Longrightarrow\\
 \exists i',q'.\ c'\,R_{i'}\,q'\ \land\
 \bigl(\lplusj{C}{q}{t}{q'}\ \lor\
       (t=\epsilon\land q'=q\land i'<i)\bigr).
\end{gathered}
\]
As in CompCert, the index is a well-founded measure: it permits finite
stuttering but rules out infinitely many source steps matched by no target
progress.  The theorem also proves that a final source state, with returned
result \(\hat v\) and final heap \(\mu\), is related
only to a final target state with a matching outcome:
\[
\begin{gathered}
 \mathsf{ret}(\hat v,\mu,\mathsf{stop})\,R_i\,q
 \quad\Longrightarrow\\
 \exists v_C,M_C.\ q=\mathsf{ret}_C(v_C,M_C,\mathsf{stop}_C)\\
 \land\Phi_{j_0}(\hat v,\mu,v_C,M_C).
\end{gathered}
\]
These obligations give the initial, step, and final matching of a forward
simulation.  They assume the forward kernel contract and admissibility;
they do not require global source safety.

The proof follows the source statement and continuation.  Ordinary
steps preserve the heap, variable, and continuation correspondence;
unmatched bookkeeping decreases the structural rank.  In the source-failure
case, each internal step shortens the remaining path to stuckness.

The memory lemmas connect cell accesses to byte accesses with the correct
scaling and struct padding.  They retain a block's representation stride
after it is freed, so dangling pointers stored elsewhere remain
representable.  The expression lemmas show that newly introduced target
temporaries preserve the source variable bindings, including while
evaluating short-circuit loop conditions.

\paragraph{Backward simulation and kernel outcomes.}
The backward direction matches each possible Clight execution.  For example,
a pending-signal check may return either zero or one.  The source relation
permits both replies, so the proof chooses the one matching the target
step.  It reasons directly about the kernel specification, independently
of how an executable oracle represents its state.

The \emph{backward kernel contract} complements the forward one.  Each
Clight external outcome must either have a matching permitted source reply,
or arise at a call whose logical source specification is
undefined.  Matching preserves the trace, returned value, heap relation,
table globals, and extension of the representation map.
For the backward simulation, assume both kernel contracts and
admissibility of the initial source state.  Observations must be total as
above and must depend only on the operation, arguments, result, and before
and after heaps.

The backward theorem uses CompCert's local safety condition.  Starting
from a source machine state \(c\), every machine state \(u\) reachable
using only silent steps must be final or able to take another step.
Call this property \(\operatorname{safe}(c)\).  Its formal definition is
\[
\begin{gathered}
 \operatorname{safe}(c)\ \equiv\
 \forall u.\ \lstepsj{E}{c}{\epsilon}{u}
 \Longrightarrow\\
 \bigl(\exists\hat v,\mu.\
          u=\mathsf{ret}(\hat v,\mu,\mathsf{stop})\bigr)
 {}\lor\bigl(\exists t,u'.\
                  \lstepj{E}{u}{t}{u'}\bigr).
\end{gathered}
\]
This condition rules out failure reachable silently from \(c\).  It
permits failure after a future observable event; it is weaker than safety
of every complete execution.

We construct relations \(S_i\) between source and target machine states;
\(c\,S_i\,q\) means that they correspond at index \(i\).
Here the natural-number index \(i\) bounds target bookkeeping that may
be matched while leaving the source unchanged.  From a related pair with
a locally safe source, each target step must either match one or more
source steps with the same trace, or be silent and strictly decrease the
index.  The resulting source machine state and index are \(c'\) and
\(i'\).  Formally,
\[
\begin{gathered}
 c\,S_i\,q\land\operatorname{safe}(c)
             \land \lstepj{C}{q}{t}{q'}
 \quad\Longrightarrow\\
 \exists i',c'.\ c'\,S_{i'}\,q'\ \land\
 \bigl(\lplusj{E}{c}{t}{c'}\ \lor\
       (t=\epsilon\land c'=c\land i'<i)\bigr).
\end{gathered}
\]
Here the index likewise prevents infinite target stuttering without source
progress, following CompCert's well-founded simulation diagrams.
The theorem also matches initial machine states: \(C_0\) is related to
the initial source state \(E_0\).  At every related pair with a locally safe
source, the target is final or can step.  Under the same relation and
local-safety hypotheses, if
\(q=\mathsf{ret}_C(v_C,M_C,\mathsf{stop}_C)\), there exist
\(\hat v,\mu\) such that the source can silently
reach \(\mathsf{ret}(\hat v,\mu,\mathsf{stop})\).  Its result and heap satisfy
\[
 \Phi_{j_0}(\hat v,\mu,v_C,M_C).
\]
These are the progress and final
matching obligations of the backward simulation.  Global source safety
is not an assumption of this theorem.

To prove the backward diagram, we account for every intermediate Clight
step implementing a source operation.  During target bookkeeping, the
source stays at the corresponding machine state and the index decreases.
When Clight reaches the operation itself, the source advances to match
it and the proof establishes a new related pair.  Deterministic internal
computations reuse the forward proof.  At a kernel call, the backward
contract supplies a matching reply or establishes that the source
specification is undefined.
The construction also covers related final machine states and source
states that silently reach stuckness; the latter cannot satisfy
the local safety guard.  We prove the backward diagram at kernel calls
directly, without invoking CompCert's generic forward-to-backward
conversion.  Kernel choices may remain unrecorded in the trace, so the
observed trace need not determine the chosen outcome.

\paragraph{Executable oracles.}
The contracts belong to the oracle's response type.  Every successful
invocation therefore yields a realizable public source step, and every
valid call receives a response.  Erasing the oracle maps executable states
to the source states above while preserving control, heaps, observations,
and the trace preceding a failure.  This argument applies to every private
state type and every certified oracle.  A particular oracle may select only
a subset of the public outcomes; the backward simulation concerns the
relational semantics, which ranges over all realizable responses.

\subsubsection{Behavior Equivalence and Undefined Behavior}
\label{sec:embc:supp-results}

We now pass from matching steps to matching complete executions.  Throughout
this subsection, keep the fixed program and related initial call above.
Assume the initial source state is admissible, both kernel contracts hold,
and observations are total.
Theorems~\ref{thm:embc:supp-equivalence} and~\ref{thm:embc:supp-ub} are
parts~(1) and~(2) of Theorem~\ref{thm:embc-clight} of the main paper.
As there, the source semantics takes, at each kernel call, any reply that
the contract allows.

\paragraph{Complete behaviors.}
We use \(B\) for the behavior of a complete execution.  Termination
records a finite trace \(t\), returned result \(r\), and final state
\(h\).  For EmbC the result and state are an optional value and a typed heap;
for Clight they are a value and a byte memory.
Silent divergence records a finite trace \(t\) followed by infinitely
many silent steps.  Reactive divergence produces the infinite trace
\(T\): its finite execution prefixes emit prefixes of \(T\) with
unboundedly many events.  Going wrong records the finite trace \(t\)
before reaching a non-final machine state with no successor.  We denote these cases by
\(\mathsf{Term}\), \(\mathsf{Div}\), \(\mathsf{React}\), and
\(\mathsf{Wrong}\), respectively:
\[
 B ::= \mathsf{Term}(t,r,h)\mid\mathsf{Div}(t)
       \mid\mathsf{React}(T)\mid\mathsf{Wrong}(t).
\]
We write
\(\behavesj{E}{c}{B}\) when an EmbC execution starting at \(c\) has
behavior \(B\), and \(\behavesj{C}{q}{B}\) for Clight.

Behaviors match when they have the same traces and, on termination,
matching final outcomes.  Extend the notation \(\rep{j_0}\) to behaviors
using \(\Phi_{j_0}\) for that outcome comparison:
\[
\begin{aligned}
 \mathsf{Term}(t,\hat v,\mu)\rep{j_0}
     \mathsf{Term}(t',v_C,M_C)
     &\iff t=t'\land\Phi_{j_0}(\hat v,\mu,v_C,M_C),\\
 \mathsf{Div}(t)\rep{j_0}\mathsf{Div}(t')
     &\iff t=t',\\
 \mathsf{React}(T)\rep{j_0}\mathsf{React}(T')
     &\iff T=T'.
\end{aligned}
\]
All other pairs are unrelated.  In particular, the matching relation
compares defined behaviors; the separate result below describes failure.

Let \(\mathcal B_E\) and \(\mathcal B_C\) be the sets of behaviors of
the initial EmbC and Clight calls.  These sets are
\[
 \mathcal B_E=\{B\mid\behavesj{E}{E_0}{B}\},
 \qquad
 \mathcal B_C=\{B\mid\behavesj{C}{C_0}{B}\}.
\]
Both include every environment choice permitted by their semantics.

\begin{theorem}[Behavior equivalence on safe inputs]
\label{thm:embc:supp-equivalence}
Suppose the initial EmbC call never goes wrong, for any finite trace:
\[
 \forall t.\quad
 \neg\bigl(\behavesj{E}{E_0}{\mathsf{Wrong}(t)}\bigr).
\]
Then every source behavior has a matching target behavior, and every
target behavior has a matching source behavior:
\[
\begin{gathered}
 \forall B_E\in\mathcal B_E.\ \exists B_C\in\mathcal B_C.\ B_E\rep{j_0}B_C,\\
 \forall B_C\in\mathcal B_C.\ \exists B_E\in\mathcal B_E.\ B_E\rep{j_0}B_C.
\end{gathered}
\]
\end{theorem}
This is equivalence of complete behaviors modulo memory representation:
it matches termination, both forms of divergence, all observable events,
and returned values and heaps.  It also implies that Clight never goes
wrong on these inputs.  The forward simulation supplies the first
direction, and the backward simulation supplies the second.
Global source safety excludes failure throughout both arguments.

For the failure result, let \(t_E\) and \(t_C\) be the traces before
source and target failure, respectively.  Write \(t_E\sqsubseteq t_C\)
when \(t_E\) is a prefix of \(t_C\).  The source may fail earlier because
its semantics is conservative.

\begin{theorem}[Reflection of target undefined behavior]
\label{thm:embc:supp-ub}
Every Clight execution that goes wrong has an EmbC execution that goes
wrong after a prefix of the target trace:
\[
 \behavesj{C}{C_0}{\mathsf{Wrong}(t_C)}
 \;\Longrightarrow\;
 \exists t_E.\
 \behavesj{E}{E_0}{\mathsf{Wrong}(t_E)}\land t_E\sqsubseteq t_C.
\]
This implication requires no global source-safety assumption.
\end{theorem}
The proof follows the backward simulation along the failing target
execution.  If the local source-safety condition fails, the source can
silently reach failure.  If it continues to hold, target progress and
final matching exclude a non-final target machine state with no successor.
Consequently, a proof that the EmbC checker never gets stuck, whatever
replies the kernel contracts allow, rules out undefined behavior in its
Clight embedding.
The cross-block pointer comparison illustrates
why source failure can also occur when the target remains defined.

\section{The Refinement Proof}
\label{app:refine}

This appendix supports \S\ref{sec:refine} and \S\ref{sec:c-proof} of the paper. It shows
how \Keel{} stores a certificate, gives the ideas of the proofs of
Lemmas~\ref{lem:check-rule} and~\ref{lem:compaction}, states the lemmas for
loading, comparing and freeing that the proof of Theorem~\ref{thm:refine:formal} uses,
and proves Lemma~\ref{lem:simulation} and Theorem~\ref{thm:refine:formal}.
Every lemma is proved in Lean against the EmbC semantics. The notation is that of the paper:
$\poolat{i}$ is the term that \cid{pool} holds at index $i$ in the memory $\mu$ before a call,
$\mu \crep \cfg{\pool}{\concl}$ is the state correspondence of
Definition~\ref{def:correspondence} with its clauses (D1)--(D5), $I$ is the footprint invariant
of Definition~\ref{def:footprint}, and a result of \cid{NOID} or 0 is a rejection.

\subsection{How \Keel{} Stores a Certificate}
\label{app:refine:storage}

\begin{figure}[t]
\small
\centering
\begin{tabular}{@{}r l l c l l@{}}
\toprule
& \multicolumn{4}{c}{Header unit} & Operand units \\
\cmidrule(lr){2-5}
Position & Operator & Type & Operands & Parameters & position: contents \\
\midrule
0 & variable & $\mathsf{BV}$ & 0 & width 32 & \\
1 & variable & $\mathsf{BV}$ & 0 & width 32 & \\
2 & constant & $\mathsf{BV}$ & 1 & width 32 & 3: 800 (the value) \\
4 & $-$ & $\mathsf{BV}$ & 2 & width 32 & 5: 2,\; 6: 0 \\
7 & $+$ & $\mathsf{BV}$ & 2 & width 32 & 8: 1,\; 9: 4 \\
10 & $\le$ & $\mathsf{Bool}$ & 2 & -- & 11: 7,\; 12: 2 \\
\bottomrule
\end{tabular}
\caption{The layout of a certificate's pool, on the terms of the running example: the variables
$\sym$ and $o$ at 0 and 1, the constant 800 at 2, $800 - \sym$ at 4, $t = o + (800 - \sym)$ at 7 and
$t \le 800$ at 10. Each term occupies one header unit, at its position, with the four fields
shown, followed by one unit per operand that holds the position of the operand's header; a
constant holds its value there instead. The positions skip these units, so after the header at
2 and its value at 3 the next term starts at 4. A term is stored once, however many terms use
it: 800 at 2 is an operand of the terms at 4 and 10.}
\label{fig:pool}
\end{figure}

\paragraph{The certificate format.}
A BCF certificate (\S\ref{sec:background}) starts with a header that gives the sizes of its
expression pool and its list of steps. In the pool, one unit per operand holds the operand's
\emph{position}, i.e., the index of the unit where the operand's header is stored. A variable has no operands and is identified by its
position, and a bit-vector constant stores its value where the operands would be, in one unit or
two. Figure~\ref{fig:pool} shows this layout on a small pool, the one that holds the term
$t \le 800$ of the running example. A step is stored the same way: a header that names the rule
and gives the number of premises and of parameters, followed by its premises, which are the
numbers of earlier steps, and its parameters, which are positions of terms in the pool or, for
some rules, plain numbers.

\paragraph{The pool.}
\Keel{}'s array \cid{pool} uses the same layout, with two differences. An operand unit holds the
index in \cid{pool} of the operand's header, not its position in the certificate. A variable has
one operand unit, which holds its position in the certificate, so that two variables of the same
type remain two terms. The two numberings differ because \cid{pool} starts with $\top$ and $\bot$,
because of these extra units, and because a term that the certificate contains twice is held once.

\paragraph{The copy of the certificate.}
As soon as the loader has added the term at position $p$ to \cid{pool}, at index $r$, it overwrites
that term's header in \cid{cert} with $r + 1$ and the term's operand units with 0. A parameter $p$ resolves to $\cid{cert}[p] - 1$. A parameter that is out of range, or that points
into the middle of a term, reads as 0 and the step is rejected.

\begin{figure}[t]
\begin{lstlisting}[style=paperc]
struct ck {
  u32 *pool;  u32 plen;    // the terms; pool[plen] is the first free u32
  u32 *htab;  u32 hcap;    // the hash table: 0 = empty, i + 1 = the term at index i
  u32 *facts; u32 flen;    // the conclusions: facts[j] = index of the conclusion of step j
  u32 *lr;                 // last_cit: for each step, the last step that cites it
  u32 *prem;               // the indices of the premises of the current step
  u32 *stk;   u32 sp;      // a scratch stack
  u32 tt, ff;              // the indices of true and false in pool
  ...
};
// if t <= 800 is at index 10, t at index 7 and 800 at index 2:
//   pool[10] = BOOL_ULE | 2 << 8;   pool[11] = 7;   pool[12] = 2;
// a step j that concludes t <= 800:
//   facts[j] = 10;
\end{lstlisting}
\caption{The arrays that hold a configuration, abridged.}
\label{fig:refine:pool}
\end{figure}

\begin{figure}[t]
\begin{lstlisting}[style=paperc]
u32 check_C(...) {                             // Keel's entry point
  u32 *cert = kvcalloc(...);                   // allocate the copy of the certificate
  if (cert == NULL || copy_from_bpfptr(cert, ubuf, ...)) ...  // free it and reject
  struct ck *st = kzalloc(sizeof(struct ck));  // allocate the state
  u32 ok = st != NULL && run(st, cert, ...);   // the main loop, below
  free_state(st);                              // free the arrays, then the struct
  kvfree(cert);
  return ok ? ACCEPT : REJECT;
}

u32 run(struct ck *st, ...) {                  // returns 1 to accept, 0 to reject
  if (!init(st, nwords))          return 0;    // hash table, premises, budgets, true and false
  if (!load(st, cert, nwords))    return 0;    // add the certificate's terms to the pool
  if (!last_citations(st))        return 0;    // one ASSUME step; conclusions; last citations
  if (!compare(st, phi))          return 0;    // the assumed formula is phi
  while (off < nsteps) {                       // one iteration checks one proof step
    cond_resched();
    if (signal_pending(current)) return 0;     // the user interrupted the load
    c = check_step(st, off);                   // the rule's check: index of the conclusion
    if (c == NOID) return 0;                   // the check failed
    st->facts[st->flen++] = c;                 // record the conclusion
    off = next_step(st, off);
    if (off < nsteps && too_big(st) && !drop(st)) return 0;  // drop the terms that are not live
  }
  return st->flen > 0 && st->facts[st->flen - 1] == st->ff;  // the last conclusion is false
}
\end{lstlisting}
\caption{\Keel{}'s entry point and main loop, as a sketch with descriptive names.
\cid{st->facts} is the conclusions array and \cid{st->ff} the index of $\bot$.}
\label{fig:refine:code}
\end{figure}

Figure~\ref{fig:refine:pool} shows these arrays, and Figure~\ref{fig:refine:code} shows how
\Keel{} uses them.

\subsection{Proof of \texorpdfstring{Lemma~\ref{lem:check-rule}}{the Lemma}: Checking a Rule}
\label{app:refine:rule}

The lemma has two parts: adding a term refines the function that adds it to the pool, and each
rule's check refines $\mathit{check}_R$. Both are in the sense of
Definition~\ref{def:refinement} of the paper. Adding a term is defined when the term is well
typed, as Definition~\ref{def:accept} requires.

\paragraph{Adding a term: idea of the proof.}
The caller makes room in \cid{pool}, which may grow it through \cid{kvrealloc}. It then writes the
term $u = \mathrm{op}(\vec u)$ at \cid{pool[plen]}: the header, followed by the indices
$i_1, \ldots, i_m$ of its arguments, with $\poolat{i_\ell} = u_\ell$. Growing the array keeps every
index, and the units past \cid{plen} are not part of any term, so all clauses still hold.

The function first checks that $u$ is well typed, and rejects otherwise. If the table is half
full, the function allocates one twice as large, inserts every term into it, and frees the old one,
which keeps (D3). So the table has an empty slot.

The search starts at the slot of the hash of the units at \cid{pool[plen]} and compares them with
the units of the term that each occupied slot records. Its loop invariant is that every slot it has
passed holds another term. Two terms with the same units are the same term: the header gives the
same operator and parameters, and the same indices give the same arguments. In the other direction,
if $u$ is held at some index $r$, then the units at $r$ are exactly the units written at
\cid{pool[plen]}, because (D1) gives each argument only one index. A variable's position and a
constant's value are compared like the other units.

If the search finds $u$ at $r$, the result is $r$ and \cid{plen} does not move, so no term changes.
By (D1), $u \in \pool$ already.

If the search reaches an empty slot, then (D3) says that $u$ is not held, since a term that is held
is reachable from the slot of its hash through occupied slots. The function records
$\cid{plen} + 1$ in that slot and advances \cid{plen} past the term. Its index $r$ is then the old
\cid{plen}. This gives (D1), since $u$ is new and all earlier terms are unchanged; (D2) and (D4),
since no earlier index changes its term; and (D3), since the slot was the first empty one on the
search path. Each operand index is below \cid{plen}, so the new term keeps the topological invariant
of the pool. Every step of this may reject because a budget or an allocation fails, and the
definition allows that.

\paragraph{The rule checks: idea of the proof.}
A check reads the headers and operand indices of its premises and parameters, and either rejects or
returns an index. It compares terms by comparing indices, which (D1) justifies. All these reads are
in bounds because each index holds a term by (D1), and the header of that term gives the number of its
operands. A step has at most 255 premises, and \cid{prem} has 256 entries. The \NumRules{} checks
fall into four groups.
\begin{itemize}
\item \emph{Checks that return a term already held,} such as \textsc{eq\_resolve} (below) and
  \textsc{and\_elim}. They write no memory, so $\mu' = \mu$. The parameter of \textsc{and\_elim}
  is a plain number, which the check compares with the number of operands of the premise.
\item \emph{Checks that build their conclusion,} such as \textsc{cong} and \textsc{and\_intro}. They
  push indices on the scratch stack \cid{stk} and add terms. Each addition is the first part of this
  lemma, and the last conjunct of Definition~\ref{def:refinement} keeps the earlier indices
  valid across the later additions. This is why these proofs compose.
\item \emph{Checks of a given equation:} \textsc{aci\_norm}, \textsc{poly\_norm} and
  \textsc{bitblast}. The parameter is the equation $a = b$ to check. \textsc{aci\_norm} and
  \textsc{poly\_norm} compute the normal form of each side on \cid{stk} and compare the two;
  \textsc{bitblast} compares the two sides directly. The result is the parameter itself, and no term
  is built. These checks write only \cid{stk} and budgets, which no clause mentions.
\item \emph{\textsc{rewrite}.} The first parameter is an entry number. The table \cid{rw\_tbl} is the
  encoding of the Lean table, and it is read-only. For an entry other than 0, the check resolves the
  remaining parameters, checks them against the parameter types of the entry, and instantiates the
  entry's condition. It compares conjunct $\ell$ of the condition with premise $\ell$, and evaluates
  every conjunct past the premises to a constant that must be true. It then instantiates both
  sides and adds their equation. There is no matching: the arguments come from the step, as in
  \S\ref{sec:checker}. Instantiation is proved once, by induction on the pattern, so this proof
  covers all \NumRewritesKeel{} entries. Entry 0 instead checks the shape of a given equation and returns it.
\end{itemize}

\paragraph{A check in full: \textsc{eq\_resolve}.}
Before it calls the check, the main loop copies the indices of the conclusions of the premises from
\cid{facts} into \cid{prem}, and the parameters are read in place from \Keel{}'s copy of the
certificate. Step 32 of the running example uses \textsc{eq\_resolve}, with the premises 31 and 29,
whose conclusions are $t > 800$ and $(t > 800) = \bot$, so it concludes $\bot$. The check of
\textsc{eq\_resolve} is the following function.

\noindent\begin{minipage}{\linewidth}
\begin{lstlisting}[style=paperc]
u32 r_eq_resolve(struct ck *st, u32 np, u32 na) {  // np premises, na parameters
  if (np != 2 || na != 0) return NOID;   // two premises and no parameters
  u32 q = st->prem[1];                   // the second premise: (t > 800) = false
  if ((st->pool[q] & 255) != BOOL_EQ)    // its operator must be =
    return NOID;
  if (st->pool[q + 1] != st->prem[0])    // its left side must be the first premise: t > 800
    return NOID;
  return st->pool[q + 2];                // the conclusion is its right side: false
}
\end{lstlisting}
\end{minipage}

\noindent The check of \textsc{eq\_resolve} takes two conclusions $\psi_1$ and $\psi_2$. It requires
$\psi_2$ to be an equation $a_1 = a_2$, where $a_1$ and $a_2$ are the two arguments of the equation,
and it requires $a_1$ to be $\psi_1$. It concludes $a_2$:
\[
  \mathit{check}_{\textsc{eq\_resolve}}(\pool;\, \psi_1, \psi_2;\,)
    = (\pool, a_2) \qquad \text{if } \psi_2 = (a_1 = a_2) \text{ and } a_1 = \psi_1 .
\]

\begin{proof}[Proof for \textsc{eq\_resolve}]
We instantiate Definition~\ref{def:refinement}, taking for $F$ the function \cid{r\_eq\_resolve}
and for $f$ the check $\mathit{check}_{\textsc{eq\_resolve}}$. \cid{r\_eq\_resolve} receives the
indices of $\psi_1$ and $\psi_2$ in \cid{prem[0]} and \cid{prem[1]}. We must show
\[
\begin{aligned}
  &\{\mu \crep \cfg{\pool}{\concl} \,\wedge\,
     \poolat{\cid{prem[0]}} = \psi_1 \,\wedge\, \poolat{\cid{prem[1]}} = \psi_2\}\;\;
     \cid{r\_eq\_resolve} \\
  &\{\hat v = \cid{NOID} \;\vee\;
     (\mathit{check}_{\textsc{eq\_resolve}}(\pool;\, \psi_1, \psi_2;\,) = (\pool, a_2) \,\wedge\,
      \poolat[\mu']{\hat v} = a_2 \\
  &\qquad\qquad\qquad\;\, \wedge\, \mu' \crep \cfg{\pool}{\concl} \,\wedge\,
      \forall i \in \mathrm{dom}(\cid{pool}_\mu).\ \poolat[\mu']{i} = \poolat{i})\} .
\end{aligned}
\]
The function writes no memory, so $\mu' = \mu$, and the correspondence and the
terms at all indices are unchanged. What remains is
\[
  \hat v = \cid{NOID} \;\vee\; \bigl(\psi_2 = (a_1 = a_2) \,\wedge\, a_1 = \psi_1 \,\wedge\,
  \poolat{\hat v} = a_2\bigr) .
\]
When \cid{r\_eq\_resolve} returns \cid{NOID}, this is satisfied. The function returns something
else only in its last line, which returns \cid{pool[q+2]} with $\cid{q} = \cid{prem[1]}$, and only
when the conditions of its two \cid{if} statements are false, i.e.\ when
$\cid{pool[q] \& 255} = \cid{BOOL\_EQ}$ and $\cid{pool[q+1]} = \cid{prem[0]}$. By the definition of
$\poolat{\cdot}$, the operator of $\psi_2 = \poolat{\cid{q}}$ is in bits 0--7 of
\cid{pool[q]}, which \cid{pool[q] \& 255} keeps, and its arguments are
$\poolat{\cid{pool[q+1]}}$ and $\poolat{\cid{pool[q+2]}}$. \cid{BOOL\_EQ} is the code of the
operator $=$.
{\small
\begin{align*}
  &\cid{pool[q] \& 255} = \cid{BOOL\_EQ}
    &&\Longrightarrow\; \text{the operator of } \psi_2 \text{ is } =
    && \text{definition of } \poolat{\cid{q}} \\
  &&&\Longrightarrow\; \psi_2 = (a_1 = a_2)
    && {=} \text{ has two arguments} \\
  &&&\phantom{\Longrightarrow\;} a_1 = \poolat{\cid{pool[q+1]}},\ a_2 = \poolat{\cid{pool[q+2]}}
    && \text{definition of } \poolat{\cid{q}} \\
  &\cid{pool[q+1]} = \cid{prem[0]}
    &&\Longrightarrow\; a_1 = \poolat{\cid{prem[0]}} = \psi_1
    && \text{the precondition}
\end{align*}}
So the last line returns $\cid{pool[q+2]}$, an index of $a_2$, as required.
\end{proof}

\subsection{Proof of \texorpdfstring{Lemma~\ref{lem:compaction}}{the Lemma}: Compaction}
\label{app:refine:compact}

The main loop calls compaction, \cid{drop}, only between steps and only when the pool exceeds a
margin that grows with its pinned part, its live part and the number of steps so far. The Lean proof is about
3,900 lines. Here we give its idea. Write $k$ for the number of steps checked, and call a term
\emph{moved} if its index is at least \cid{pin}. The function allocates a forwarding table
\cid{fw} with one entry per unit from \cid{pin} to \cid{plen}, all 0, and then runs seven loops.

\begin{center}
\small
\begin{tabular}{@{}l p{0.82\linewidth}@{}}
\toprule
Loop & What it does \\
\midrule
Roots & For every step $j < k$ with $\cid{last\_cit}[j] \ge k$, if $\cid{facts}[j]$ is the index of a moved term, mark that term and push it on \cid{stk}. \\
Marking & Pop a term. Unless it is a variable or a constant, mark each moved operand not yet marked and push it. Ends when \cid{stk} is empty. \\
Moving & Go through the moved terms from the lowest index, with a write cursor $w$ that starts at \cid{pin}. Copy each marked term $i$ to $w$, replace each moved operand $i_\ell$ with $\cid{fw}[i_\ell]$ unless the term is a variable or a constant, and set $\cid{fw}[i] \gets w$. Finally $\cid{plen} \gets w$. \\
Facts & Replace each $\cid{facts}[j]$ that is the index of a moved term with its new index, or with \cid{NOID} if the term was dropped. \\
Counting & Count the terms below \cid{pin}. \\
Sizing & Double a size from 1024 until it is at least four times the number of terms. \\
Rehashing & Free the hash table, allocate one of that size, and insert every term of the compacted pool, including those below \cid{pin}. \\
\bottomrule
\end{tabular}
\end{center}

Then \cid{fw} is freed. Any of these loops may reject when a budget or an allocation fails: \cid{push} may
have to grow \cid{stk}, for example, and the new hash table may not be available. A rejection frees
\cid{fw} and ends the run, so the lemma claims nothing about the memory in that case.

\emph{Idea of the proof.} Define $\sigma(i) = \cid{fw}[i]$ for a marked term $i$, and $\sigma(i) = i$
below \cid{pin}. Three facts follow from the loops. The marked terms include every moved term that is
a still-cited conclusion, and every moved operand of a marked term. Since $i_\ell < i$ for each
operand, the moving loop has already set $\cid{fw}[i_\ell]$ when it reaches $i$, and
$\sigma(i_\ell) < \sigma(i)$, so the topological invariant holds in the new pool. And $w \le i$ throughout,
so a copy never overwrites a term that the loop has not yet reached.

By strong induction on $i$, every marked term keeps its meaning:
$\poolat[\mu']{\sigma(i)} = \poolat[\mu]{i}$. The operands of $i$ are smaller, so the induction
hypothesis gives $\poolat[\mu']{\sigma(i_\ell)} = \poolat[\mu]{i_\ell}$, and the header is copied
unchanged. Variables and constants have their units copied unchanged. The function $\sigma$ is one-to-one,
and every term of the new pool is the image of a marked term or lies below \cid{pin}. So
$\mu' \crep \cfg{\pool}{\concl}$ for the same configuration:
\[
\begin{array}{@{}l@{\quad}p{0.86\linewidth}@{}}
  \text{(D1)} & Every term held is an old term of $\pool$, by the induction, and is held once, since $\sigma$ is one-to-one. \\
  \text{(D2)} & The terms below \cid{pin} do not move, and neither do \cid{tt} and \cid{ff}. The still-cited conclusions and their subterms are marked, so they are held. \\
  \text{(D3)} & The hash table is rebuilt by inserting every term held, using the hashes of the new units, so each term is reachable from the slot of its hash. \\
  \text{(D4)} & Take a step $j$ with $\cid{last\_cit}[j] \ge k$, and let $f$ be the old value of $\cid{facts}[j]$. The facts loop writes $\sigma(f)$ there, and the induction gives
     $\poolat[\mu']{\sigma(f)} = \poolat[\mu]{f} = \psi_j$. A dropped conclusion is one that no later step cites. \\
  \text{(D5)} & Compaction writes nothing into \cid{cert} or \cid{last\_cit}, and the index $r < \cid{pin}$ of a certificate term does not move.
\end{array}
\]
For the footprint invariant, \cid{fw} is a new block that is freed on every path out, the hash table
freed was \Keel{}'s and the one allocated is new, and \cid{stk} grows the way \cid{pool} does. While
compaction runs, \cid{fw} is one live block that is not part of the checker's state, so the Lean
form of the invariant for the functions it calls sets that block aside.

\subsection{Loading, Comparing and Freeing}
\label{app:refine:other}

The proof of Theorem~\ref{thm:refine:formal} uses three more specifications. As everywhere,
they are stated under the footprint invariant.

\paragraph{Loading.}
\[
  \{\mathit{true}\}\;\; \text{init, load, last citations}\;\;
  \{\hat v = 0 \,\vee\, \mu' \crep \cfg{\pool_0}{\emptyset}\}
\]
where, in the second case, the copied bytes encode a certificate $\cert$ with pool $\pool_\cert$
whose terms are well typed and with exactly one \textsc{assume} step, and
$\pool_0 = \pool_\cert \cup \{\top, \bot\}$.
These calls allocate the hash table and \cid{prem}, add $\top$ and $\bot$, grow \cid{pool} while they add
the certificate's terms, and allocate \cid{facts} and \cid{last\_cit}. The copy of the certificate
exists before they run. The pass that records last citations does not check that a premise is an earlier
step. The main loop does, when it checks the step, and rejects otherwise.
\emph{Idea of the proof.} \Keel{} reads the certificate in one pass: it decodes each term, adds it
to the pool, and overwrites the term's header in its copy of the certificate with the term's index.
The decoder of the paper's \S\ref{sec:checker} instead decodes the whole file first and then adds its terms. We
wrote a second decoder in Lean that works in one pass like \Keel{}, proved that it gives the same
configuration as the decoder of \S\ref{sec:checker}, and proved that loading refines it.

\paragraph{Comparing with $\varphi$.}
\[
  \begin{aligned}
    &\{\mu \crep \cfg{\pool}{\concl}\}\;\; \text{compare} \\
    &\{\hat v = 0 \,\vee\, (\mu' \crep \cfg{\pool}{\concl} \,\wedge\,
      \text{the formula of the \textsc{assume} step of } \cert \text{ is } \varphi \text{ up to renaming})\}
  \end{aligned}
\]
for every configuration $\cfg{\pool}{\concl}$, where the verifier's array holds $\varphi$ at the index it passes.
\emph{Idea of the proof.} The function goes through the verifier's array once, in order, and
matches each term of $\varphi$ with a term of the certificate: a variable with a variable of the
same type, any other term with a term that has the same operator and matched arguments. The loop
invariant is that every pair matched so far is checked. A certificate variable can be matched only
once and each term of $\varphi$ with only one index, so a successful sweep gives a one-to-one
map between the variables of the two formulas. A separate Lean lemma shows that matched terms have
the same value under corresponding assignments. The function keeps its working tables in \cid{pool} past \cid{plen}, which
may grow the array, and it writes no term, so the correspondence is kept.

\paragraph{Freeing.}
\[
\begin{aligned}
  &\{I\}\;\; \text{free the state, the struct and the copy of the certificate} \\
  &\{\forall b < |\mu_{\mathrm{in}}|.\ \mu'(b) = \mu_{\mathrm{in}}(b) \,\wedge\,
    \forall b \ge |\mu_{\mathrm{in}}|.\ \mu'(b) = \mathsf{dead}\}
\end{aligned}
\]
\emph{Idea of the proof.} \Keel{} frees each array of \cid{struct ck}, then the struct, then the copy
of the certificate. These are the only live blocks at or above $|\mu_{\mathrm{in}}|$, so afterwards all of
those blocks are dead. By the invariant, every block below $|\mu_{\mathrm{in}}|$ is as it was.

\subsection{Proof of \texorpdfstring{Lemma~\ref{lem:simulation}}{the Lemma}: Simulation}
\label{app:refine:simulation}

The main loop of Figure~\ref{fig:refine:code} checks one proof step per iteration, and
the lemma for one iteration is the simulation step of Figure~\ref{fig:sim-overview}. The body
of the loop does four things, in this order:
\begin{enumerate}
\item it rejects if a signal is pending, i.e.\ if the user interrupted the load;
\item it reads the step from \Keel{}'s copy of the certificate (D5): it copies the indices of the
  conclusions of the premises from \cid{facts} into \cid{prem}, rejecting a premise that is not an
  earlier step or is past its last citation, and calls $\mathit{check}_{R,\mathrm{C}}$ for the rule
  $R$ of the step;
\item it writes the result $\hat v$ of the check into \cid{facts[j]};
\item if \cid{pool} has grown too large, it drops the terms that are not live.
\end{enumerate}
Each part has a specification of its own, and the simulation lemma combines them.

\begin{proof}
Each part meets its specification, so it does not get stuck, and the footprint invariant holds
after it, also when it rejects. A part that rejects ends the iteration. Otherwise, we give the
condition that holds after each part, writing $\mu$ for the current memory. Let
$\psi_{j_1}, \ldots, \psi_{j_m}$ be the conclusions of the premises of the step.
{\small
\begin{align*}
  &\mu \crep \cfg{\pool}{\concl}
    && \text{before the iteration} \\
  \text{1.}\quad &\text{the same}
    && \text{the signal check writes no memory} \\
  \text{2.}\quad &\ldots \wedge\, \forall \ell.\ \poolat{\cid{prem[}\ell\cid{]}} = \psi_{j_\ell}
    && \text{(D4) and (D5) for the premises; (D5) for the parameters} \\
  &\mu \crep \cfg{\pool'}{\concl}
     \,\wedge\, \poolat{\hat v} = \psi_j
    && \text{Lemma~\ref{lem:check-rule}} \\
  &\qquad \text{where } \cfg{\pool}{\concl} \stepto \cfg{\pool'}{\concl \cup \{\psi_j\}}
    && \text{Definition~\ref{def:step}} \\
  \text{3.}\quad &\mu \crep \cfg{\pool'}{\concl \cup \{\psi_j\}}
    && \text{(D4) for the new step} \\
  \text{4.}\quad &\mu' \crep \cfg{\pool'}{\concl \cup \{\psi_j\}}
    && \text{Lemma~\ref{lem:compaction}, or } \mu' = \mu \text{ if nothing is dropped}
\end{align*}}
Part 3 writes $\hat v$ into \cid{facts[j]}, and part 4 never runs after the last step, so then
$\poolat[\mu']{\cid{facts[n-1]}} \allowbreak = \psi_{n-1}$.
\end{proof}

\subsection{Proof of \texorpdfstring{Theorem~\ref{thm:refine:formal}}{the Theorem}: Refinement and Safety}
\label{app:refine:theorem}

The theorem makes three claims about a run of $\checkC$: it never gets stuck, and, if it returns,
parts (1) and (2) of the theorem hold, which we call safety and refinement. Here $\cert$ is the
certificate that the copied bytes decode to, and \emph{$\cert$ assumes $\varphi$} means that the
formula of its \textsc{assume} step is $\varphi$ up to renaming.
Before the main loop, \Keel{} loads the certificate and compares the formula of its
\textsc{assume} step with $\varphi$. We write $\mu_0$ for the memory before the loop and
$\mu_{j+1}$ for the memory after the iteration for step $j$. The specifications of loading,
comparing and freeing are those of \S\ref{app:refine:other}.

\begin{proof}

\emph{No stuck state.} The run of $\checkC$ is the sequence of calls of
Figure~\ref{fig:refine:code}. When a call in \cid{run} rejects, \cid{run} returns to $\checkC$,
which frees the state. A function that meets its specification does not get stuck when it is
called in a state where its precondition holds. Therefore, it is enough that we show each call
starts in such a state:
{\small
\begin{align*}
  &\text{loading:} && \mathit{true}
    && \text{nothing to show} \\
  &\text{comparing:} && \mu \crep \cfg{\pool_0}{\emptyset}
    && \text{the lemma for loading} \\
  &\text{iteration for step } 0: && \mu_0 \crep \cfg{\pool_0}{\emptyset}
    && \text{the lemma for comparing} \\
  &\text{iteration for step } j+1: && \mu_{j+1} \crep \cfg{\pool_{j+1}}{\concl_{j+1}}
    && \text{Lemma~\ref{lem:simulation} for step } j \\
  &\text{freeing:} && \text{the footprint invariant}
    && \text{Lemma~\ref{lem:footprint} for the calls before it}
\end{align*}}
The footprint invariant, the other half of every precondition, holds at each call by
Lemma~\ref{lem:footprint} applied to the calls before it. It holds at the start, since no
block exists at $b \ge |\mu_{\mathrm{in}}|$.

\emph{Safety.} Take a run that returns, accepting or rejecting. By
Lemma~\ref{lem:footprint} for the calls of \cid{run}, the footprint invariant holds when
\cid{run} returns. $\checkC$ then frees every block \Keel{} owns, the struct and the copy of the
certificate, and the lemma for freeing gives both conjuncts of safety. A run that rejects before
\cid{run} has allocated at most the copy of the certificate, and frees it.

\emph{Refinement.} Take a run that returns $\accept$. Every call in
Figure~\ref{fig:refine:code} returned without rejecting, and the loop ran one iteration for each
of the $n$ steps of $\cert$. We show, by induction along the simulation, that for every
$j \le n$ the proof certificate checking process reaches a configuration $\cfg{\pool_j}{\concl_j}$ that the memory matches:
\[
  \cfg{\pool_0}{\emptyset} \replayto \cfg{\pool_j}{\concl_j}
  \;\wedge\; \mu_j \crep \cfg{\pool_j}{\concl_j} .
\]
{\small
\begin{align*}
\intertext{\normalsize Base case:}
  &\mu_0 \crep \cfg{\pool_0}{\emptyset} \;\wedge\; \cert \text{ assumes } \varphi
    && \text{the lemmas for loading and comparing}
\intertext{\normalsize Inductive case:}
  &\mu_j \crep \cfg{\pool_j}{\concl_j}
    \;\Longrightarrow\; \cfg{\pool_j}{\concl_j} \stepto \cfg{\pool_{j+1}}{\concl_{j+1}}
    && \text{Lemma~\ref{lem:simulation}} \\
  &\phantom{\mu_j \crep \cfg{\pool_j}{\concl_j} \;\Longrightarrow\;}
    \wedge\; \mu_{j+1} \crep \cfg{\pool_{j+1}}{\concl_{j+1}}
    &&
\intertext{\normalsize Acceptance, with $\psi_{n-1}$ the conclusion of the last step:}
  &\poolat[\mu_n]{\cid{facts[n-1]}} = \psi_{n-1}
    && \text{Lemma~\ref{lem:simulation}, last step} \\
  &\cid{facts[n-1]} = \cid{ff}
    && \text{the loop returned } \accept \\
  &\psi_{n-1} = \poolat[\mu_n]{\cid{ff}} = \bot
    && \text{(D2)} \\
  &\checkL(\cert) = \accept
    && \text{Definition~\ref{def:accept}}
\end{align*}}
For the last line, the lemma for loading gives the single \textsc{assume} step and the well-typed
terms, and each iteration checked that the premises come from earlier steps. Together with the
base case, which gives that $\cert$ assumes $\varphi$, this is refinement.
\end{proof}

\subsection{The Fixed Limits of the C Checker}
\label{app:limits}

Besides a failed allocation, a pending signal and an exhausted budget, \Keel{} rejects a
certificate that exceeds one of the following limits of its C code. The Lean checker has none
of them.
\begin{itemize}
\item a constant wider than 64 bits, which \Keel{}'s evaluator does not compute;
\item a walk over a term deeper than \MaxDepth{} levels;
\item a normal form of more than \MaxSortTerms{} terms to sort;
\item a pool of more than \MaxPoolGiB{}~GiB.
\end{itemize}
\Keel{} also rejects a step that cites a conclusion after the last citation that its prepass
recorded for it. On the corpus, none of these limits rejects a certificate that the Lean
checker accepts.

\section{The Cost Bound}
\label{app:cost}

This appendix gives the constants of Theorem~\ref{thm:cost} of the paper and the proofs that the paper defers.

\subsection{The Constants}
\label{app:constants}

Table~\ref{tab:constants} lists every constant of the cost theorem of the paper (\S\ref{sec:cost}) and
where it comes from. The budget rates and bases are a policy: we chose them, and the theorem
holds for the values we chose. With them,
$W_T(B) = \WorkPerByte \cdot B + \WorkBase$ is the initial value of \cid{work},
and $W_M(B)$ exceeds the initial value of \cid{mem} by the $|\cert|$ bytes of the
copy of the certificate, hence the \MemPerByteProved{}, and the \MemFixed{} bytes of fixed
tables. The other constants follow from the code.

\begin{table}[H]
\caption{The constants of the cost theorem of the paper, in bytes.}
\label{tab:constants}
\small
\begin{tabular}{@{}llp{0.52\linewidth}@{}}
\toprule
Constant & Value & Where it comes from \\
\midrule
work per input byte & \WorkPerByte{} &
  policy: a power of two, about $3\times$ the corpus maximum (\WorkUsedMax) \\
work base & $\WorkBase$ &
  policy: \Keel{}'s former per-step limit; the largest test needs \WorkBaseNeedK{} thousand \\
memory per byte of $\varphi$ & \MemPerByte{} &
  policy: a power of two, about $3\times$ the corpus maximum (\MemUsedMax) \\
memory per certificate byte & \MemPerByteProved{} &
  \MemPerByte{}, plus 1 for the copy of the certificate \\
memory base and fixed tables & $\MemBase + \MemFixed$ &
  policy: $\MemBase$ matches the work base, and no corpus certificate uses it; plus the
  fixed tables: hash table (\HashTableBytes), premise array (\PremArrayBytes), state (\StructBytes) \\
rate, $R$ & \CostR{} & the ticks one proof step may take without touching the counter (below) \\
\bottomrule
\end{tabular}
\end{table}

\subsection{Where $R$ Comes From}
\label{app:cost:rate}
A decrement by $k$ is not $k$ ticks, as \S\ref{sec:cost} says. We say that a
decrement \emph{covers} the code that follows it. For each decrement, the ticks of the covered
code are at most a fixed multiple of $k$, and this multiple depends only on the program text.
We define $R$ to be the largest of these multiples:
\[
  R \;=\; \max_d \frac{t_d}{k_d} .
\]
Here $d$ ranges over the different kinds of decrements of \cid{work} in \Keel{}.
$k_d$ is the amount by which $d$ decrements \cid{work}.
$t_d$ is the most ticks that the code covered by $d$ takes. For example:
\begin{itemize}
\item For a loop of $k$ iterations whose body takes at most $j$ ticks, $k_d = k$ and
  $t_d = j \cdot k$. The multiple is $j$.
\item Adding a term with $n$ operands follows a decrement by $k_d = 1 + n$, the number of
  units the term occupies in the pool, one for its header and one per operand
  (\S\ref{sec:background}). Building the term writes each of these units. Suppose writing a unit
  takes at most $h$ ticks; then $t_d = h \cdot (1 + n)$. The multiple is $h$.
\end{itemize}
Since $R$ is defined as the maximum of such multiples, every decrement by $k$ covers at most $R \cdot k$ ticks.
Over a whole call from $\sigma$ to $\sigma'$, the ticks of covered code therefore add up to
at most $R \cdot (\sigma.\cid{work} - \sigma'.\cid{work})$, as Definition~\ref{def:bounded-fn}
requires, and the proof of Lemma~\ref{lem:stmt-cost} checks that all code whose ticks depend on
the input is covered.

\paragraph{The value of $R$.}
The proof certificate checking loop decrements \cid{work} by one per proof step, so that decrement by one must
cover every tick a step may take without touching the counter. That is the largest constant of a rule, \CostRuleAllow{} ticks for
\textsc{bitblast}; plus \CostPremAllow{} per premise and \CostArgAllow{} per argument, up to
255 of each; plus \CostStepAllow{} for the step's own ticks and \CostLoopAllow{} for the loop.
The driver's own constants, for loading the certificate and matching $\varphi$, add
\CostDriverAllow{}. Hence $R = \CostR$.

\subsection{Proof of \texorpdfstring{Lemma~\ref{lem:stmt-cost}}{the Statement Lemma}: Cost of a Statement}
\label{app:cost:stmt}

\begin{proof}[Proof of Lemma~\ref{lem:stmt-cost}]
By induction over the structure of $s$, whose forms are those of
Figure~\ref{fig:embc:syntax} of the main paper. The base cases are the statements that contain no other
statement. The inductive cases are the forms that contain statements: sequences,
conditionals and loops. Below, $w$ and $w'$ are shorthands for $\sigma.\cid{work}$ and $\sigma'.\cid{work}$,
$A$ with an index is the constant of a part, and $c$ is the number of ticks that the parts
of a statement outside the code a decrement covers take, such as the call of \cid{spend} and
the test of a loop; the text of the statement fixes it. The right column names the form and
what its line uses.
{\footnotesize
\begin{align*}
\intertext{\normalsize\emph{Base cases.} No statement increases the counter.}
  T + R\,w' &\le R\,w + 1
    && \text{statement of one tick: } T = 1,\ w' \le w \\
  T + R\,w' &\le R\,w + A^T_G + 2
    && \text{call of } G\text{: the premise, plus 2 ticks to call and return}
\intertext{\normalsize\emph{Inductive cases.} Each follows from the inequalities of its parts.
  In a sequence, the value of \cid{work} between the parts cancels.}
  T_1 + T_2 + R\,w' &\le R\,w + A_1 + A_2
    && \text{sequence } s_1; s_2\text{: the two inequalities, added} \\
  T + R\,w' &\le R\,w + \max(A_1, A_2)
    && \text{conditional: the inequality of the branch taken}
\intertext{\normalsize Code whose ticks depend on the input follows a decrement
  (\S\ref{sec:cost}) and is either a loop, of one of three forms in \Keel{}, or a kernel
  call. The four lines below cover these four cases. Where the right column gives the
  ratio of the ticks of the covered code to the decrement, $R$ is at least that ratio by its
  definition, and substituting $R$ for the ratio gives the line.}
  T + R\,w' &\le j\,k + c + R\,(w - k) \le R\,w + c
    && \text{loop after \cid{spend}}(k)\text{, body of} \le j \text{ ticks: ratio } j \\
  T + R\,w' &\le 2k + c + 1 + R\,(w - k) \le R\,w + c + 1
    && \text{kernel call after \cid{spend}}(k)\text{, } \le 1 + 2k \text{ ticks: ratio } 2 \\
  T + R\,w' &\le R\,w + 255\,A_{\mathrm{body}}
    && \text{loop over a term's operands, no decrement: body} \le 255 \text{ times} \\
  T + R\,w' &\le R\,w + A^T_G + c
    && \text{loop calling } G \text{ each iteration: \S\ref{app:cost:deferred}}
\end{align*}}

\emph{Memory.} The memory inequality has the same cases with $R = 1$. The one new case is an
allocation of $n$ bytes, which follows a decrement of \cid{mem} by $n$, so
$M + \mathit{mem}' = n + (\mathit{mem} - n) = \mathit{mem}$.

\emph{Termination.} Every loop ends. A loop after $\cid{spend}(k)$ runs $k$ times, a loop over
the operands of a term at most 255 times, and in a loop calling $G$ every iteration decrements
the counter, which cannot go below zero. Every call ends by the premise. So every execution of
$s$ ends.
\end{proof}

\subsection{The Deferred Case of \texorpdfstring{Lemma~\ref{lem:stmt-cost}}{the Statement Lemma}}
\label{app:cost:deferred}
The proof of Lemma~\ref{lem:stmt-cost} (cost of a statement) defers one case: a loop that calls a function
$G$ in each iteration. The bound of Definition~\ref{def:bounded-fn} for $G$ is not enough for it. Over $n$
iterations it leaves $n \cdot (A^T_G + c)$ ticks uncovered, with $c$ the ticks of one
iteration outside the call of $G$, and $n$ grows with the input. The case is proved with a
stronger induction hypothesis for the functions that loops call:
\[
  \text{every run of } G \text{ that decrements \cid{work} satisfies }
  T \;\le\; R \cdot (\sigma.\cid{work} - \sigma'.\cid{work} - 1) + A^T_G ,
\]
the inequality of Definition~\ref{def:bounded-fn} with $R$ to spare.

\emph{Why the hypothesis holds.} It does not follow from the definitions; it is how \Keel{}
is written. Each function that a loop calls begins by decrementing \cid{work}, and it
decrements one more than its own code needs, the extra one being for the iteration of the
loop that called it. The step function, for instance, spends one more than the number of
premises and parameters of the step it checks, and \cid{tick} and \cid{spend} decrement but
cover no code of their own, since the loop they stand before belongs to their caller. The
proof checks this for each such function. A run that rejects before its first decrement is
outside the hypothesis. For a run that decrements, the ticks of $G$ are those of the entry
decrement, a constant that $A^T_G$ absorbs, plus those of the rest of its body, which Lemma~\ref{lem:stmt-cost}
bounds by $R$ times the remaining decrement, $\sigma.\cid{work} - \sigma'.\cid{work} - 1$,
plus a constant. This is the inequality with $R$ to spare, and it is established in the
induction of Lemma~\ref{lem:fn-cost} (cost of a function) alongside the bound itself.

\emph{The derivation.} Let the loop run $n$ times, let $w_0$ be the counter before it and
$w_i$ the counter after iteration $i$, and let $c$ be as above; the code of the loop fixes
$c$. If iteration $i$ makes $T_i$ ticks in $G$, then
{\small
\begin{align*}
  T_i + R\,(w_i + 1) &\le R\,w_{i-1} + A^T_G
    && \text{the hypothesis for } G \\
  T_i + c + R\,w_i &\le R\,w_{i-1} + (A^T_G + c - R) \;\le\; R\,w_{i-1}
    && R \ge A^T_G + c\text{, by the definition of } R \\
  \textstyle\sum_{i \le n} (T_i + c) + R\,w_n &\le R\,w_0
    && \text{added over the iterations; middle values cancel}
\end{align*}}
And in the case that $G$ rejects, the loop stops, and that last iteration adds at most
$A^T_G + c$. So the loop satisfies $T + R\,w' \le R\,w + A^T_G + c$, the line of the proof of
Lemma~\ref{lem:stmt-cost}.

\subsection{Proof of \texorpdfstring{Lemma~\ref{lem:fn-cost}}{the Function Lemma}: Cost of a Function}
\label{app:cost:fn}

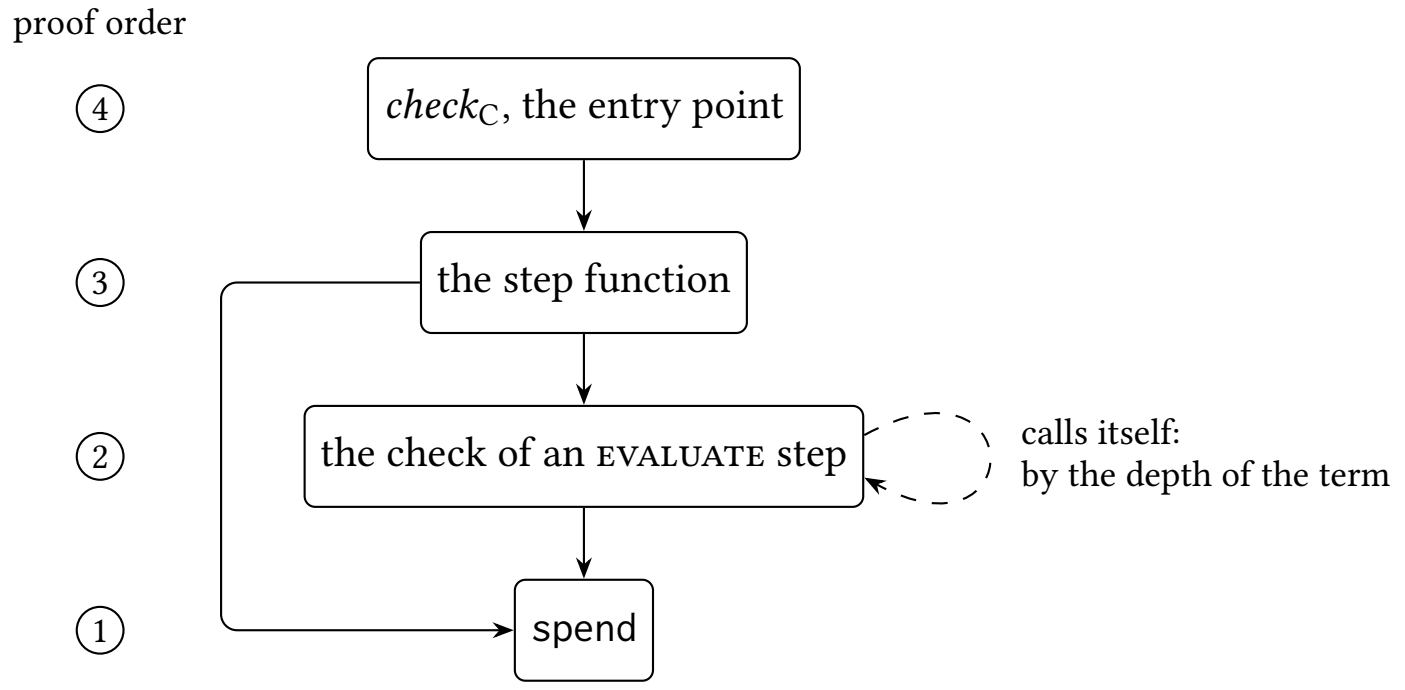
\begin{figure}[t]
\centering
\begin{tikzpicture}[
  font=\footnotesize,
  fn/.style={draw, rounded corners=2pt, minimum height=1.9em, inner sep=3pt, align=center, fill=white},
  call/.style={-{Stealth[length=4pt]}},
  ord/.style={circle, draw, inner sep=1.2pt, font=\scriptsize}]
  \node[fn] (top)   at (0,0)    {$\checkC$, the entry point};
  \node[fn] (step)  at (0,-1.15) {the step function};
  \node[fn] (eval)  at (0,-2.3) {the check of an \textsc{evaluate} step};
  \node[fn] (spend) at (0,-3.45) {\cid{spend}};
  \draw[call] (top) -- (step);
  \draw[call] (step) -- (eval);
  \draw[call] (eval) -- (spend);
  \draw[call, rounded corners=3pt] (step.west) -- (-2.4,-1.15) -- (-2.4,-3.45) -- (spend.west);
  \draw[call, dashed] ([yshift=4pt]eval.east) .. controls +(1.1,0.6) and +(1.1,-0.6)
    .. ([yshift=-4pt]eval.east);
  \node[font=\scriptsize, align=left, anchor=west] at ([xshift=26pt]eval.east)
    {calls itself:\\by the depth of the term};
  \node[ord] at (-3.2,-3.45) {1};
  \node[ord] at (-3.2,-2.3)  {2};
  \node[ord] at (-3.2,-1.15) {3};
  \node[ord] at (-3.2,0)     {4};
  \node[font=\scriptsize, align=center] at (-3.2,0.55) {proof order};
\end{tikzpicture}
\caption{Example of the order in which Lemma~\ref{lem:fn-cost} proves functions bounded, on 4 of
\Keel{}'s \PrintedCFunctions{} functions. Arrows go from a function to the functions it
calls. A function is proved bounded after all the functions it calls, starting from those
that call none. A function that calls itself, like the check of an \textsc{evaluate} step, is
proved by induction over the depth of the term it walks.}
\label{fig:cost:order}
\end{figure}

\begin{proof}[Proof of Lemma~\ref{lem:fn-cost}]
\Keel{} has \PrintedCFunctions{} functions. Order them so that each comes after the functions
it calls, as in Figure~\ref{fig:cost:order}. Apart from the functions that call themselves,
the call graph has no cycles, so such an order exists. We prove the functions bounded in this
order. Let $s_F$ be the body of $F$. A call of $F$ runs $s_F$ plus $c_F$ one-tick transitions
that enter and leave it, so
\[
  T \;\le\; T_{s_F} + c_F \;\le\; R \cdot (\sigma.\cid{work} - \sigma'.\cid{work}) + A^T_{s_F} + c_F
  \qquad \text{by Lemma~\ref{lem:stmt-cost} for } s_F ,
\]
and $F$ is bounded with $A^T_F = A^T_{s_F} + c_F$ and $A^M_F = A^M_{s_F}$.
The first functions in the order call no function, so there is nothing to show.
A later function calls only functions that come
before it, which are bounded, and itself only on a smaller term, for which an induction over
the depth of the term gives the bound. A call of $F$ ends because its body does.
\end{proof}

\subsection{Proof of \texorpdfstring{Theorem~\ref{thm:cost}}{the Cost Theorem}}
\label{app:cost:thm}

\begin{proof}[Proof of Theorem~\ref{thm:cost}]
Theorem~\ref{thm:refine:formal} rules out a stuck state, so by Lemma~\ref{lem:fn-cost} for
$\checkC$ the run exists. Let $w_0$ and $\mathit{mem}_0$ be the values the entry point gives
the counters, $w'$ and $\mathit{mem}'$ their final values, and recall that the entry point
allocates the copy of $\cert$ and the fixed tables before it sets \cid{mem}. Then
\begin{align*}
  T &\le R\,(w_0 - w') + A^T_{\checkC}
    && \text{Lemma~\ref{lem:fn-cost} for } \checkC \\
    &\le R\,w_0 + A^T_{\checkC}
    && w' \ge 0\text{, the counter is unsigned} \\
    &= R\,W_T(B) + A^T_{\checkC}
    && \text{Definition~\ref{def:budgets}} \\[4pt]
  M &\le (\mathit{mem}_0 - \mathit{mem}') + A^M_{\checkC}
    && \text{Lemma~\ref{lem:fn-cost} for } \checkC \\
    &\le \mathit{mem}_0 + A^M_{\checkC}
    && \mathit{mem}' \ge 0 \\
    &= W_M(B)
    && \text{Definition~\ref{def:budgets}: } A^M_{\checkC} \text{ is the copy of } \cert \text{ and the fixed tables}
\end{align*}
The allowance $A^T_{\checkC}$ is folded into the rate: $W_T(B) \ge \cid{WORK\_BASE} \ge 1$,
so $R\,W_T(B) + A^T_{\checkC} \le (R + A^T_{\checkC})\,W_T(B)$, and the $R$ of the theorem
is this larger constant.
\end{proof}

\subsection{How Far the Corpus Stays Below the Bound}
\label{app:cost:corpus}
No corpus certificate is rejected for running out of budget, and the most work or memory any
of them uses is about \BudgetUsedFrac{} of its budget. The rate $R$ is looser still: the
theorem allows some \BoundSlack{} times more ticks per input byte than any certificate
takes, because the constants in the proof were chosen to make the lemmas easy rather than
tight. What the kernel relies on at run time is the budget \Keel{} enforces; the theorem's
contribution is that no run can escape it.

\section{Details of the Evaluation}
\label{app:eval}

This appendix gives the full evaluation behind \S\ref{sec:eval}
of the paper: a case study for each bug, our fuzzing experiments, the
comparison with the BCF checker, the sensitivity to \Keel{}'s design
decisions, the cost of extending the checker, and the measurement
details. Section, theorem and figure numbers without a letter, such as
\S\ref{sec:background} or Theorem~\ref{thm:sound:formal},
refer to the paper.

\subsection{Bugs Discovered in the BCF Checker}
\label{sec:eval:soundness}

During our effort to build \tool, using a combination of manual audit
comparing BCF's checker to \tool and fuzzing tests, we uncovered
\BCFDefects{} bugs in the BCF checker, including unchecked rules,
unsound rule semantics, and memory safety faults
(Table~\ref{tab:eval:defects}).  In
\S\ref{sec:eval:bug1}--\S\ref{sec:eval:bug13}, we describe the root
cause of these bugs. Table~\ref{tab:eval:defects} lists every bug that
we discovered in the BCF checker and the process of discovery. The
last two rows in Table~\ref{tab:eval:defects} are fuzzing findings
that we omit in the main paper because they do not concern a inference
rule.

\begin{table}[h]
\caption{Problems in BCF's checker. ``Proof'': a soundness lemma of \Keel{}'s table could not
be proved for BCF's entry; ``audit'': reading BCF's C against \Keel{}'s specification;
``fuzzing'': \S\ref{app:fuzz}. ``Bug'': the case study that presents it.
``Certificate'': whether we built a certificate that triggers the problem.}
\label{tab:eval:defects}
\footnotesize
\setlength{\tabcolsep}{3pt}
\begin{tabular}{@{}p{1.4cm} p{4.0cm} p{3.2cm} l c p{1.5cm}@{}}
\toprule
Class & Problem & Rules & Found by & Bug & Certificate \\
\midrule
unchecked & conclusion accepted with no check at all & \textsc{rewrite} with entry 0, \textsc{poly\_norm} & audit & 1 & yes \\
unsound & products can wrap around in a comparison rewrite & rewrites 73, 74 & proof & 2 & yes \\
unsound & merged constant computed in 32 bits, overflow dropped & rewrite 80 & proof & 3 & yes, also in the kernel \\
unsound & concatenation sorted as if commutative & \textsc{aci\_norm} & fuzzing & 4 & yes \\
unsound & a shared term is modified while sorting & \textsc{aci\_norm} & audit & 5 & no \\
unsound & operands after the second ignored & \textsc{xor\_elim}, \textsc{not\_xor\_elim}, \textsc{cnf\_xor\_pos/neg} & audit & 6 & yes \\
unsound & shifts by 64 or more computed wrong & \textsc{evaluate} & audit & 7 & no \\
missing check & non-Boolean operands accepted & \textsc{equiv\_elim}, \textsc{not\_equiv\_elim}, \textsc{cnf\_equiv\_pos/neg}, \textsc{cnf\_ite\_pos/neg} & audit, fuzzing & 8 & fuzzer input \\
missing check & lists accepted as distinct values & \textsc{distinct\_values} & audit & 9 & no \\
missing check & operator header trusted; null-pointer read & \textsc{cong} & audit, fuzzing & 10 & fuzzer input \\
missing check & references to terms not validated; null-pointer read & \textsc{rewrite} & audit, fuzzing & 11 & fuzzer input \\
memory & search stack of 128 entries in an array of 64: write past the end, on the kernel stack & \textsc{absorb} & audit & 12 & yes \\
memory & parameter printed before it is checked: null-pointer read & \textsc{bitblast} (logger) & fuzzing & 13 & fuzzer input \\
undefined behavior & widths not bounded: shift by a negative amount or by 64 or more & type checker & fuzzing & -- & fuzzer input \\
kernel warning & malformed terms reach internal sanity checks, at \WarnSites{} places & several & fuzzing & -- & fuzzer input \\
\bottomrule
\end{tabular}
\end{table}

\textbf{Structure of the case studies for the bugs discovered in the
  BCF checker}.
Each row of Table~\ref{tab:eval:defects} represents a bug in the BCF
checker that we found. Each case study for a bug provides a short
summary, where the bug occurs, what the BCF checker does, and why it
is wrong. We also provide concrete inputs where the bug manifests. We
also describe how \tool avoids this bug. For some bugs, we also
provide a malicious proof certificate where one false fact makes the
BCF checker accept an unsafe eBPF program with such a certificate.
The bugs can be broadly classified into four groups (first column of
Table~\ref{tab:eval:defects}). In Bug\#1, two different kinds of proof
steps are not checked at all. In Bugs\#2--7, a rule is checked, but
the check is too weak, so a step can conclude a false fact; one false
fact is enough to make the checker accept a refutation of any
condition, as Bug\#2 shows. In Bugs\#8--11, a check is missing, but
the bug alone lets no false fact through: the checker builds terms
that its own format forbids, and two of these bugs let a certificate
crash it. In Bugs\#12 and 13, the checker writes or reads memory that
it does not own.

We describe a few facts about terms and bit-vector, which will be
useful to understand the description in the case study.

\textbf{Types}. Every term has a type. It is either a formula~(\ie,
\texttt{true} or \texttt{false}, or a bit-vector of some width $w$
(\ie, a string of $w$ bits).  We write $\top$ for \texttt{true} and
$\bot$ for \texttt{false}, and $\wedge$, $\vee$ and $\neg$ for
``and'', ``or'' and ``not''. For two formulas $A$ and $B$, we write $A
\leftrightarrow B$, read ``$A$ if and only if $B$'', for the formula
that says that $A$ and $B$ are both true or both false. We use capital
letters such as $A$ and $B$ stand for formulas, and lowercase $u$ and
$v$ stand for terms whose values are numbers or bit-vectors.

\textbf{Positions}. Every term in the pool has a number, its
\emph{position}. A term refers to its operands, and a step refers to
terms, by their positions. Each term also has a \emph{header}, which
records which operator it applies, what its type is and how many
operands it has, followed by the positions of its operands. A term
that several facts use is stored only one time, and all of them refer
to the same position. \S\ref{app:refine:storage} shows the exact
layout.

\textbf{Signed and unsigned interpretation of a bit-string}.  The
\emph{unsigned} interpretation of a $w$-bit vector gives every number
from 0 to $2^w - 1$. The \emph{signed} interpretation gives every
number from $-2^{w-1}$ to $2^{w-1} - 1$, because the leftmost bit, the
\emph{top bit} or \emph{sign bit}, has the negative weight
$-2^{w-1}$. \emph{Sign extension} adds copies of the sign bit on the
left and \emph{zero extension} adds zeros, which makes a vector wider
without changing its signed or its unsigned value, respectively.

\textbf{Arithmetic and wrapping around}. Adding or multiplying two
$w$-bit vectors produces a $w$-bit vector. However, the exact result
can need more than $w$ bits.  In this case, only the low $w$ bits are
kept and the others are dropped. For example, in 8 bits,
$\texttt{11111111} + \texttt{00000001}$ is exactly $256$, which is the
9 bits \texttt{100000000}. The low 8 bits are \texttt{00000000}, so
the 8-bit sum is 0. We say that a result \emph{wraps around} when it
differs from the exact result of the same arithmetic on ordinary
integers.

\textbf{Constants}. We write $c{:}w$ for the constant $c$ of width
$w$, so that $0{:}w$ is the zero of width $w$, the $w$-bit vector
whose bits are all 0.

\textbf{The definition of the operators}.  Two programs that both
compute ``shift left'' or ``divide'' on bit-vectors could still give
different results on some inputs. This is because they could, for
example, disagree on what a shift by more than $w$ bits gives, or what
happens when we divide by 0. SMT-LIB, a public standard used by SMT
solvers such as CVC5, provides the specification of bit-vector
operators. For every operator, it states the exact result on every
input. BCF's format uses SMT-LIB's specification. A fact in a
certificate is therefore true or false according to these definitions,
and a checker is correct only if it computes every operator exactly as
SMT-LIB defines it. Bug\#7 is a place where BCF's checker does not.

\textbf{Description of the C code of the BCF checker}.  BCF's checker
is one C file in the kernel. The case studies of Bugs\#3, 5, 12 and 13
show short excerpts of it. We leave out lines with
``\texttt{...}''. We provide our comments in the code snippets after
\texttt{//}.  Original comments from BCF are between \texttt{/*} and
\texttt{*/} comment blocks.
  \begin{itemize}
  \item \emph{Terms.} A term in memory is a \texttt{struct bcf\_expr}. Its header has three
    fields: \texttt{code}, which gives the operator and the type, \texttt{vlen}, the number of
    operands, and \texttt{params}, the parameters of the operator, such as a width. The array
    \texttt{args} follows, with the positions of the operands. For a term \texttt{e},
    \texttt{e->args[0]} is the position of its first operand. \texttt{id\_to\_expr(st, i)}
    returns the term at position \texttt{i}. Here \texttt{st} is the state of the checker, which
    almost every function receives. \texttt{expr\_equiv(st, e0, e1)} returns 1 when \texttt{e0}
    and \texttt{e1} are the same term.
  \item \emph{Tests.} A function whose name starts with \texttt{is\_} tests the field
    \texttt{code} of a term, i.e., its operator and its type.
  \item \emph{Rejecting.} A function returns 0 when it succeeds and a negative \emph{error
    code}, such as \texttt{-EFAULT} or \texttt{-EINVAL}, when it fails. The line
    \texttt{if (err) return err;} passes a failure on to the caller, and a failure anywhere
    makes the checker reject the certificate.
  \item \texttt{ENSURE(c)} rejects the certificate unless the condition \texttt{c} holds.
  \item \texttt{if (WARN\_ON\_ONCE(c)) return -EFAULT;} also rejects the certificate when
    \texttt{c} holds, and in addition writes a warning to the kernel's log the first time.
  \end{itemize}

\subsection{Bug\#1: Two Unchecked Proof Steps}
\label{sec:eval:bug1}
We discovered that BCF's checker does not check two proof steps at
all. It trusts the equations provided in the proof certificate, so a
certificate can state any equation through them, true or false. The
first is \textsc{poly\_norm}. A \textsc{poly\_norm} step states an
equation $u = v$ between two bit-vector terms built from numbers,
variables, $+$, $-$ and multiplication by a constant. It claims that
both sides have the same \emph{normal form} as polynomials. The normal
form of a side is what is left after two operations:
\begin{enumerate}
\item multiply out every product, so that $2 \cdot (y + 1)$ becomes $2y + 2$;
\item add up the terms that differ only in their constant factor, so that $3y + 5y$ becomes
  $8y$, and $\sym + (-\sym)$ becomes $0$.
\end{enumerate}
Two sides with the same normal form are equal for every value of their
variables. Step 1 of the running example
(Figure~\ref{fig:example}(c)) is such a step. It states $\sym +
(800 + {-\sym}) = 800$. On the left, $\sym$ and $-\sym$ add up to 0,
so the normal form of the left side is $800$, which is also the normal
form of the right side. This step is the heart of the running
example. It is the fact that the start and the length of the second
write add up to 800, which the verifier could not see.

The second is a \textsc{rewrite} step with entry 0. The rewrite table
has no entry 0. CVC5 uses many more rewrite rules than the table
contains, and it prints entry 0 for a step that uses one of them
(\S\ref{sec:background}). Such a step states an equation and
claims that it is an instance of some rewrite rule, without saying
which.

\textbf{Bug}. BCF's checker accepts both kinds of step without
checking them. For both, it records the stated equation as a
conclusion and prints a warning to the verifier's log. It never
computes a normal form, and it never looks for a rule. For entry 0, it
does not even check that the stated term is a formula.

\textbf{Concrete instance of the bugs}. A single such step can state
any equation. A \textsc{poly\_norm} step can state $y + 1 = y$, whose
two sides have different normal forms and which is false for every
$y$. A \textsc{rewrite} step with entry 0 can state that the 8-bit
constants 0 and 1 are equal, i.e., $0{:}8 = 1{:}8$. BCF's checker
records either one as a fact. One false fact is enough to refute any
condition, and Bug\#2 shows how.

\textbf{How \tool addresses it?} \Keel{} checks both proof steps. For
\textsc{poly\_norm}, it multiplies out both sides, adds up the terms
that differ only in their constant factor, and compares the results.
It accepts the step only if the two results are the same. For entry 0,
\Keel{} supports more rewrite rules, covering all rewrite rules that
occur in the evaluation, and rejects any rewrite that it does not
support yet.

\subsection{Bug\#2: Unsound Rewrite 73, a Product That May Wrap}
\label{sec:eval:bug2}

The rewrite rule entry 73 replaces a comparison of two products with
three simpler conditions. This is correct for mathematical integers,
but BCF lets the products be computed in too few bits, so they can
wrap around, and then the two sides of the entry disagree.
A \textsc{rewrite} step works as follows
(\S\ref{sec:background}).  BCF's checker has a fixed table of
equations, called the rewrite table. Each entry of the table is an
equation between two terms that contain \emph{holes}, i.e., named
parameters such as $x$ or $n$. A \textsc{rewrite} step says which
entry it uses and which term or number to put in each hole. The
checker puts them in and records the resulting equation as the step's
conclusion. An entry may also have a \emph{side condition}, a formula
about the holes.  The side condition must be proven before the entry
may be used.

The rewrite entry 73, which comes from CVC5's rule
\texttt{bv-mult-slt-mult-1}, is about a comparison of two products
that share a factor $a$:
\[
  y \cdot a \;<\; x \cdot a .
\]
In this case, $<$ and $>$ always compare signed values. For ordinary
integers (\S\ref{sec:eval}), the comparison holds exactly when
three conditions are all true: $y \ne x$; $a \ne 0$; and ``$y < x$''
and ``$a > 0$'' are both true or both false.

For bit-vectors, this holds only if the products are exact (\ie, they
do not wrap around). A product can need as many bits as its two
factors combined. For example, the product of two 8-bit numbers can
need 16 bits. The entry therefore does not multiply $y$, $x$ and $a$
themselves. It multiplies wider copies of them, made by \emph{sign
extension}. We write $\mathit{sext}(n, t)$ for the vector $t$ with $n$
copies of its sign bit, i.e., its top bit, added on the left. Sign
extension makes a vector wider without changing its signed value. For
example, the 8-bit \texttt{11111110} is $-2$. $\mathit{sext}(4,
\texttt{11111110})$ is the 12-bit \texttt{111111111110}, which is
$-2048 + 1024 + 512 + \cdots + 2 = -2$, still $-2$.

The entry has seven holes. Three hold bit-vectors: $y$ and $x$, the
two numbers being compared, and $a$, the shared factor. Four hold
numbers: $n$ is how many bits $y$ and $x$ are extended by, $m$ is how
many bits $a$ is extended by, $tn$ is the width of $y$ and $x$, and
$an$ is the width of $a$. ($tn$ and $an$ are each the name of one
hole, not products.) We write $\mathit{size}(x)$ for the width of
$x$. In full, the entry states
\begin{align*}
  & \mathit{sext}(n, y) \cdot \mathit{sext}(m, a) \;<\; \mathit{sext}(n, x) \cdot \mathit{sext}(m, a) \\
  &\quad \leftrightarrow\; (y - x \ne 0{:}tn) \wedge (a \ne 0{:}an) \wedge \big((y < x) \leftrightarrow (a > 0{:}an)\big)
  \qquad \text{if } tn = \mathit{size}(x) \wedge an = \mathit{size}(a) .
\end{align*}
The left side is the comparison of the two products, computed on the widened copies. The right
side is the three conditions, in order. The first is written $y - x \ne 0{:}tn$, which says the
same as $y \ne x$. The third uses $\leftrightarrow$, so it holds when both formulas are true
or both are false. The side condition, after ``if'', requires $tn$
and $an$ to be the actual widths of $x$ and $a$.

\textbf{The bug.} The equation holds for mathematical integers. For
bit-vectors, it holds only if the widened copies are wide enough for
the products to be exact.
\begin{itemize}
\item $\mathit{sext}(n, y)$ has $tn + n$ bits, and $\mathit{sext}(m, a)$ has $an + m$ bits.
\item Two bit-vectors can be multiplied only when they have the same width. So every instance
  of the left side has $tn + n = an + m$. We call this common width $W$. Both products are
  computed in $W$ bits.
\item The exact product of a $tn$-bit number and an $an$-bit number can need up to $tn + an$
  bits. So the products are exact whenever $W \ge tn + an$.
\item Since $W = tn + n$, the condition $W \ge tn + an$ is the same as $n \ge an$. In words,
  $y$ and $x$ must be extended by at least as many bits as $a$ has.
\end{itemize}
For example, with 8-bit $y$, $x$ and $a$, extending everything by 8
bits gives $W = 16$, and the products are exact. However, the side
condition does not require $n \ge an$. It fixes $tn$ and $an$ to the
widths of $x$ and $a$ and requires nothing about $n$ and $m$. Thus,
extending by zero bits becomes legal, and BCF's checker applies the
entry as written.

\textbf{Concrete instance of the bug}. Take 8-bit values with $n = m =
0$, so that nothing is extended, and $y = 64$, $x = 0$, $a = 2$. The
left side computes $64 \cdot 2 = 128$, which does not fit in a signed
8-bit vector. Its low 8 bits are \texttt{10000000}, which is $-128$
read as signed, and $0 \cdot 2 = 0$. So the comparison $-128 < 0$ is
true. On the right, $y < x$ is $64 < 0$, which is false, while $a > 0$
is true, so the last conjunct fails and the right side is false. The
entry equates a true formula with a false one.

\textbf{Forging a refutation}. A false fact such as this one is enough
to make BCF's checker accept a refutation of any condition $\varphi$,
even a satisfiable one. When the checker accepts a refutation of
$\varphi$, the verifier concludes that $\varphi$ cannot hold, i.e.,
that the check it could not prove never fails on this path. It then
continues as if the check had passed (\S\ref{sec:background}). An
attacker who can make the checker accept a refutation of a satisfiable
$\varphi$ can therefore get a program past a check that actually can
fail.

Table~\ref{tab:eval:forge} shows a certificate that does this. The
same certificate works for every $\varphi$. We ran it with two
conditions that are clearly satisfiable: $\top$, which is always true,
and $r_0 > 100$, which $r_0 = 200$ satisfies. A correct checker must
reject the certificate in both cases. BCF's checker accepts it in both
cases.

We write $\mathit{lhs}$ and $\mathit{rhs}$ for the two
sides of entry 73 with the holes filled in as in the instance above. $\mathit{lhs}$ is the
formula
\[
  \mathit{sext}(0, 64{:}8) \cdot \mathit{sext}(0, 2{:}8) \;<\; \mathit{sext}(0, 0{:}8) \cdot \mathit{sext}(0, 2{:}8) ,
\]
and $\mathit{rhs}$ is the formula
\[
  (64{:}8 - 0{:}8 \ne 0{:}8) \wedge (2{:}8 \ne 0{:}8) \wedge \big((64{:}8 < 0{:}8) \leftrightarrow (2{:}8 > 0{:}8)\big) .
\]
By the instance, $\mathit{lhs}$ is true and $\mathit{rhs}$ is false.

Steps 11 and 12 connect step 0 to the end. BCF's checker requires
every step except the last to be cited by a later step, and no step
cites step 0 yet.  Every step except step 5 is a correct use of a
sound rule. The certificate reaches $\bot$ only because step 5 states
a false fact, and it never uses $\varphi$ to conclude. This is why the
same certificate can be used to refute every condition.

The same construction works for the unchecked rules of Bug\#1 and the
other unsound rules of Bugs\#3--7, so we do not repeat it for
them. Each of these rules can conclude a false fact. When that fact
mentions only constants, the same few sound steps turn it into $\bot$
for any condition. \textsc{evaluate} computes the constants, and
\textsc{eq\_resolve}, or \textsc{distinct\_values}, which states that
two different constants are unequal, and \textsc{contra}, which
concludes $\bot$ from a formula and its negation, finish the
refutation. For those cases, we only show the false fact that the rule
can produce.

\textbf{How \tool addresses it?} We discovered this unsoundness when
the soundness lemma of the entry as transcribed from BCF could not be
proved. For mitigation, \Keel{}'s entry 73 is the same equation with
one more conjunct in its condition (\S\ref{sec:checker}), $n \ge
an$.  Together with $tn + n = an + m$, which every instance of the
left side satisfies, it gives $W \ge tn + an$, and the products cannot
wrap. \Keel{} correctly rejects the certificate of
Table~\ref{tab:eval:forge} at step 5. Entry 74, CVC5's
\texttt{bv-mult-slt-mult-2}, is the same rule for the unsigned
comparison, with zero extension in place of sign extension. It has the
same bug, and \Keel{} adds the same conjunct to its condition.

\begin{table}[t]
\caption{A certificate forged through rewrite 73. Only step 5 is false, and steps 1--10 never
mention $\varphi$, so the same certificate refutes every condition. $\mathit{lhs}$ and $\mathit{rhs}$ are the two
sides of entry 73 with the holes filled in as in the instance.}
\label{tab:eval:forge}
\small
\begin{tabular}{@{}clll@{}}
\toprule
Step & Rule & Premises & Concludes \\
\midrule
0 & \textsc{assume} & & $\varphi$ \\
1 & \textsc{evaluate} & & $\mathit{size}(x) = 8$ \\
2 & \textsc{symm} & 1 & $8 = \mathit{size}(x)$ \quad (first half of the side condition) \\
3 & \textsc{evaluate} & & $\mathit{size}(a) = 8$ \\
4 & \textsc{symm} & 3 & $8 = \mathit{size}(a)$ \quad (second half of the side condition) \\
5 & \textsc{rewrite} 73 & 2, 4 & $\mathit{lhs} \leftrightarrow \mathit{rhs}$ \quad (false: $\mathit{lhs}$ is true, $\mathit{rhs}$ is false) \\
6 & \textsc{evaluate} & & $\mathit{lhs} \leftrightarrow \top$ \quad ($\mathit{lhs}$ is true) \\
7 & \textsc{true\_elim} & 6 & $\mathit{lhs}$ \\
8 & \textsc{eq\_resolve} & 7, 5 & $\mathit{rhs}$ \\
9 & \textsc{evaluate} & & $\mathit{rhs} \leftrightarrow \bot$ \quad ($\mathit{rhs}$ is false) \\
10 & \textsc{eq\_resolve} & 8, 9 & $\bot$ \\
11 & \textsc{and\_intro} & 0, 10 & $\varphi \wedge \bot$ \\
12 & \textsc{and\_elim} & 11 & $\bot$ \\
\bottomrule
\end{tabular}
\end{table}

\subsection{Bug\#3: Unsound Rewrite 80, a Constant Merged in 32 Bits}
\label{sec:eval:bug3}

The rewrite rule entry 80 merges two neighboring constants of a
concatenation into one. BCF computes the merged value in 32 bits even
when it needs more, so its high bits are silently lost.

The rewrite rule entry 80 is about \emph{concatenation}. The
concatenation of two bit-vectors places their bits side by side, the
first operand on the left, i.e., in the high bits, and the second on
the right, in the low bits. For example, $\mathit{concat}(1{:}8,
0{:}8)$ is the 16-bit vector
\[
  \underbrace{\texttt{00000001}}_{1{:}8}\;\underbrace{\texttt{00000000}}_{0{:}8}
  \;=\; \texttt{0x0100} ,
\]
which is the number 256. A concatenation can have any number of operands. When two of them
next to each other are constants, as $1{:}8$ and $0{:}8$ are here, they can be replaced by a
single constant, $256{:}16$ in the example.

To state the entry, we need a formula for the merged constant. Take
two neighboring constants $n_1$ of width $w_1$ and $n_2$ of width
$w_2$. In the merged constant, the bits of $n_1$ sit $w_2$ places
further left than in $n_1$ itself, because the $w_2$ bits of $n_2$ are
to their right. Moving bits $w_2$ places to the left is the same as
multiplying by $2^{w_2}$, just as appending $k$ zeros to a decimal
number multiplies it by $10^k$. The bits of $n_2$ then fill the $w_2$
low bits, which the move left at zero, so the merged value is $n_1
\cdot 2^{w_2} + n_2$. In the example, $1 \cdot 2^8 + 0 = 256$. The
merged constant has width $ww = w_1 + w_2$. Writing $\mathit{xs}$ and
$\mathit{zs}$ for any operands before and after the two constants, the
entry states
\[
  \mathit{concat}(\mathit{xs}, n_1{:}w_1, n_2{:}w_2, \mathit{zs})
  \;=\; \mathit{concat}(\mathit{xs}, (n_1 \cdot 2^{w_2} + n_2){:}ww, \mathit{zs})
  \qquad \text{if } ww = w_1 + w_2 .
\]
Its holes are $\mathit{xs}$, $\mathit{zs}$, the numbers $n_1$ and $n_2$, and the widths $w_1$,
$w_2$ and $ww$. Unlike entry 73, the equation itself is correct.

\textbf{The bug}. The bug is in how BCF's checker computes the number
$n_1 \cdot 2^{w_2} + n_2$ on the right side. Every number that a step
puts into a hole is a 32-bit bit-vector, so the checker computes this
number in 32 bits, and a result that needs more bits wraps around. The
checker's evaluator, which computes the value of a term without
variables, notices the wraparound. It returns the value together with
a flag, \texttt{overflow}, which is true when the value
wrapped. However, the code that builds the merged constant never reads
the flag. Two lines earlier, the same code computes the width $ww$
with a helper that does read it:
\begin{lstlisting}[style=paperc,basicstyle=\ttfamily\scriptsize]
err = pack_bv_params(st, expr, size_expr_id, true);        // computes ww; rejects if it wrapped
err = err ?: eval_const_expr(st, val_expr_id, &eval_res);  // computes n1 * 2^w2 + n2
...
set_bv_val(expr, eval_res.bv_res);                         // stores it; .overflow is never read
\end{lstlisting}
The only check on the merged value, in the lines left out, is that it fits in $ww$ bits. A value
cut to 32 bits always fits in $ww$ bits when $ww$ is more than 32, so this check cannot catch
the wrap.

\textbf{Concrete instance of the bug}. Take $n_1 = 2^{31}$ with $w_1 =
32$, and $n_2 = 0$ with $w_2 = 8$, so that $ww = 40$. The merged
constant should be $2^{31} \cdot 2^8 + 0 = 2^{39}$. In hexadecimal,
$2^{31}$ is \texttt{0x80000000}, a single 1 followed by 31
zeros. Moving it 8 places to the left gives \texttt{0x8000000000},
which needs 40 bits. Its low 32 bits are all zero, so the checker
builds the constant $0{:}40$. The rule then concludes
$\mathit{concat}(2^{31}{:}32, 0{:}8) = 0{:}40$. This is false: the
left side has a 1 in its top bit, and the right side has no 1
anywhere.

\textbf{How \tool addresses it?} Like rewrite rule entry 73, we
discovered this when the soundness lemma of the entry as transcribed
could not be proved, and the case we could not prove was the overflow.
\Keel{}'s entry 80 is the same equation with one more conjunct in its
condition, $n_1 \cdot 2^{w_2} < 2^{32}$. It says that the moved
constant fits in the 32 bits that compute it. Adding $n_2$ cannot
overflow either. $n_2$ has $w_2$ bits, so it fits in the $w_2$ low
bits that the move left at zero, and adding it only fills those bits.

\subsection{Bug\#4: \textsc{aci\_norm} Sorts a Concatenation}
\label{sec:eval:bug4}

To check whether two terms differ only in grouping, order and
repetition, BCF reorders their operands. It also does this for
concatenation, where the order matters, so it can be made to accept
$\mathit{concat}(x, y) = \mathit{concat}(y, x)$.
Some operators have properties that let the same term be written in
different ways. Conjunction is one. For example, $A_1 \wedge (A_2
\wedge A_1)$ and $A_2 \wedge A_1$ are equivalent, because conjunction
has the three properties below.  (When the operands are formulas, $=$
means $\leftrightarrow$.)
\begin{itemize}
\item \emph{Associativity} says that the grouping does not matter:
  $u \circ (v \circ w) = (u \circ v) \circ w$. For conjunction,
  $A_1 \wedge (A_2 \wedge A_3) \leftrightarrow (A_1 \wedge A_2) \wedge A_3$.
\item \emph{Commutativity} says that the order does not matter: $u \circ v = v \circ u$. For
  conjunction, $A_1 \wedge A_2 \leftrightarrow A_2 \wedge A_1$.
\item \emph{Idempotence} says that a repeated operand does not matter: $u \circ u = u$. For
  conjunction, $A_1 \wedge A_1 \leftrightarrow A_1$.
\end{itemize}
Not every operator has all three. Concatenation (Bug\#3) is associative:
$\mathit{concat}(x, \mathit{concat}(y, z))$ and $\mathit{concat}(\mathit{concat}(x, y), z)$
both place the bits of $x$, $y$ and $z$ side by side, in this order. However, it is neither commutative
nor idempotent. $\mathit{concat}(1{:}8, 0{:}8)$ is \texttt{0x0100}, but
$\mathit{concat}(0{:}8, 1{:}8)$ is \texttt{0x0001}, and $\mathit{concat}(x, x)$ has twice as
many bits as $x$.

The rule \textsc{aci\_norm} (ACI stands for Associativity, Commutativity and Idempotence)
proves that two terms are equal when applying the ACI properties can turn one into the other.
It has no premises. Its one parameter is an equation $u = v$, and its conclusion is that same
equation, $u = v$. For example, a step of \textsc{aci\_norm} with the parameter
$A_1 \wedge (A_2 \wedge A_1) \leftrightarrow A_2 \wedge A_1$ concludes this equivalence. The
step is correct when the two sides are equal by applying the properties. To
test this, the checker builds a normal form of each side, as a new term outside the pool, in
three transformations, one per property:
\begin{enumerate}
\item flatten: replace nested applications of the operator by one application with all their
  operands, so that $A_1 \wedge (A_2 \wedge A_1)$ becomes $A_1 \wedge A_2 \wedge A_1$, a single
  conjunction with three operands (associativity);
\item drop repeated operands, so that $A_1 \wedge A_2 \wedge A_1$ becomes $A_1 \wedge A_2$
  (idempotence);
\item sort the operands (commutativity). As mentioned in the setup list, a term is a
  \texttt{struct bcf\_expr}, whose field \texttt{args} is an array that holds the
  positions of its operands. So,
  we can sort the operands because they are stored in an array, and sorting them means sorting the
  array \texttt{args}.
\end{enumerate}
If both sides end up as the same term, the checker accepts the step, and $u = v$ becomes a
fact. Otherwise it rejects the step. Each transformation is valid only for an operator that has
the matching property. Conjunction has all three, so all three transformations are valid for
it. For concatenation, only flattening is valid. Since every term has an array \texttt{args},
the checker can sort the operands of any term, but this keeps the value of the term only when
its operator is commutative.

\textbf{The bug}. Before it sorts the operands of the two new terms,
BCF's checker tests whether the operator allows sorting, i.e., whether
it is commutative. This check wrongly includes concatenation. The
checker therefore sorts the operands of a concatenation, which changes
the term's value.

\textbf{Concrete instance of the bug}. Take the equation
$\mathit{concat}(1{:}8, 0{:}8) = \mathit{concat}(0{:}8, 1{:}8)$. The
checker goes through it as follows.
\begin{itemize}
\item Flattening and dropping repeated operands change nothing, so the two new terms are
  copies of the two sides.
\item They differ, and the check that allows sorting accepts concatenation, so the checker
  sorts their operands.
\item Both new terms have the same two operands, so sorting them by position gives the same
  term for both. The checker accepts the equation.
\end{itemize}
The equation is false. Its left side is \texttt{0x0100}, i.e., 256, and its right side is
\texttt{0x0001}, i.e., 1. The same happens for any $x$ and $y$ in place of the two constants:
the checker accepts $\mathit{concat}(x, y) = \mathit{concat}(y, x)$, which places the same bits
at different positions.

\textbf{How \tool addresses it?} \Keel{} sorts the operands of the
commutative operators only.  This is the one unsound rule that we
found only by fuzzing (\S\ref{sec:eval:fuzz}).

\subsection{Bug\#5: \textsc{aci\_norm} Changes a Shared Term}
\label{sec:eval:bug5}

In one special case, the reordering of Bug\#4 is applied not to a
private copy but to a term that other facts share. The step can then
conclude a false fact and, at the same time, change the meaning of
earlier facts, including the condition being refuted.
This case is about the same rule as Bug\#4, \textsc{aci\_norm}. Two
details of how BCF's checker builds the normal form matter for this
case.
\begin{itemize}
\item BCF's checker sorts the array in place, i.e., it rearranges the operands inside the
  array itself, rather than building a sorted copy.
\item When the new term ends up with a single operand, the checker uses that operand itself
  in its place. For instance, after the repeated operand is dropped, $A \wedge A$ becomes a
  conjunction with the one operand $A$, and the checker then uses $A$ itself.
\end{itemize}

\textbf{The bug}. The bug is in \texttt{check\_aci\_norm}, the
function that checks an \textsc{aci\_norm} step, and it happens
between flattening and comparing.  At that point, the variables
\texttt{an} and \texttt{bn} are pointers to two new terms outside the
pool: the two sides, flattened and with repeated operands
dropped. Since these terms are outside the pool, changing them, for
example by sorting their operands, affects no other terms. The test
\texttt{reducible\_variadic(code)} is true for the operators, such as
conjunction, whose application to a single operand equals that
operand, and \texttt{is\_assoc(code)} is the check of Bug\#4 that
decides whether sorting is allowed.
\begin{lstlisting}[style=paperc,basicstyle=\ttfamily\scriptsize]
// in check_aci_norm: an and bn point to new terms, the flattened a and b
if (an->vlen == 1 && reducible_variadic(an->code))   // one operand left in the new term for a:
    an = id_to_expr(st, an->args[0]);                 //   an now points to that operand, in the pool
if (bn->vlen == 1 && reducible_variadic(bn->code))   // the same for b
    bn = id_to_expr(st, bn->args[0]);
if (a->code == b->code) {                       // tests the operator of a and b: conjunction
    if (expr_equiv(st, an, bn) == 1)
        return 0;
    if (is_assoc(a->code)) {                    // true for conjunction
        sort(an->args, an->vlen, sizeof(u32), __cmp_u32, NULL);   // may sort a term in the pool
        sort(bn->args, bn->vlen, sizeof(u32), __cmp_u32, NULL);
        ENSURE(expr_equiv(st, an, bn) == 1);
    }
    ...
}
\end{lstlisting}
The two \texttt{if}s at the top change what \texttt{an} and \texttt{bn} are.
\begin{itemize}
\item Before them, \texttt{an} points to the new term for \texttt{a}, outside the pool.
  Sorting its operands changes nothing outside the step.
\item After them, if the new term had a single operand $A$, \texttt{an} points to $A$ itself,
  i.e., the term at $A$'s position in the pool, which every other term that mentions $A$ refers to.
  Thus, sorting \texttt{an->args} will reorder the operands of the real $A$ in the pool. If
  the operator of $A$ is not commutative, such as $<$, this changes the value of $A$ for every
  term that refers to it.
\end{itemize}

\textbf{Concrete instance of the bug}. Take two variables $y$ and $z$,
where $y$ has a smaller position than $z$ in the pool, and the two
formulas $A = (y < z)$ and $B = (z < y)$. Consider the step that
claims $(A \wedge A) \leftrightarrow (B \wedge B)$ with
\textsc{aci\_norm}.  The checker goes through it as follows.
\begin{itemize}
\item After the repeated operand is dropped, the left side $A \wedge A$ becomes a conjunction
  with the one operand $A$, and then $A$ itself. The right side becomes $B$ itself in the same
  way.
\item $A$ and $B$ are different terms, so the first comparison fails. Both sides are
  conjunctions, so the checker sorts.
\item It sorts the operands $y, z$ of $A$. They are already in order, so $A$ stays $y < z$.
\item It sorts the operands $z, y$ of $B$. This puts $y$ first, and the term at $B$'s position
  now reads $y < z$.
\item The two sides are now the same term, $y < z$, so the checker accepts the step.
\end{itemize}
Two things have gone wrong. First, the accepted equivalence is false: for $y = 0$ and $z = 1$,
$y < z$ holds but $z < y$ does not. Second, $B$ itself has changed in the pool, so every fact
that mentions $B$ now means something else. BCF's checker compares $\varphi$ with the kernel's
condition only once, before the proof certificate checking process. If $B$ is part of $\varphi$, the certificate goes on to
refute a formula other than the one the kernel asked about.

\textbf{How \tool addresses it?} We found this bug by reading BCF's
code against \Keel{}'s specification. \Keel{}'s normal form is a
separate value, the positions of the operands, that it computes and
compares. It never writes to a term in the pool, so every term keeps
its meaning for the whole proof certificate checking process.

\subsection{Bug\#6: The XOR Rules Read Only Two Operands of an XOR}
\label{sec:eval:bug6}

Four rules about XOR are correct only for an XOR of two operands. BCF
also applies them to longer XORs and ignores every operand after the
second. \emph{Exclusive or}, or XOR, written $\oplus$, is true when an
odd number of its operands are true. With two operands, $A \oplus B$
is true when exactly one of $A$ and $B$ is true. In BCF, an XOR can
have 2 to 255 operands. For example, $A_1 \oplus A_2 \oplus A_3$ is
true when one or all three of $A_1$, $A_2$, $A_3$ are true.

Several rules conclude \emph{clauses}, i.e., disjunctions of formulas and negated formulas, such
as $A \vee \neg B$. Four of them turn an XOR into clauses, and each clause is built from the
XOR's two operands $A$ and $B$. For example, \textsc{cnf\_xor\_neg} concludes
$(A \oplus B) \vee \neg A \vee B$, which always holds: if $A \oplus B$ is false, then $A$ and $B$
are both true or both false, so $A$ is false or $B$ is true. Such a clause is correct only for an
XOR of exactly two operands. The value of $A_1 \oplus A_2 \oplus A_3$ also depends on $A_3$,
which a clause built from $A_1$ and $A_2$ does not mention.

\textbf{The bug}. BCF's checker tests that the term is an XOR, but not
that it has two operands. It reads the first two operands and ignores
the rest, so it treats $A_1 \oplus A_2 \oplus A_3$ as if it were $A_1
\oplus A_2$.  All four rules above have this bug.

\textbf{Concrete instance of the bug}. Take the XOR $\top \oplus \bot
\oplus \top$. Two of its operands are true, which is an even number,
so it is false. \textsc{cnf\_xor\_neg} on it takes $A = \top$ and $B =
\bot$, its first two operands, and builds the clause
\[
  (\top \oplus \bot \oplus \top) \;\vee\; \neg\top \;\vee\; \bot .
\]
Its first disjunct is the XOR itself, which is false. Its second disjunct, $\neg\top$, is false.
Its third disjunct, $\bot$, is false. A disjunction of three false formulas is false, so the
rule concludes a false fact.

\textbf{How \tool addresses it?} All four rules accept an XOR only
when it has exactly two operands, and their soundness lemmas rely on
that. The two CNF rules occur in real certificates from CVC5, in
\CnfXorSteps{} steps of the corpus. They cannot be dropped, and the
operand count is the one check they lack.

\subsection{Bug\#7: \textsc{evaluate} Shifts by 64 or More}
\label{sec:eval:bug7}

BCF computes shifts with C's shift on 64-bit integers, which does not
agree with the definition of bit-vector shifts when the shift amount
is 64 or more. Here, \textsc{evaluate} has no premises and has one
parameter, which is a term $u$ without variables, and it concludes $u
= c$, where $c$ is the value of $u$.  The correctness of checking
\textsc{evaluate} hinges on the checker's evaluator, the code that
computes $c$. This case is about how the evaluator computes
\emph{shifts}. A shift moves the bits of a bit-vector $x$ by $s$
places, where the amount $s$ is itself a bit-vector of the same width
$w$ as $x$:
\begin{itemize}
\item $x \ll s$ moves the bits $s$ places to the left. The $s$ top bits fall off, and $s$ zeros
  come in on the right. For example, in 8 bits, $\texttt{00000001} \ll 3$ is
  \texttt{00001000}.
\item $x \gg s$ moves the bits $s$ places to the right. The $s$ low bits fall off, and $s$ zeros
  come in on the left.
\item $x \gg_a s$ also moves the bits to the right, but copies of the sign bit come in on the
  left. This keeps the sign of the signed reading.
\end{itemize}
SMT-LIB also defines what happens when $s \ge w$: every bit falls off. So $x \ll s$ and
$x \gg s$ are 0, and $x \gg_a s$ consists of copies of the sign bit only. For example, in 8
bits, $\texttt{00000001} \ll 8$ is \texttt{00000000}.

\textbf{The bug}. BCF's evaluator holds every value in a 64-bit C integer, computes the
shift on that integer, and keeps the low $w$ bits of the result. This gives the SMT-LIB value
for every amount below 64, even one of $w$ or more, because the bits that fall off end up above
the low $w$ bits or past the end of the integer.
Amounts of 64 or more do occur: the amount is itself a $w$-bit vector, so for
a width of 7 bits or more it can reach 64. For those amounts, the evaluator is wrong in two
different ways.
\begin{itemize}
\item For $x \ll s$, it calls \texttt{check\_shl\_overflow}, a kernel helper that refuses to
  shift by 64 or more. It leaves the value unshifted and sets the overflow flag of Bug\#3,
  which \textsc{evaluate} never reads.
\item For the two right shifts, it uses plain C shifts of a 64-bit integer, and C leaves a shift
  by 64 or more \emph{undefined}. That is, the C standard gives such a shift no meaning, so the
  compiler may produce any result.
\end{itemize}

\textbf{Concrete instance of the bug}. Take 8-bit vectors, whose
amounts can go up to 255. For $1 \ll 64$, the helper leaves 1
unshifted, and the checker concludes $(1 \ll 64) = 1$, where SMT-LIB
gives 0.  For $1 \gg 64$, the processors of x86-64 and arm64 machines
take the amount of a right shift modulo 64, i.e., they shift by $64 -
64 = 0$, and $(1 \gg 64)$ evaluates to 1 instead of 0.  Either way,
\textsc{evaluate} concludes a false equation between two constants.

\textbf{How \tool addresses this bug?}  \Keel{}'s evaluator computes
the value of $u$ exactly as SMT-LIB defines it for every amount. By
Theorem~\ref{thm:refine:formal}, which says that \Keel{}'s C code
computes what its Lean specification says on every input, its C never
performs a computation that C leaves undefined
(\S\ref{sec:embc}).

\subsection{Bug\#8: Equivalence Rules That Accept Bit-Vectors}
\label{sec:eval:bug8}

Rules meant for formulas also accept bit-vectors, so BCF builds
clauses such as ``not $3$, or $3$'', which mean nothing.
BCF's format has one equality operator for both formulas and
bit-vectors. Between bit-vectors, $u = v$ says that they have the same
value. Between formulas, $A = B$ means $A \leftrightarrow B$.
Several of the rules that produce clauses (Bug\#6) are about
equivalences. For example, \textsc{equiv\_elim} has one premise, $A
\leftrightarrow B$, and concludes $\neg A \vee B$, or $A \vee \neg B$;
a parameter picks which.  \textsc{not\_equiv\_elim}, the two
\textsc{cnf\_equiv} rules, and the two \textsc{cnf\_ite} rules, which
are about an if-then-else $\mathit{ite}(D, A, B)$, conclude similar
clauses. All these clauses are correct only when $A$ and $B$ are
formulas. Negation and disjunction are defined by truth, and a
bit-vector such as $3{:}8$ is a number, neither true nor false.

\textbf{The bug}. BCF's checker tests that the premises of these rules
are equations, but not that their operands are formulas.  Since one
operator serves both types, an equation between two bit-vectors passes
the test, and the rule then negates a bit-vector and uses one as a
disjunct. The \textsc{cnf\_ite} rules likewise accept an if-then-else
whose branches are bit-vectors.

\textbf{Concrete instance of the bug}. From $3{:}8 = 3{:}8$, which a
certificate can prove, \textsc{equiv\_elim} concludes $\neg(3{:}8)
\vee 3{:}8$, which is not a formula at all.

A certificate reaches $\bot$ from clauses with the rule
\textsc{resolution}. From two clauses, one with a formula $A$ among
its disjuncts and the other with its negation $\neg A$, it concludes
the clause of all their other disjuncts. For example, from $A \vee B$
and $\neg A \vee D$ it concludes $B \vee D$, since $A$ is either
false, and then $B$ holds, or true, and then $D$ holds. A chain of
such steps reaches $\bot$ when no disjunct is left.

This clause does not lead to a false conclusion. \textsc{resolution}
cancels only a formula that the step names from the certificate's
pool. $3{:}8$ is not a formula, and $\neg(3{:}8)$ cannot be in the
pool, because the checker type-checks the pool before the proof
certificate checking process. So the clause can never shrink to
$\bot$. It is a bug nonetheless. The checker builds a term that its
own type check would reject in a certificate, and the clause is
harmless only because of what every other rule happens not to do,
which a new rule could change.

\textbf{How \tool addresses it?} Fuzzing found \textsc{equiv\_elim}'s
case in the BCF checker. We read the code found the rest. \Keel{}
type-checks every term when it builds it. A negation or a disjunction
is built only from formulas, so each of these steps is rejected at the
rule.

\subsection{Bug\#9: \textsc{distinct\_values} on Lists}
\label{sec:eval:bug9}

A rule that states that two different constants are unequal also
accepts lists, so BCF builds an equation between two lists, which its
own format forbids. So far we have said that every term is a formula
or a bit-vector. BCF's format has a third type, \emph{lists}, which we
left out because only the rewrite table uses them.  A list is written
$[u_1, \ldots, u_k]$, and its elements can be any terms, variables
included.  Lists only serve one purpose, which is to let one hole of a
table entry hold several operands, as the holes $\mathit{xs}$ and
$\mathit{zs}$ of entry 80 in Bug\#3 do. A list is never a fact or a
premise, and BCF's format does not let two lists be compared: its type
check rejects an equation whose sides are lists.  In C, a list is a
\texttt{struct bcf\_expr} like any other term (setup list). Its field
\texttt{code} gives the type ``list'', and its array \texttt{args}
holds the positions of its elements. For example, $[x]$ is a header
followed by one position, that of $x$.

The rule \textsc{distinct\_values} has no premises. Its two parameters
are two different constants $v_0$ and $v_1$ of the same type, such as
$1{:}8$ and $2{:}8$, and it concludes $\neg(v_0 = v_1)$. This is
correct because two constants that are written differently have
different values. The rule must therefore make sure that both
parameters are constants.

\textbf{The bug}. BCF's checker tests that both terms are constants,
that they have the same type, and that they are different
terms. However, its test for a constant is too weak.  The field
\texttt{code} of a term gives both its operator and its type, and the
test looks only at the operator.  In BCF's format, a list has the same
operator as a constant, so the two differ only in their type. So, a
list passes the test, and the rule builds an equation between two
lists, which BCF's format forbids.

\textbf{Concrete instance of the bug}. Take two different variables
$x$ and $y$. The rule accepts the lists $[x]$ and $[y]$ and concludes
$\neg([x] = [y])$, a term that BCF's type check would reject in any
certificate. On its own, this fact is useless to a certificate, since
no other sound rule builds an equation between two lists to contradict
it. Bug\#10, however, can build exactly such an equation, $[x] = [y]$,
from the premise $x = y$. The rule \textsc{contra}, which concludes
$\bot$ from a fact $A$ and its negation $\neg A$, compares its two
premises term by term, so it would accept these two facts. Together,
the two bugs could therefore let a certificate derive $\bot$ from $x =
y$, a satisfiable condition.

\textbf{How \tool addresses it?} \Keel{}'s \textsc{distinct\_values}
accepts only constants that are formulas or bit-vectors. \Keel{}
defines equality only between two formulas or two bit-vectors of the
same width, so neither the inequality nor the equation between two
lists can be built.

\subsection{Bug\#10: \textsc{cong} Trusts Its Operator Header}
\label{sec:eval:bug10}

To build a term, \textsc{cong} copies the header of an operator,
supplied by the certificate, into a piece of memory that all steps
share, and it trusts the number of operands that the header gives. A
wrong number can make \textsc{cong} pick up leftovers from earlier
steps, or crash the checker. \textsc{cong} applies the same operator
to equal operands. Its premises are $n$ equations $a_1 = b_1, \ldots,
a_n = b_n$, and it concludes $\mathit{op}(a_1, \ldots, a_n) =
\mathit{op}(b_1, \ldots, b_n)$. For example, from $y = 3{:}8$ and $z =
4{:}8$ it concludes $y + z = 3{:}8 + 4{:}8$. This is correct, since
equal operands give equal results. Step 23 of the running example
(Figure~\ref{fig:example}(c)) is a \textsc{cong} step.

The step must also say which operator to apply, and it does so in an unusual way. The
parameters of a step are plain 32-bit numbers. Most rules use a parameter as the position of a
term in the pool, or as a small number, such as which of two forms to conclude. \textsc{cong}
is different. It needs an operator without operands, because its operands come from its
premises. A header is exactly that. So BCF puts a header itself into the
one parameter of a \textsc{cong} step, and \textsc{cong} is the only rule whose parameter is a
header.
In the $y + z = 3{:}8 + 4{:}8$ example,
the parameter of \textsc{cong} is the header of an 8-bit $+$ with two operands. Two things about the
header must hold for the conclusion to make sense:
\begin{itemize}
\item the header must give as many operands as the step has premises, one operand per
  equation;
\item the header must name an operator, such as $+$, not a constant, a variable or a list
  (Bug\#9).
\end{itemize}

\textbf{The bug}. BCF's checker tests neither of the two conditions.
The first gap lets operands of earlier steps slip into the conclusion
of a \textsc{cong} step. The header is part of the certificate, which
comes from user space, so its author can make it say any number of
operands, for example three for a step with two
premises. \textsc{cong} trusts this number. It assembles the term
$\mathit{op}(a_1, \ldots, a_n)$ in the \emph{scratch buffer}, a piece
of memory that every step shares for assembling terms. The buffer is
itself a \texttt{struct bcf\_expr}, a header followed by operand
slots, and before each use the checker clears only its header. The
operand slots keep whatever an earlier step wrote there. \textsc{cong}
copies the step's header into the buffer and writes one operand per
premise. When the header says more operands than the step has
premises, the extra slots still hold operands of an earlier step, and
the type check, which reads as many operands as the header says, takes
them as operands of the new term. Two consequences follow:
\begin{itemize}
\item the step concludes an equation about a term whose operands come from some earlier,
  unrelated step;
\item a leftover may not be the position of any term, and looking it up then reads through a
  null pointer, which crashes the checker.
\end{itemize}

The second gap lets a header describe a list instead of an operator, so
\textsc{cong} can build an equation between two lists, which BCF's format forbids (Bug\#9).

\textbf{Concrete instance of the bug}. For the second missing check,
take a header that describes a list with one element, and the premise
$x = y$. \textsc{cong} then builds the equation $[x] = [y]$, which
Bug\#9 can contradict.

\textbf{How \tool addresses it?} Fuzzing found the null-pointer read
in the BCF checker. We read the code and found that a list is accepted
too. \Keel{}'s \textsc{cong} takes only the operator and its
parameters from the header. It builds the two terms from the premises
alone, with one operand per premise, and rejects a header whose
operator is a variable or a constant. It has no scratch buffer for a
term to inherit operands from.

\subsection{Bug\#11: \textsc{rewrite} Trusts Its References}
\label{sec:eval:bug11}

A \textsc{rewrite} step names terms by number, and BCF never checks
that these numbers are positions of terms. A bad number can make BCF
read something that is not a term as if it were one, or crash.  As
Bug\#2 describes, a \textsc{rewrite} step names an entry of the
rewrite table, and its parameters say which term to put in each hole
of the entry. A parameter names a term by its position in the pool. To
see what a valid position is, we need the layout of the pool that the
setup list sketches (\S\ref{app:refine:storage} gives it in full). The
pool is an array of 32-bit \emph{units}, numbered from 0. A term takes
up several consecutive units: first its header, then one unit per
operand, which holds the position of that operand. The position of a
term is the number of its header unit. For example, in
Figure~\ref{fig:pool}, the term $t = o + (800 - \sym)$ of the running
example takes up three units:
\begin{itemize}
\item unit 7 is its header, which says ``$+$, 32 bits, two operands'', so the term's position
  is 7;
\item unit 8 holds 1, the position of its first operand $o$;
\item unit 9 holds 4, the position of its second operand $800 - \sym$.
\end{itemize}
So 7 is a valid position, but 8 and 9 are not. They are units inside a term, and they hold
positions rather than headers. The parameters of a step, however, are plain 32-bit numbers
(Bug\#10), so a certificate can put any value there. A checker must therefore make sure that
each parameter that refers to a term is the position of a term.

\textbf{The bug}. BCF's checker makes this check for the parameters of
every rule, before the rule runs, except two. It skips \textsc{cong},
whose parameter is a header, not a position (Bug\#10). It also skips
\textsc{rewrite}, and \textsc{rewrite} does not check its parameters
anywhere else. It looks each one up with a function whose comment says
that the caller must pass a valid position (``REQUIRES: id must be
valid''), but no code on the path of a \textsc{rewrite} step actually
checks that.

\textbf{Concrete instance of the bug}. A \textsc{rewrite} step can go
wrong in two ways.
\begin{itemize}
\item If a parameter is a number past the end of the pool that belongs to no term built during
  the proof certificate checking process, the lookup returns a null pointer, and the checker then reads through it, which
  crashes the checker.
\item If a parameter is a position inside a term, such as 8 in the example above, the checker
  reads the unit stored there, the number 1, as if it were a header. The step is then about a
  term that the certificate never contained.
\end{itemize}

\textbf{How \tool addresses it?} Fuzzing found both: the null-pointer
reads, and \textsc{rewrite} steps that BCF's checker accepts with a
position inside a term, in certificates without any unchecked
step. \Keel{} looks up every reference, for every rule, through a
function that accepts only the position of a term, and rejects the
step otherwise.

\subsection{Bug\#12: \textsc{absorb}'s Search Stack}
\label{sec:eval:bug12}

A search keeps its progress in an array of 64 entries but checks
against 128, so a deeply nested term makes it write past the end of
the array, on the kernel stack. Some operators have an \emph{absorbing
element}, an operand that decides the result on its own. $\bot$ is the
absorbing element of conjunction: a conjunction with $\bot$ anywhere
among its operands is false, whatever the other operands are.
Likewise, $\top$ is the absorbing element of disjunction.  The
\textsc{absorb} rule concludes that a term equals the absorbing
element of its operator when that element occurs somewhere among its
operands, possibly nested inside further applications of the same
operator. Like \textsc{aci\_norm} (Bug\#4), \textsc{absorb} has no
premises. Its one parameter is an equation $u = e$, whose left side
$u$ is the term and whose right side $e$ is the absorbing element, and
its conclusion is that same equation. For example, a step of
\textsc{absorb} with the parameter
\[
  A_1 \wedge (A_2 \wedge (\bot \wedge A_3)) \;\leftrightarrow\; \bot
\]
concludes this equivalence, i.e., that the term on the left is false, because $\bot$ sits
inside the conjunction. The checker accepts the step only if it finds the
element $e$ in the term $u$.
Otherwise it rejects the step. The nesting can be as deep as the certificate makes it.

To find the absorbing element, BCF's checker searches the term \emph{depth first}, i.e., it follows one
operand all the way down before it tries the next. In the example, it goes down through the
three nested conjunctions until it finds $\bot$.
Each time the search goes one level down, it must remember where it was on the level above, so
that it can continue there if the level below does not contain the element. Without recursion,
the search keeps these places in its own list, called the \emph{search stack}.
The search stack is the local array \texttt{stack} of the function \texttt{check\_absorb},
declared as
\begin{lstlisting}[style=paperc,basicstyle=\ttfamily\scriptsize]
struct bcf_expr_stack_elem stack[BCF_MAX_ITER_STACK];   // BCF_MAX_ITER_STACK is 64
\end{lstlisting}
so it has 64 entries, and since it is a local variable,
it lives on the kernel stack.
Therefore, before each push, the search must check that the array has room.

\textbf{The bug}. The search does check before each push, but it
checks against the wrong limit. With \texttt{sp} as the number of
entries in use, the check is
\begin{lstlisting}[style=paperc,basicstyle=\ttfamily\scriptsize]
if (sp >= BCF_MAX_CMP_STACK)                 // BCF_MAX_CMP_STACK is 128, not 64
    return -E2BIG;
\end{lstlisting}
It compares the number of entries with 128, a constant that belongs to a different search in
BCF's checker, the one that compares two terms, whose own stack has 128 entries. The array of
\textsc{absorb} holds only 64.
A term that nests the operator more than 64 deep therefore makes \textsc{absorb}
push past the end of its array, which is a write beyond a buffer on the kernel stack.

\textbf{Concrete instance of the bug}. The exploit is a certificate
with one \textsc{absorb} step. Its term is a conjunction nested
\AbsorbDepth{} deep, with $\bot$ at the bottom. BCF's checker works as
follows on this certificate:
\begin{itemize}
\item The search pushes one entry for each level, \AbsorbDepth{} entries in all. The check
  against 128 never stops it.
\item The entries past the 64th land outside the array, on whatever the function keeps next to
  it on the stack, including its return address.
\item The checker then never returns a verdict. The compiler's stack protector, which BCF's
  kernel configuration turns on, places a known value next to the return address and checks it
  before the function returns. It detects the corruption at that point. In user space, where we
  ran the certificate, it kills the process, and in the kernel it halts the whole machine.
\end{itemize}
Without the protector, the function returns to the overwritten address, which the author of
the certificate chose.

\textbf{How \tool addresses it?} \Keel{} searches the term with a
recursive function that carries a depth counter and stops when the
counter reaches \MaxDepth{}. It keeps no array of its own, and it
rejects a term nested deeper than that. Since every recursion in
\Keel{} has such a fixed bound, the depth does not depend on the
input: measured under the kernel's own build flags, the whole checker
uses at most about \StackKiB{}~KiB of the kernel stack. On the
certificate above, \Keel{} returns an error: the term is deeper than
its bound, so it rejects the step.

\subsection{Bug\#13: The Log Prints an Unchecked Parameter}
\label{sec:eval:bug13}

When a \textsc{bitblast} step fails, the code that writes the error
message to the log looks up a term without checking that it exists,
and crashes the checker. The verifier keeps a \emph{log}, a text
record of its work, which helps a programmer see why a program was
rejected. The loader decides whether there is a log at all.  When the
loader asks the kernel to load a program, it provides a buffer for the
log, sets how detailed the log is, or turns it
off~\cite{linux-bpf-uapi}. In our threat model, the attacker controls
the certificate and may be the loader, so the attacker also controls
whether the log is on.  BCF's checker writes to the same
log~\cite{bcf-repo}. When the log is on and a step fails, it prints
the step, and to print a parameter that names a term, it looks up the
term at that position.

\textbf{The bug}. Before that lookup, the printing function checks
that the parameter is the position of a term, as the case study of
Bug\#11 requires, except for steps of the rule \textsc{bitblast}. For
those, it looks the parameter up unchecked. \textsc{bitblast} itself
rejects a parameter that is not a position, but the log line that
reports the rejection then reads through a null pointer.  The fault is
in the code that reports an error, not in the code that checks a rule.

\textbf{Concrete instance of the bug}. Take any \textsc{bitblast} step
whose first parameter is not the position of a term, for example a
number past the end of the pool, and send it with the log turned on.
\begin{itemize}
\item \textsc{bitblast} rejects the step, which is correct.
\item The checker prints the failing step to the log, and reads through a null pointer while
  printing the parameter, which crashes the checker.
\end{itemize}

\textbf{How \tool addresses it?} We found this only by
fuzzing. \Keel{} prints nothing: it reports a rejection as an error
code. Theorem~\ref{thm:refine:formal} covers all of its printed
C, so there is no code outside the proof for such a fault to hide in.

\subsection{Fuzzing}
\label{sec:eval:fuzz}
\label{app:fuzz}

To look for failures we did not anticipate, we fuzzed \Keel{} and BCF's checker side by side on
about \FuzzInputsM{} million mutated certificates, both built for user space with the address
and undefined-behavior sanitizers, with mutations that know the format and a harness that
compares the two checkers step by step. Because \Keel{} is proved sound, a step that BCF's
checker accepts and \Keel{} rejects is either a bug in BCF's checker or a valid step outside
\Keel{}'s specification, and we read each such step by hand. Fuzzing cannot add to a proof, but
it tests what the proof trusts (\S\ref{sec:embc}): that the printed C, compiled by a real
compiler and run against real memory, behaves as the EmbC semantics says. On none of the
\FuzzInputsM{} million certificates did \Keel{} crash, hang, leak memory, or trigger a
sanitizer. For BCF's checker, fuzzing found Bugs\#4, 8, 10, 11 and 13, two of
them, Bugs\#4 and 13, only this way, and it found rewrites 73 and 80 again. The
other failures it found are not in Table~\ref{tab:eval:defects}: inputs that make the type
checker compute an undefined shift in C, and inputs that reach the checker's internal warnings,
which halt the machine if the kernel is configured to panic on a warning. \Keel{} avoids each
of them for the same reason as the rest of the table: every input that does not meet the
specification is rejected with an error code, and Theorem~\ref{thm:refine:formal} covers every input.

The harness gives each mutated certificate to both checkers, built for user space with the
address and undefined-behavior sanitizers; BCF's checker is unmodified and also records which
code it covers, which guides the fuzzer towards new paths through it. The seeds are the
\FuzzCorpusSeeds{} corpus certificates smaller than 64\,KiB and six certificates captured from
the kernel, and the mutations know the format: besides flipping raw bytes, they replace whole
terms, steps and references to earlier terms. Four campaigns of 30 minutes with four workers
each, on an arm64 machine, ran about \FuzzInputsM{} million certificates in total. For each
one, the harness records crashes, sanitizer reports, runs longer than 10 seconds, leaked
memory, and the kernel warnings that BCF's checker raises internally, which the harness counts
instead of printing.
Three of the bugs let a certificate make BCF's checker read through a null pointer, at
\CrashSites{} places in its code; in the kernel, such a read is an oops, and the smallest
certificate that causes one is \CrashMinBytes{} bytes.

\subsection{What Verification Costs}
\label{sec:eval:cost}

In this subsection we measure whether there is any cost in performance
or code size in replacing BCF's checker with
\Keel{}. Table~\ref{tab:eval:compare} compares the two checkers.  In
short, \Keel{} loads and rejects exactly what BCF's checker does,
loading takes as long as before, and the checking itself is faster, at
a little more memory.

\begin{table}[t]
\caption{\Keel{} compared with BCF's checker on all \IdxObjects{} object files and the
\CorpusShipped{} corpus certificates. Time rows give \Keel{}'s speedup over BCF's checker, except load time, which is a ratio.}
\label{tab:eval:compare}
\small
\begin{tabular}{@{}lrr@{}}
\toprule
 & BCF's checker & \Keel{} \\
\midrule
Object files loaded (of \IdxObjects{}) & \IdxLoaded{} & \IdxLoaded{}, the same ones \\
Object files rejected by the verifier & \IdxRejected{} & \IdxRejected{}, the same ones \\
Certificates checked per load of all objects & \IdxCertsPerLoad{} & \IdxCertsPerLoad{} \\
Certificates accepted in user space (of \CorpusShipped{}) & \CorpusShared{} & \CorpusAccepted{} \\
\midrule
Load time with \Keel{} as a share of that with BCF's checker &
  \multicolumn{2}{r}{\IdxLoadPct\%} \\
Checking in the kernel: all certificates, the slowest \IdxSlowPct\% &
  \multicolumn{2}{r}{\IdxCheckSpeedup$\times$, \IdxSlowSpeedup$\times$} \\
Checking in user space: total, per certificate (geom.\ mean), largest &
  \multicolumn{2}{r}{\CorpusSpeedup$\times$, \CorpusSpeedupGeo$\times$, \BigSpeedup$\times$} \\
\midrule
Peak memory: median certificate, largest certificate & \MemMedBCF{}\,KB, \BigMemBCF{}\,MB & \MemMedKeel{}\,KB, \BigMemKeel{}\,MB \\
C source, lines & \BCFCheckerTotalLines{} & \PrintedCLines{} \\
Machine code in the kernel (x86-64) & \BCFText{}\,KiB & \KeelText{}\,KiB \\
\bottomrule
\end{tabular}
\end{table}

\S\ref{app:cost:corpus} reports how much of each budget the
certificates use. BCF's checker accepts the forged certificates, which
\Keel{} rejects.

\paragraph{Load time does not meaningfully change.}
Loading an object file is dominated by CVC5, which runs in user space, and by the verifier
itself. The checker takes at most \IdxShareMaxBCF\% of any load with BCF's checker and at most
\IdxShareMaxKeel\% with \Keel{}. Over the \IdxLoaded{} objects that both load, loading with
\Keel{} takes \IdxLoadPct\% of the time it takes with BCF's checker. Replacing BCF's checker with \Keel{} therefore does not change the load time.

\paragraph{Code.}
\Keel{}'s C is printed from the EmbC program: \PrintedCLines{} lines in \PrintedCFunctions{}
functions, against \BCFCheckerTotalLines{} lines for BCF's checker together with its rewrite
table and the macros that expand it. The extra lines are the checks that BCF's checker does not
make: the normalizer behind \textsc{poly\_norm}, the three shapes of a rewrite with entry 0,
the conjuncts we added to the conditions of the table, and the two budgets. The machine code goes the other way. Compiled
into the production kernel by the same compiler, \Keel{} is \KeelText{}\,KiB of code against
\BCFText{}\,KiB for BCF's checker, with the rewrite table as data in both.

\subsection{Where the Speedup and the Extra Memory Come From}
\label{sec:eval:why}

Table~\ref{tab:eval:compare} shows that \Keel{} checks faster than BCF's checker and uses a
little more memory. This subsection asks where both differences come from. We reuse the data from
\S\ref{sec:eval:cost}: the time of every check that the two kernels recorded while loading the
\IdxBcfEval{} object files of BCF's own evaluation, and the time of every corpus certificate in
user space.

\begin{figure}[t]
\centering
\pgfplotstableread[row sep=\\]{
  y name mb speedup \\
  10 bench-1076 5.6 8.14 \\
  9 ext-con-0128 5.3 17.65 \\
  8 ext-con-0064 4.1 15.63 \\
  7 ext-con-0032 3.7 14.13 \\
  6 ext-con-0016 3.4 13.19 \\
  5 bench-3791 3.3 7.35 \\
  4 bench-1075 2.0 6.39 \\
  3 bench-7917 0.8 5.96 \\
  2 bench-7924 0.8 5.95 \\
  1 bench-8021 0.7 5.84 \\
}\topten
\begin{tikzpicture}
\begin{axis}[
  width=7.5cm, height=5cm, xbar, bar width=4pt,
  xmin=0, xmax=21, ymin=0.4, ymax=10.6,
  ytick=data, yticklabels from table={\topten}{name}, yticklabel style={font=\scriptsize\ttfamily},
  xtick={0,5,10,15,20}, xticklabels={0,$5\times$,$10\times$,$15\times$,$20\times$},
  tick label style={font=\scriptsize}, label style={font=\footnotesize},
  xlabel={speedup, BCF's checker / \Keel}, grid=major, grid style={black!10},
  nodes near coords, nodes near coords style={font=\scriptsize, /pgf/number format/.cd, fixed, precision=1, /tikz/.cd},
  point meta=explicit, every node near coord/.append style={anchor=west}]
\addplot[fill=keelblue, draw=none] table[x=speedup, y=y, meta=speedup] {\topten};
\end{axis}
\end{tikzpicture}
\caption{The speedup of \Keel{} over BCF's checker on the ten largest corpus certificates that both
checkers accept, from \BigCertMB{}\,MB (top) to 0.7\,MB (bottom).}
\label{fig:eval:topten}
\end{figure}
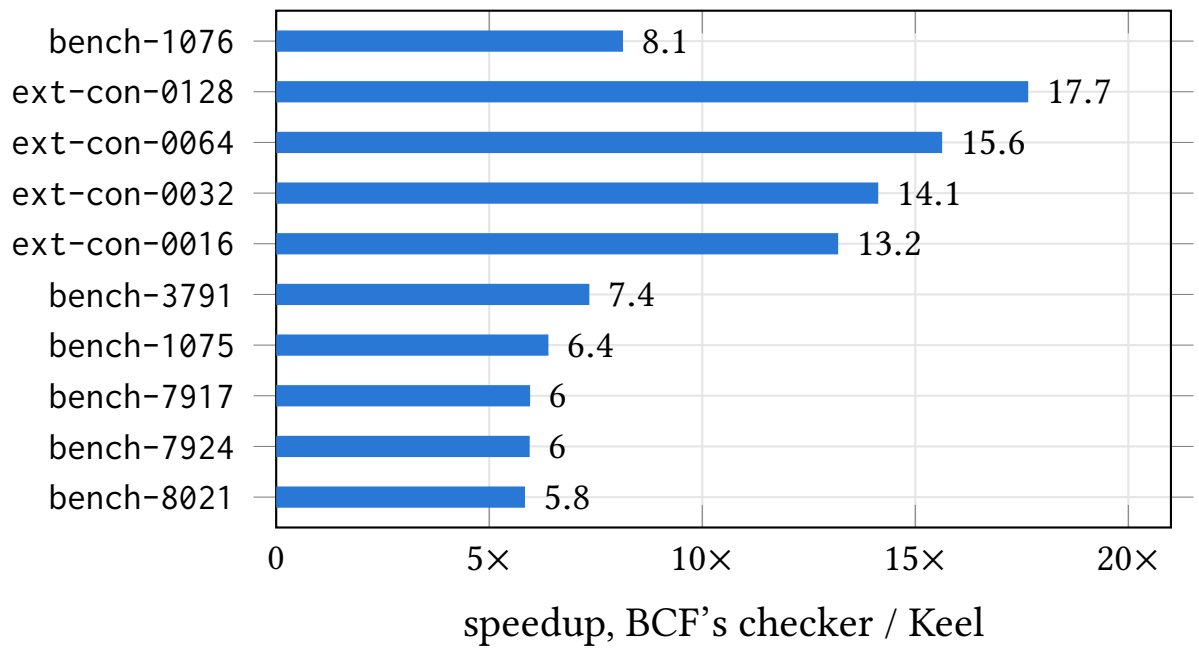

\paragraph{Where.}
The certificates that the verifier sends while loading the object files are small:
\XCertMedBytes{} bytes at the median and \XCertMaxKB{}\,KB at most. Below 2\,KB both checkers
take about the same time, which goes into copying the certificate out of user memory and
setting up the checker's state. At \PfKernelSizedLo{}--\PfKernelSizedHi{}\,KB, \Keel{} is
\WhyKernelLargeSpeedup$\times$ faster. The corpus certificates are larger, up to
\BigCertMB{}\,MB, and on the ten largest \Keel{} is \WhyTopTenLo{} to \WhyTopTenHi$\times$
faster (Figure~\ref{fig:eval:topten}).

\begin{table}[t]
\caption{Where BCF's checker spends its time: the share of its profile samples by activity, in user space, on the ten largest corpus certificates and on five corpus certificates
of \PfKernelSizedLo{}--\PfKernelSizedHi{}\,KB.}
\label{tab:eval:profile}
\small
\begin{tabular}{@{}lrr@{}}
\toprule
 & ten largest & \PfKernelSizedLo{}--\PfKernelSizedHi{}\,KB \\
\midrule
comparing two terms node by node & \PfBcfEquiv\% & \PfBcfEquivK\% \\
finding a term by its position in the \texttt{xarray}, with its lock & \PfBcfXarray\% & \PfBcfXarrayK\% \\
allocating, reference-counting and freeing terms & \PfBcfAlloc\% & \PfBcfAllocK\% \\
checking the rules & \PfBcfRulesAll\% & \PfBcfRulesAllK\% \\
reading the certificate & \PfBcfInput\% & \PfBcfInputK\% \\
\bottomrule
\end{tabular}
\end{table}

\paragraph{Why.}
To understand why \Keel{} is faster than BCF's checker,
we profiled both checkers on the ten largest certificates and on
five corpus certificates of \PfKernelSizedLo{}--\PfKernelSizedHi{}\,KB (Table~\ref{tab:eval:profile}).
Most of the difference comes from two decisions about how to store terms, which the two
checkers make differently. In both, \Keel{}'s choice makes it faster but costs it some memory.

First, the checkers differ in whether equal terms share a position. BCF's checker stores every
new term that a rule builds at a new position, even when an equal term is already stored. Equal
terms can therefore have different positions. To check whether two terms are equal, BCF's
checker must compare their structure. It does so with a depth-first search that descends into
their subterms, and every call of this search first clears a \WhyVarMapKB{}\,KB scratch table.
On the largest certificates, these calls take \PfBcfEquiv\% of BCF's checker's time. On each of
the ten largest, they alone take longer than \Keel{}'s whole check, \WhyEquivVsKeelLo{} to
\WhyEquivVsKeelHi{} times as long. On \texttt{bench\_1076} alone, BCF's checker makes
\WhyEquivCallsM{} million such calls. Clearing the scratch table is a
small detail that BCF's checker could change, but storing equal terms at different positions is
part of its design.

\Keel{} never stores a term twice. Before it stores a new term, it looks the term up in a hash
table and reuses the position of an equal term (\S\ref{app:refine:storage}). Two terms are
therefore equal exactly when their positions are equal, and \Keel{} checks equality with one
comparison of two integers. What this costs is the memory of the hash table, which BCF's
checker does not have. On the median certificate, the table takes \WhyHtabKB{}\,KB of
\Keel{}'s \MemMedKeel{}\,KB.

Second, the checkers differ in how they allocate terms, and therefore in how they free them.
BCF's checker allocates every term it builds separately and keeps track of it in the
\texttt{xarray}, a general-purpose kernel search tree from numbers to memory addresses. Because
each term is a separate allocation, BCF's checker can free a term as soon as it is not part of a
conclusion that a later step still cites. However, this design is costly in time. Finding a term by its
position walks down the tree through several nodes, each a separate memory access. Adding or
removing a term takes the \texttt{xarray}'s lock, because the \texttt{xarray} is built to be
shared between threads, although the checker uses it from one thread only. In addition, each
term is allocated, reference-counted and freed on its own. On the certificates of
\PfKernelSizedLo{}--\PfKernelSizedHi{}\,KB, the \texttt{xarray} and the allocations take about
two thirds of BCF's checker's run time (Table~\ref{tab:eval:profile}).

\Keel{} keeps all its terms in one array (\S\ref{app:refine:storage}). Finding a term is one read of
that array, and adding one takes no lock. This difference also explains why, on the smallest
certificates, the corpus shows a larger speedup than the loads.
In user space, BCF's checker runs on a port that replaces the kernel's allocator and the
\texttt{xarray} with a slower library. \Keel{} needs only plain allocation, so it runs alike in
both.

The array, however, costs \Keel{} memory on large certificates. \Keel{} does not free terms one at
a time, because a freed term would leave unused units in the middle of the array. Instead, it
frees them in batches. When it compacts the array, it drops every term that is not part of a
conclusion that a later step still cites (\S\ref{app:refine:compact}). Until then, these terms stay
in memory, and compaction itself needs some extra room while it moves the remaining terms.
These terms and this room account for \Keel{}'s extra memory on large certificates. We measure in
\S\ref{sec:eval:sensitivity} how much memory compaction saves.

Finally, BCF's checker accepts \ProgSetTrustedPct\% of the steps the kernel sees without
checking them (\S\ref{sec:eval:bug1}), while \Keel{} checks them. Thus, the comparison
in fact understates \Keel{}'s advantage per checked step.

Overall, both decisions make \Keel{} faster at some cost in memory, and we believe the tradeoff is
worth it. On the largest certificate, \Keel{} uses \BigMemExtraPct\% more memory than BCF's
checker and checks it \BigSpeedup$\times$ faster.

\subsection{Sensitivity to Design Decisions}
\label{sec:eval:sensitivity}

\paragraph{Compaction.}
Compaction (\S\ref{app:refine:compact}) keeps \Keel{}'s memory on the same scale as BCF's
checker. It runs only when a certificate's terms outgrow a fixed threshold, which happens on
\CompRuns{} of the \CorpusShared{} corpus certificates, all of at least \CompRunsMinKB{}\,KB.
On every smaller certificate the two versions of \Keel{} behave the same.
With compaction, \Keel{}'s peak stays \CompExtraMinMB{} to \CompExtraMaxMB{}\,MB above BCF's
checker's at every size (Figure~\ref{fig:eval:compaction} of the main paper), for the reasons given in
\S\ref{sec:eval:why}.
On the \CompBigN{} certificates above 2\,MB, it keeps \Keel{}'s peak within
\CompBigKeelBCF$\times$ of BCF's checker's. Without it, the peak grows to
\CompBigOffBCF$\times$, \CompMemOff{}\,MB against \BigMemBCF{}\,MB on the largest
(the same figure). Overall, compaction actually saves time. The \CompRuns{}
certificates take \CompNetSavePct\% less time with it, because the terms still in use stay packed
together in memory, which reduces the cache misses on the largest certificate. It does add run time to smaller certificates.
On the \CompMidN{} certificates of 0.3--0.8\,MB, the compaction adds about \CompMidExtraMs{}\,ms to a check of
\CompMidKeelMs{}\,ms, and \Keel{} is still \CompMidSpeedup$\times$ faster than BCF's checker
there. On the two smallest, compaction raises the peak.
This is because when the certificates are small, the memory saved by compaction doesn't offset the
cost of compaction itself. However, the cost is small at only a few hundred kilobytes, so the effects are
negligible.

\paragraph{The counters.}
The counters cost very little. On the certificates of 1 to 256\,KB, removing them changes
the time by about \CountersSmallPct\%, which is within the \RecompilePct\% that merely
recompiling \Keel{} with a different constant causes.

\subsection{Extending a Verified Checker}
\label{sec:eval:rules}

The proof cost for changing verified code is a common concern since every change to the code needs
a change to its proof. In our case, CVC5 indeed adds new rewrite rules over time, which \Keel{} must be able
to follow. In \Keel{}, a rewrite is an entry
of a table that the C code interprets, and each entry comes with a soundness lemma
(\S\ref{sec:checker}). To see what adding rules costs, we added \NumRewritesAdded{} rewrites
from CVC5's rewrite rules that BCF's table lacks: Boolean De~Morgan and distributivity laws, a
simplification of exclusive or, multiplication by a power of two, and five laws of unsigned
division and remainder.

\paragraph{The table.}
\Keel{}'s table has \NumRewritesKeel{} entries: the format's \NumRewritesBCF{}, transcribed
from BCF's header by a script so that they are faithful by construction, with the lemma
statements and the dispatch from entry number to lemma generated as well, and the
\NumRewritesAdded{} rules above. The soundness proof never looks inside the table; it uses
only the soundness of each entry (\S\ref{sec:checker}), and the C code interprets the same
table as data, so adding an entry means proving that it is sound and nothing else. The format
also reserves entry 0 for any rewrite outside the table (\S\ref{sec:background}); \Keel{}
accepts it for three shapes of equation, each a verified lemma: splitting a bitwise \emph{and}
at a bit position, unfolding repeat into a concatenation, and turning distinct into a negated
equality. These are the shapes that occur in practice: of the \RwZeroSteps{} entry-0 steps in
the corpus, \RwZeroShapes{} have one of the three, and most of the rest are instances of the
\NumRewritesAdded{} added rules.

\paragraph{Cost.}
The refinement proof needed no new lemmas, only two one-line edits,
one of which adds the two division operators to the operators the pattern language accepts.
The cost proof mentions the table in one lemma, that every entry fits in a fixed size, which is
checked by computation over the whole table and needs no edit when an entry is added. A change to
the table rebuilds nearly every proof, since all of them depend on it, and rechecking everything
took \RebuildMinutes{} minutes.

\subsection{Proof Effort}
\label{sec:eval:effort}

\Keel{} is \LeanCheckerLines{} lines of Lean for the checker and its soundness proof
(Theorem~\ref{thm:sound:formal}), about \LeanCheckerCodeLines{} of them code and
\LeanCheckerProofLines{} proof, half of the proof for the rewrite table and about \BitblastProofLines{} lines for the
bit-blaster; \EmbCFrameworkLines{} for EmbC's syntax, semantics, printer and
cost semantics, and \EmbCProgramLines{} for the checker as an EmbC program, which prints to
\PrintedCLines{} lines of C in \PrintedCFunctions{} functions. The refinement and cost proofs are together \ProofToProgramRatio{} times the size of the
program they are about. A clean build of everything takes under \BuildTotalMin{} minutes on an Apple M4 Pro
with 12 cores, \BuildDefaultSec{} seconds for the Lean checker, its soundness proof, the
Clight correspondence and the EmbC program, and \BuildProofsMin{} minutes for the refinement and cost proofs; the proofs use
only Lean's three standard axioms.

\subsection{Measurement Details}
\label{app:method}

The \IdxObjects{} object files are each compiled with several versions of clang at several
optimization levels. Unless stated otherwise, performance measurements run on one CloudLab server
with an \XCPU{} (16 cores) and 128\,GB of memory, under Ubuntu 22.04 with Linux 5.15.

Each object is loaded with BCF's own load script unchanged: with the program type listed in
BCF's index, pinned to one core at the highest priority, once to warm up and then five times,
reporting the median of the five. A load is stopped after \IdxCapMin{} minutes, the cap BCF's
own evaluation used; no load reaches it. The \IdxSingleRun{} objects that the verifier takes
more than ten minutes to reject were loaded once instead of six times. \XVMs{} virtual
machines run in parallel on the \XCPU{}, each pinned to one physical core, and each core loads
the same objects on both kernels one after the other, so that interference from the other cores
affects both checkers alike. Outside the kernel, BCF's own benchmark times only the call of the
checker, twice to warm up and then ten times; \Keel{} is compiled from the same C file that
runs in the kernel with stand-ins for the kernel APIs. The compiled C is also checked against
the EmbC semantics directly: on all \CorpusShipped{} certificates, and on \InterpAllocRuns{}
runs in which kernel allocations fail at chosen points, the compiled C and the EmbC interpreter
give the same results.

The run that loads programs with forged certificates (\S\ref{sec:eval}) is separate: an
arm64 kernel v7.0-rc1 with BCF's patch sets 1--5 in QEMU. The loader replaces CVC5's certificates for
two of BCF's example programs, \texttt{trace\_sys\_enter\_execve} and our running example
\texttt{test\_get\_stack\_rawtp}, with certificates forged through rewrite 80 (Bug\#3). The kernel
with BCF's checker loads both programs. The kernel with \Keel{} rejects the forged certificates and
accepts CVC5's.

\end{document}